\documentclass[fleqn,twoside]{article}
\usepackage{espcrc1}

\newcommand{\AmS}{{\protect\the\textfont2
  A\kern-.1667em\lower.5ex\hbox{M}\kern-.125emS}}
\title{Morphological methods for design of  modular systems (a survey)
}
\author{Mark Sh. Levin
\thanks{
 E-mail address: mslevin@acm.org.
  ~Http://www.mslevin.iitp.ru/
  }
 }

\begin{document}

\maketitle

\begin{abstract}
The article addresses
%
 morphological approaches to design of modular systems.
%
 The following methods are briefly described:
 (i) basic version of morphological analysis (MA),
 (ii) modification of MA as method of closeness to ideal point(s),
 (iii) reducing of MA to linear programming,
 (iv)
 multiple choice problem,
 (v) quadratic assignment problem,
 (vi) Pareto-based MA
  (i.e., revelation of Pareto-efficient solutions),
%
%
 (vii) Hierarchical Morphological
 Multicriteria Design (HMMD) approach, and
 (viii) Hierarchical Morphological
 Multicriteria Design (HMMD) approach based on
 fuzzy estimates.
 The above-mentioned methods are illustrated by schemes, models,
 and
 illustrative examples.
 An additional realistic example (design of GSM network)
 is presented to illustrate main considered methods.

~~~~~~~~~~~


\vspace{1pc}
\end{abstract}

\newcounter{cms}
\setlength{\unitlength}{1mm}


\section{Introduction}


%
%
%
%
%

%
%
%

 Morphological analysis (MA) was firstly suggested by F. Zwicky in 1943
 for design of aerospace systems.
 Morphological analysis is a well-known general powerful method to
 synthesis of modular systems in various domains
 (e.g., \cite{ayr69}, \cite{jon81}, \cite{lev98}, \cite{ritchey06}, \cite{zwi69}).
 MA is based on {\it divide and conquer} technique.
 A hierarchical structure of the designed system is a basis for
 usage of the method.
 The following basic partitioning techniques can be used to obtain the
 required system hierarchical model:
 (a) partitioning by system component/parts,
 (b) partitioning by system functions,
 (c) partitioning by system properties/attributes, and
 (d) integrated techniques.
 In this article, system hierarchy
  of system
 components (parts, subsystems) is considered as a basic one.
 This case corresponds to modular systems which are widely used
 in many domains of engineering, information technology, and management
 (e.g., \cite{bald00}, \cite{hua98}, \cite{kus99}, \cite{lev98}, \cite{lev06},
  \cite{pah88}, \cite{ul95}).
%
%
%
%
%
%
 Many years the usage of morphological analysis in system design
 was very limited by the reason that the method leads to a very
 large combinatorial domain of possible solutions.
 On the other hand, contemporary computer systems can solve
 very complex computational problems and
 hierarchical system models can be used as a basis for
 partitioning/decomposition solving frameworks.
 Recent
 trends in the study, usage, and modification/extension
 of morphological analysis may be considered as the following:
 (1) hierarchical systems modeling,
 (2) optimization models,
 (3) multicriteria decision making, and
 (4) taking into account uncertainty
 (i.e.,
 probabilistic and/or
 fuzzy estimates).

 In the article,
 the following system design methods are briefly described:
 (i) basic morphological analysis (as
 morphological generation of admissible composite solutions),
 (ii) modification of MA as method of closeness to ideal point(s),
 (iii) reducing of morphological analysis to optimization model as
 linear programming,
 (iv) multiple choice problem,
 (v) quadratic assignment problem,
 (vi) multicriteria analysis of morphological decisions with revelation
 of Pareto-efficient solutions,
%
%
 (vii) Hierarchical Morphological
 Multicriteria Design (HMMD) approach, and
 (viii) version of Hierarchical Morphological
 Multicriteria Design (HMMD) approach based on
 fuzzy estimates.
 The above-mentioned methods are illustrated by
 solving schemes, mathematical models,
 and
 illustrative numerical examples.
%
 Preliminary materials (a description of the morphological
 methods and an example for GSM network)
 were published in
 \cite{lev09card}, \cite{levvis07}.
%
%



\section{Basic Configuration Problem}


 Generally, morphological system design approaches are targeted
 to design of system configuration
 as a selection of alternatives for systems parts
 (e.g., \cite{lev09}).
 Fig. 1 illustrates this problem.
 Here a composite (modular)
 system consists of \(m\) system parts:
 \( \{  P(1), ... ,P(i), ... ,P(m)  \} \).
 For each system part (i.e., \(\forall i, ~i=\overline{1,m}\))~
 there are corresponding alternatives
 \( \{ X^{i}_{1}, X^{i}_{2}, ... , X^{i}_{q_{i}} \} \), where
 \(q_{i}\) is the number of alternatives for part \(i\).
 Thus, the problem is:

 {\it Select an alternative for each system part
 while taking into account some
 local and/or global
 objectives/preferences and constraints.}
 Evidently, the objective/prereferences and constraints
 are based on (correspond to)
 quality of the selected alternatives and quality of compatibility among
 the selected alternatives.
 In \cite{lev09},
  some other system configuration problems
 are described as well (e.g., reconfiguration, selection and
 allocation).

\begin{center}
\begin{picture}(74,51)

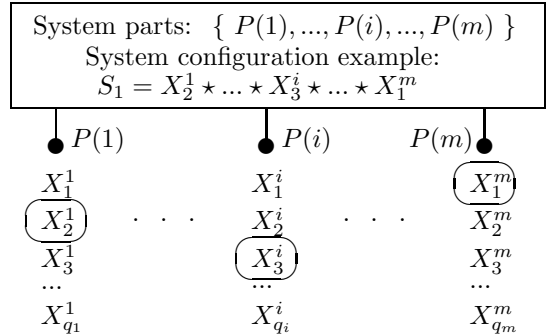
\put(0,00){\makebox(0,0)[bl] {Fig. 1. System configuration problem
(selection)
}}

\put(4,5){\makebox(0,8)[bl]{\(X^{1}_{q_{1}}\)}}
\put(4,11){\makebox(0,8)[bl]{\(. . .\)}}
\put(4,13){\makebox(0,8)[bl]{\(X^{1}_{3}\)}}
\put(4,18){\makebox(0,8)[bl]{\(X^{1}_{2}\)}}
\put(4,23){\makebox(0,8)[bl]{\(X^{1}_{1}\)}}
\put(06,20){\oval(8,5.6)}

\put(32,5){\makebox(0,8)[bl]{\(X^{i}_{q_{i}}\)}}
\put(32,11){\makebox(0,8)[bl]{\(. . .\)}}
\put(32,13){\makebox(0,8)[bl]{\(X^{i}_{3}\)}}
\put(32,18){\makebox(0,8)[bl]{\(X^{i}_{2}\)}}
\put(32,23){\makebox(0,8)[bl]{\(X^{i}_{1}\)}}
\put(34,15){\oval(8,5.6)}

\put(61,5){\makebox(0,8)[bl]{\(X^{m}_{q_{m}}\)}}
\put(61,11){\makebox(0,8)[bl]{\(. . .\)}}
\put(61,13){\makebox(0,8)[bl]{\(X^{m}_{3}\)}}
\put(61,18){\makebox(0,8)[bl]{\(X^{m}_{2}\)}}
\put(61,23){\makebox(0,8)[bl]{\(X^{m}_{1}\)}}
\put(63,25){\oval(8,5.6)}

\put(6,30){\circle*{2}} \put(34,30){\circle*{2}}
\put(63,30){\circle*{2}}

\put(06,31){\line(0,1){4}} \put(34,31){\line(0,1){4}}
\put(63,31){\line(0,1){4}}

\put(08,29){\makebox(0,8)[bl]{\(P(1)\) }}
\put(36,29){\makebox(0,8)[bl]{\(P(i)\) }}
\put(53,29){\makebox(0,8)[bl]{\(P(m)\) }}


\put(15,20){\makebox(0,8)[bl]{ .~ .~ .}}
\put(43,20){\makebox(0,8)[bl]{ .~ .~ .}}


\put(00,35){\line(1,0){71}} \put(00,49){\line(1,0){71}}
\put(00,35){\line(0,1){14}} \put(71,35){\line(0,1){14}}

\put(03,44){\makebox(0,8)[bl]{System parts:~  \( \{ ~ P(1), ...
,P(i), ... ,P(m) ~ \} \) }}


\put(10,40){\makebox(0,8)[bl]{System configuration example:}}

\put(08,36){\makebox(0,8)[bl]{~ ~\(S_{1}=X^{1}_{2}\star ... \star
X^{i}_{3} \star ... \star X^{m}_{1}\)}}

\end{picture}
\end{center}



\section{Morphological Design Approaches}

 Our basic list of morphological design approaches consists of the
 following:
%
%
 (1) the basic version of morphological analysis
 (MA)
 (e.g., \cite{bela00}, \cite{boon05}, \cite{jon81},
 \cite{ritchey06},
 \cite{zwi69});
 (2) the modification of morphological analysis as searching for
  an admissible (by compatibility) element combination
  (one representative from each morphological class, i.e.,
  a set of alternatives for system part/component)
  that is the closest
  to a combination consisting of the best elements
  (at each morphological class)
  (e.g., \cite{ayr69}, \cite{dubov86});
 (3) modification of morphological analysis via reducing to
  linear programming (MA\&linear programming)
   \cite{kol86};
 (4) modification of morphological analysis via reducing to
  multiple choice problem (MCP) or
  multicriteria multiple choice problem
  (e.g.,
   \cite{levsaf10}, \cite{sousa06});
 (5) modification of morphological analysis via reducing to
  quadratic assignment problem (QAP)
  (e.g., \cite{cela98}, \cite{lev98}, \cite{lev09});
 (6) the multicriteria modification of morphological analysis as follows
 (Pareto-based MA):
  ~(a) searching for all admissible (by compatibility) elements combinations
      (one representative from each morphological class),
  ~(b) evaluation of the found combinations upon a set of
  criteria,
      and
   ~(c) selection of the Pareto-efficient solutions
      (e.g., \cite{emel72}, \cite{gaft82});
%
%
%
%
 (7) hierarchical morphological multicriteria design
 (HMMD) approach
  (\cite{lev98}, \cite{lev06}); and
 (8) version of hierarchical morphological multicriteria design
 approach with probabilistic and/or fuzzy estimates
 (HMMD\&uncertainty)
  \cite{lev98}.
 Table 1 contains some properties of the approaches above.

\begin{center}
\begin{picture}(153,48)
\put(47,43){\makebox(0,0)[bl]{Table 1. Description of
 approaches}}

\put(00,0){\line(1,0){153}} \put(00,35){\line(1,0){153}}
\put(00,41){\line(1,0){153}}

\put(00,0){\line(0,1){41}} \put(44,00){\line(0,1){41}}
\put(68,00){\line(0,1){41}} \put(92,00){\line(0,1){41}}
\put(131,0){\line(0,1){41}} \put(153,0){\line(0,1){41}}

\put(1,37){\makebox(0,0)[bl]{Method}}
\put(45,37){\makebox(0,0)[bl]{Scale for DAs}}
\put(69,37){\makebox(0,0)[bl]{Scale for IC}}
\put(93,37){\makebox(0,0)[bl]{Quality for decision}}
\put(132,37){\makebox(0,0)[bl]{References}}


\put(1,30){\makebox(0,0)[bl]{1.MA}}
\put(45,30){\makebox(0,0)[bl]{None}}
\put(69,29.5){\makebox(0,0)[bl]{\(\{0,1\}\)}}
\put(93,30){\makebox(0,0)[bl]{Admissibility}}
\put(132,30){\makebox(0,0)[bl]{\cite{jon81},
 \cite{ritchey06},
  \cite{zwi69}}}


\put(1,26){\makebox(0,0)[bl]{2.Ideal-point method}}
\put(45,26){\makebox(0,0)[bl]{None}}
\put(69,25.5){\makebox(0,0)[bl]{\(\{0,1\}\)}}
\put(93,26){\makebox(0,0)[bl]{Closeness to ideal point}}
\put(132,26){\makebox(0,0)[bl]{\cite{ayr69}, \cite{dubov86}}}


\put(01,22){\makebox(0,0)[bl]{3.MA\&linear programming}}
\put(45,22){\makebox(0,0)[bl]{Quantitative}}
\put(69,21.5){\makebox(0,0)[bl]{\(\{0,1\}\)}}
\put(93,22){\makebox(0,0)[bl]{Additive function}}
\put(132,21.5){\makebox(0,0)[bl]{\cite{kol86} }}


\put(01,18){\makebox(0,0)[bl]{4.Multiple choice problem}}
\put(45,18){\makebox(0,0)[bl]{Quantitative}}
\put(69,18){\makebox(0,0)[bl]{None}}
\put(93,18){\makebox(0,0)[bl]{Additive function}}
\put(132,17.5){\makebox(0,0)[bl]{\cite{levsaf10},
 \cite{sousa06}
}}


\put(01,14){\makebox(0,0)[bl]{5.QAP}}
\put(45,14){\makebox(0,0)[bl]{Quantitative}}
\put(69,14){\makebox(0,0)[bl]{Quantitative}}
\put(93,14){\makebox(0,0)[bl]{Additive function}}
\put(132,13.5){\makebox(0,0)[bl]{\cite{cela98}, \cite{lev09}}}


\put(01,10){\makebox(0,0)[bl]{6.Pareto-based MA }}
\put(45,10){\makebox(0,0)[bl]{None}}
\put(69,9.5){\makebox(0,0)[bl]{\(\{0,1\}\)}}
\put(93,10){\makebox(0,0)[bl]{Multicriteria}}
\put(132,09.5){\makebox(0,0)[bl]{\cite{emel72}, \cite{gaft82}}}




\put(01,06){\makebox(0,0)[bl]{7.HMMD}}
\put(45,06){\makebox(0,0)[bl]{Ordinal}}
\put(69,06){\makebox(0,0)[bl]{Ordinal}}
\put(93,06){\makebox(0,0)[bl]{Multicriteria}}
\put(132,05.5){\makebox(0,0)[bl]{\cite{lev98}, \cite{lev06}}}

\put(01,02){\makebox(0,0)[bl]{8.HMMD \& uncertainty }}
\put(45,01.4){\makebox(0,0)[bl]{Ordinal/fuzzy}}
\put(69,01.4){\makebox(0,0)[bl]{Ordinal/fuzzy}}
\put(93,02){\makebox(0,0)[bl]{Multicriteria}}
\put(132,01.5){\makebox(0,0)[bl]{\cite{lev98}}}


\end{picture}
\end{center}


\subsection{Morphological Analysis}

%
 The MA approach consists of the following stages:

 {\it Stage 1.} Building a system structure as a set of system
 parts/components.

 {\it Stage 2.} Generation of design alternatives (DAs) for each system
 part (i.e., a morphological class).

 {\it Stage 3.} Binary assessment of
 compatibility for each DAs pair (one DA from one morphological
 class, other DA from another morphological class).
 Value of compatibility 1 corresponds to compatibility of two
 corresponding DAs, value 0 corresponds to incompatibility.

 {\it Stage 4.} Generation of all admissible compositions
 (one DA for each system part)
 while taking into account compatibility for each two
 DAs in the composition obtained.

 The method above is an enumerative one.
 Fig. 2 illustrates MA (binary compatibility estimates are
 depicted in Table 2).


\begin{center}
\begin{picture}(80,42)

\put(18,00){\makebox(0,0)[bl] {Fig. 2. Illustration for MA}}

\put(4,5){\makebox(0,8)[bl]{\(X^{1}_{5}\)}}
\put(4,10){\makebox(0,8)[bl]{\(X^{1}_{4}\)}}
\put(4,15){\makebox(0,8)[bl]{\(X^{1}_{3}\)}}
\put(4,20){\makebox(0,8)[bl]{\(X^{1}_{2}\)}}
\put(4,25){\makebox(0,8)[bl]{\(X^{1}_{1}\)}}

\put(6,31){\line(-1,0){6}}

\put(0,31){\line(0,-1){24}}

\put(0,27){\line(1,0){2}} \put(2.5,27){\circle{1}}
\put(0,22){\line(1,0){2}} \put(2.5,22){\circle{1}}
\put(0,17){\line(1,0){2}} \put(2.5,17){\circle{1}}
\put(0,12){\line(1,0){2}} \put(2.5,12){\circle{1}}
\put(0,7){\line(1,0){2}} \put(2.5,7){\circle{1}}


\put(09,22.5){\line(2,1){22}}

\put(31,33.5){\line(1,0){10}}

\put(41,33.5){\line(2,-1){11}}

\put(11,7){\line(2,3){14}}

\put(11,22){\line(3,-1){14}}

\put(11,17){\line(3,-2){14}}


\put(32,5){\makebox(0,8)[bl]{\(X^{i}_{5}\)}}
\put(32,10){\makebox(0,8)[bl]{\(X^{i}_{4}\)}}
\put(32,15){\makebox(0,8)[bl]{\(X^{i}_{3}\)}}
\put(32,20){\makebox(0,8)[bl]{\(X^{i}_{2}\)}}
\put(32,25){\makebox(0,8)[bl]{\(X^{i}_{1}\)}}

\put(34,31){\line(-1,0){6}}

\put(28,31){\line(0,-1){24}}

\put(28,27){\line(1,0){2}} \put(30.5,27){\circle{1}}
\put(28,22){\line(1,0){2}} \put(30.5,22){\circle{1}}
\put(28,17){\line(1,0){2}} \put(30.5,17){\circle{1}}
\put(28,12){\line(1,0){2}} \put(30.5,12){\circle{1}}
\put(28,7){\line(1,0){2}} \put(30.5,7){\circle{1}}



\put(39,7){\line(2,3){14}}

\put(39,17){\line(1,0){14}}

\put(37,17){\line(3,2){14}}


\put(61,15){\makebox(0,8)[bl]{\(X^{m}_{3}\)}}
\put(61,20){\makebox(0,8)[bl]{\(X^{m}_{2}\)}}
\put(61,25){\makebox(0,8)[bl]{\(X^{m}_{1}\)}}

\put(63,31){\line(-1,0){6}}

\put(57,31){\line(0,-1){14}}

\put(57,27){\line(1,0){2}} \put(59.5,27){\circle{1}}
\put(57,22){\line(1,0){2}} \put(59.5,22){\circle{1}}
\put(57,17){\line(1,0){2}} \put(59.5,17){\circle{1}}



\put(6,31){\circle*{2}} \put(34,31){\circle*{2}}
\put(63,31){\circle*{2}}

\put(06,31){\line(0,1){4}} \put(34,31){\line(0,1){4}}
\put(63,31){\line(0,1){4}}

\put(08,29){\makebox(0,8)[bl]{\(P(1)\) }}
\put(36,29){\makebox(0,8)[bl]{\(P(i)\) }}
\put(65,29){\makebox(0,8)[bl]{\(P(m)\) }}


\put(17,32){\makebox(0,8)[bl]{ .~ .~ .}}
\put(45,32){\makebox(0,8)[bl]{ .~ .~ .}}


\put(00,35){\line(1,0){71}} \put(00,41){\line(1,0){71}}
\put(00,35){\line(0,1){06}} \put(71,35){\line(0,1){06}}




\put(04,36){\makebox(0,8)[bl]{Example:  ~~~\(S_{1}=X^{1}_{2}\star
... \star X^{i}_{3} \star ... \star X^{m}_{1}\)}}

\end{picture}
%
\begin{picture}(57,55)
\put(05,50){\makebox(0,0)[bl]{Table 2. Binary compatibility}}

\put(00,00){\line(1,0){57}} \put(00,42){\line(1,0){57}}
\put(00,48){\line(1,0){57}}

\put(00,0){\line(0,1){48}} \put(06,0){\line(0,1){48}}
\put(57,0){\line(0,1){48}}


\put(12,42){\line(0,1){6}} \put(18,42){\line(0,1){6}}
\put(24,42){\line(0,1){6}} \put(30,42){\line(0,1){6}}
\put(36,42){\line(0,1){6}} \put(43,42){\line(0,1){6}}
\put(50,42){\line(0,1){6}}

\put(0.7,37){\makebox(0,8)[bl]{\(X^{1}_{1}\)}}
\put(0.7,33){\makebox(0,8)[bl]{\(X^{1}_{2}\)}}
\put(0.7,29){\makebox(0,8)[bl]{\(X^{1}_{3}\)}}
\put(0.7,25){\makebox(0,8)[bl]{\(X^{1}_{4}\)}}
\put(0.7,21){\makebox(0,8)[bl]{\(X^{1}_{5}\)}}
\put(0.7,17){\makebox(0,8)[bl]{\(X^{i}_{1}\)}}
\put(0.7,13){\makebox(0,8)[bl]{\(X^{i}_{2}\)}}
\put(0.7,09){\makebox(0,8)[bl]{\(X^{i}_{3}\)}}
\put(0.7,05){\makebox(0,8)[bl]{\(X^{i}_{4}\)}}
\put(0.7,01){\makebox(0,8)[bl]{\(X^{i}_{5}\)}}


\put(08,38){\makebox(0,8)[bl]{\(0\)}}
\put(14,38){\makebox(0,8)[bl]{\(0\)}}
\put(20,38){\makebox(0,8)[bl]{\(0\)}}
\put(26,38){\makebox(0,8)[bl]{\(0\)}}
\put(32,38){\makebox(0,8)[bl]{\(0\)}}

\put(38,38){\makebox(0,8)[bl]{\(0\)}}
\put(45,38){\makebox(0,8)[bl]{\(0\)}}
\put(52,38){\makebox(0,8)[bl]{\(0\)}}


\put(08,34){\makebox(0,8)[bl]{\(0\)}}
\put(14,34){\makebox(0,8)[bl]{\(0\)}}
\put(20,34){\makebox(0,8)[bl]{\(1\)}}
\put(26,34){\makebox(0,8)[bl]{\(0\)}}
\put(32,34){\makebox(0,8)[bl]{\(0\)}}

\put(38,34){\makebox(0,8)[bl]{\(1\)}}
\put(45,34){\makebox(0,8)[bl]{\(0\)}}
\put(52,34){\makebox(0,8)[bl]{\(0\)}}


\put(08,30){\makebox(0,8)[bl]{\(0\)}}
\put(14,30){\makebox(0,8)[bl]{\(0\)}}
\put(20,30){\makebox(0,8)[bl]{\(0\)}}
\put(26,30){\makebox(0,8)[bl]{\(0\)}}
\put(32,30){\makebox(0,8)[bl]{\(1\)}}

\put(38,30){\makebox(0,8)[bl]{\(0\)}}
\put(45,30){\makebox(0,8)[bl]{\(0\)}}
\put(52,30){\makebox(0,8)[bl]{\(0\)}}


\put(08,26){\makebox(0,8)[bl]{\(0\)}}
\put(14,26){\makebox(0,8)[bl]{\(0\)}}
\put(20,26){\makebox(0,8)[bl]{\(0\)}}
\put(26,26){\makebox(0,8)[bl]{\(0\)}}
\put(32,26){\makebox(0,8)[bl]{\(0\)}}

\put(38,26){\makebox(0,8)[bl]{\(0\)}}
\put(45,26){\makebox(0,8)[bl]{\(0\)}}
\put(52,26){\makebox(0,8)[bl]{\(0\)}}


\put(08,22){\makebox(0,8)[bl]{\(1\)}}
\put(14,22){\makebox(0,8)[bl]{\(0\)}}
\put(20,22){\makebox(0,8)[bl]{\(0\)}}
\put(26,22){\makebox(0,8)[bl]{\(0\)}}
\put(32,22){\makebox(0,8)[bl]{\(0\)}}

\put(38,22){\makebox(0,8)[bl]{\(0\)}}
\put(45,22){\makebox(0,8)[bl]{\(0\)}}
\put(52,22){\makebox(0,8)[bl]{\(0\)}}


\put(38,18){\makebox(0,8)[bl]{\(0\)}}
\put(45,18){\makebox(0,8)[bl]{\(0\)}}
\put(52,18){\makebox(0,8)[bl]{\(0\)}}

\put(38,14){\makebox(0,8)[bl]{\(0\)}}
\put(45,14){\makebox(0,8)[bl]{\(0\)}}
\put(52,14){\makebox(0,8)[bl]{\(0\)}}

\put(38,10){\makebox(0,8)[bl]{\(1\)}}
\put(45,10){\makebox(0,8)[bl]{\(0\)}}
\put(52,10){\makebox(0,8)[bl]{\(1\)}}

\put(38,06){\makebox(0,8)[bl]{\(0\)}}
\put(45,06){\makebox(0,8)[bl]{\(0\)}}
\put(52,06){\makebox(0,8)[bl]{\(0\)}}

\put(38,02){\makebox(0,8)[bl]{\(1\)}}
\put(45,02){\makebox(0,8)[bl]{\(0\)}}
\put(52,02){\makebox(0,8)[bl]{\(0\)}}


\put(07,43){\makebox(0,8)[bl]{\(X^{i}_{1}\)}}
\put(13,43){\makebox(0,8)[bl]{\(X^{i}_{2}\)}}
\put(19,43){\makebox(0,8)[bl]{\(X^{i}_{3}\)}}
\put(25,43){\makebox(0,8)[bl]{\(X^{i}_{4}\)}}
\put(31,43){\makebox(0,8)[bl]{\(X^{i}_{5}\)}}
\put(37,43){\makebox(0,8)[bl]{\(X^{m}_{1}\)}}
\put(44,43){\makebox(0,8)[bl]{\(X^{m}_{2}\)}}
\put(51,43){\makebox(0,8)[bl]{\(X^{m}_{3}\)}}

\end{picture}
\end{center}

   Here the following morphological classes are examined:
 (a) morphological class \(1\):
 \(\{ X^{1}_{1}, X^{1}_{2}, X^{1}_{3}, X^{1}_{4}, X^{1}_{5} \}\),
 (b) morphological class \(i\):
 \(\{ X^{i}_{1}, X^{i}_{2}, X^{i}_{3}, X^{i}_{4}, X^{i}_{5} \}\),
 and
 (c) morphological class \(m\):
 \(\{ X^{m}_{1}, X^{m}_{2}, X^{m}_{3} \}\).
 Further a simplified case is considered for three system
 parts (and corresponding morphological classes).
 The resultant (admissible) solution (composition or composite design alternative) is:
 ~\(S_{1}=X^{1}_{2}\star ... \star X^{i}_{3} \star ... \star X^{m}_{1}\).

\subsection{Method of Closeness to Ideal Point}

 First, modification of MA as method of closeness to ideal point was suggested
 (e.g., \cite{ayr69}, \cite{dubov86}).
  Illustration for method of closeness  to ideal point is shown in
  Fig. 3 (binary compatibility estimates are contained in Table
  3).

\begin{center}
\begin{picture}(80,46)

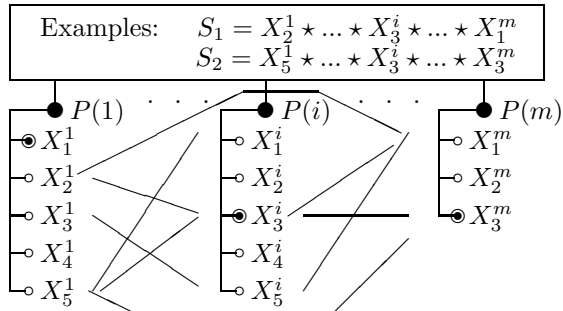
\put(4,00){\makebox(0,0)[bl] {Fig. 3. Illustration for MA with
ideal point}}

\put(4,5){\makebox(0,8)[bl]{\(X^{1}_{5}\)}}
\put(4,10){\makebox(0,8)[bl]{\(X^{1}_{4}\)}}
\put(4,15){\makebox(0,8)[bl]{\(X^{1}_{3}\)}}
\put(4,20){\makebox(0,8)[bl]{\(X^{1}_{2}\)}}
\put(4,25){\makebox(0,8)[bl]{\(X^{1}_{1}\)}}

\put(6,31){\line(-1,0){6}}

\put(0,31){\line(0,-1){24}}

\put(0,27){\line(1,0){2}} \put(2.5,27){\circle*{1}}
\put(2.5,27){\circle{1.7}}

\put(0,22){\line(1,0){2}} \put(2.5,22){\circle{1}}

\put(0,17){\line(1,0){2}} \put(2.5,17){\circle{1}}

\put(0,12){\line(1,0){2}} \put(2.5,12){\circle{1}}

\put(0,7){\line(1,0){2}} \put(2.5,7){\circle{1}}


\put(09,22.5){\line(2,1){22}}

\put(31,33.5){\line(1,0){10}}

\put(41,33.5){\line(2,-1){11}}

\put(11,7){\line(2,3){14}}

\put(11,22){\line(3,-1){14}}

\put(11,17){\line(3,-2){14}}

\put(12,7){\line(4,3){13}}

\put(10.5,7){\line(2,-1){6}} \put(16.5,4){\line(1,0){26.5}}
\put(43,4){\line(1,1){10}}


\put(32,5){\makebox(0,8)[bl]{\(X^{i}_{5}\)}}
\put(32,10){\makebox(0,8)[bl]{\(X^{i}_{4}\)}}
\put(32,15){\makebox(0,8)[bl]{\(X^{i}_{3}\)}}
\put(32,20){\makebox(0,8)[bl]{\(X^{i}_{2}\)}}
\put(32,25){\makebox(0,8)[bl]{\(X^{i}_{1}\)}}

\put(34,31){\line(-1,0){6}}

\put(28,31){\line(0,-1){24}}

\put(28,27){\line(1,0){2}} \put(30.5,27){\circle{1}}
\put(28,22){\line(1,0){2}} \put(30.5,22){\circle{1}}

\put(28,17){\line(1,0){2}} \put(30.5,17){\circle*{1}}
\put(30.5,17){\circle{1.7}}

\put(28,12){\line(1,0){2}} \put(30.5,12){\circle{1}}
\put(28,7){\line(1,0){2}} \put(30.5,7){\circle{1}}



\put(39,7){\line(2,3){14}}

\put(39,17){\line(1,0){14}}

\put(37,17){\line(3,2){14}}


\put(61,15){\makebox(0,8)[bl]{\(X^{m}_{3}\)}}
\put(61,20){\makebox(0,8)[bl]{\(X^{m}_{2}\)}}
\put(61,25){\makebox(0,8)[bl]{\(X^{m}_{1}\)}}

\put(63,31){\line(-1,0){6}}

\put(57,31){\line(0,-1){14}}

\put(57,27){\line(1,0){2}} \put(59.5,27){\circle{1}}

\put(57,22){\line(1,0){2}} \put(59.5,22){\circle{1}}

\put(57,17){\line(1,0){2}} \put(59.5,17){\circle*{1}}
\put(59.5,17){\circle{1.7}}



\put(6,31){\circle*{2}} \put(34,31){\circle*{2}}
\put(63,31){\circle*{2}}

\put(06,31){\line(0,1){4}} \put(34,31){\line(0,1){4}}
\put(63,31){\line(0,1){4}}

\put(08,29){\makebox(0,8)[bl]{\(P(1)\) }}
\put(36,29){\makebox(0,8)[bl]{\(P(i)\) }}
\put(65,29){\makebox(0,8)[bl]{\(P(m)\) }}


\put(17,32){\makebox(0,8)[bl]{ .~ .~ .}}
\put(45,32){\makebox(0,8)[bl]{ .~ .~ .}}


\put(00,35){\line(1,0){71}} \put(00,45){\line(1,0){71}}
\put(00,35){\line(0,1){10}} \put(71,35){\line(0,1){10}}



\put(04,40){\makebox(0,8)[bl]{Examples:  ~~~\(S_{1}=X^{1}_{2}\star
... \star X^{i}_{3} \star ... \star X^{m}_{1}\)}}

\put(20,36){\makebox(0,8)[bl]{ ~~~\(S_{2}=X^{1}_{5}\star ... \star
X^{i}_{3} \star ... \star X^{m}_{3}\)}}

\end{picture}
%
\begin{picture}(57,55)


\put(05,50){\makebox(0,0)[bl]{Table 3. Binary compatibility}}

\put(00,00){\line(1,0){57}} \put(00,42){\line(1,0){57}}
\put(00,48){\line(1,0){57}}

\put(00,0){\line(0,1){48}} \put(06,0){\line(0,1){48}}
\put(57,0){\line(0,1){48}}


\put(12,42){\line(0,1){6}} \put(18,42){\line(0,1){6}}
\put(24,42){\line(0,1){6}} \put(30,42){\line(0,1){6}}
\put(36,42){\line(0,1){6}} \put(43,42){\line(0,1){6}}
\put(50,42){\line(0,1){6}}

\put(0.7,37){\makebox(0,8)[bl]{\(X^{1}_{1}\)}}
\put(0.7,33){\makebox(0,8)[bl]{\(X^{1}_{2}\)}}
\put(0.7,29){\makebox(0,8)[bl]{\(X^{1}_{3}\)}}
\put(0.7,25){\makebox(0,8)[bl]{\(X^{1}_{4}\)}}
\put(0.7,21){\makebox(0,8)[bl]{\(X^{1}_{5}\)}}
\put(0.7,17){\makebox(0,8)[bl]{\(X^{i}_{1}\)}}
\put(0.7,13){\makebox(0,8)[bl]{\(X^{i}_{2}\)}}
\put(0.7,09){\makebox(0,8)[bl]{\(X^{i}_{3}\)}}
\put(0.7,05){\makebox(0,8)[bl]{\(X^{i}_{4}\)}}
\put(0.7,01){\makebox(0,8)[bl]{\(X^{i}_{5}\)}}


\put(08,38){\makebox(0,8)[bl]{\(0\)}}
\put(14,38){\makebox(0,8)[bl]{\(0\)}}
\put(20,38){\makebox(0,8)[bl]{\(0\)}}
\put(26,38){\makebox(0,8)[bl]{\(0\)}}
\put(32,38){\makebox(0,8)[bl]{\(0\)}}

\put(38,38){\makebox(0,8)[bl]{\(0\)}}
\put(45,38){\makebox(0,8)[bl]{\(0\)}}
\put(52,38){\makebox(0,8)[bl]{\(0\)}}


\put(08,34){\makebox(0,8)[bl]{\(0\)}}
\put(14,34){\makebox(0,8)[bl]{\(0\)}}
\put(20,34){\makebox(0,8)[bl]{\(1\)}}
\put(26,34){\makebox(0,8)[bl]{\(0\)}}
\put(32,34){\makebox(0,8)[bl]{\(0\)}}

\put(38,34){\makebox(0,8)[bl]{\(1\)}}
\put(45,34){\makebox(0,8)[bl]{\(0\)}}
\put(52,34){\makebox(0,8)[bl]{\(0\)}}


\put(08,30){\makebox(0,8)[bl]{\(0\)}}
\put(14,30){\makebox(0,8)[bl]{\(0\)}}
\put(20,30){\makebox(0,8)[bl]{\(0\)}}
\put(26,30){\makebox(0,8)[bl]{\(0\)}}
\put(32,30){\makebox(0,8)[bl]{\(1\)}}

\put(38,30){\makebox(0,8)[bl]{\(0\)}}
\put(45,30){\makebox(0,8)[bl]{\(0\)}}
\put(52,30){\makebox(0,8)[bl]{\(0\)}}


\put(08,26){\makebox(0,8)[bl]{\(0\)}}
\put(14,26){\makebox(0,8)[bl]{\(0\)}}
\put(20,26){\makebox(0,8)[bl]{\(0\)}}
\put(26,26){\makebox(0,8)[bl]{\(0\)}}
\put(32,26){\makebox(0,8)[bl]{\(0\)}}

\put(38,26){\makebox(0,8)[bl]{\(0\)}}
\put(45,26){\makebox(0,8)[bl]{\(0\)}}
\put(52,26){\makebox(0,8)[bl]{\(0\)}}


\put(08,22){\makebox(0,8)[bl]{\(1\)}}
\put(14,22){\makebox(0,8)[bl]{\(0\)}}
\put(20,22){\makebox(0,8)[bl]{\(1\)}}
\put(26,22){\makebox(0,8)[bl]{\(0\)}}
\put(32,22){\makebox(0,8)[bl]{\(0\)}}

\put(38,22){\makebox(0,8)[bl]{\(0\)}}
\put(45,22){\makebox(0,8)[bl]{\(0\)}}
\put(52,22){\makebox(0,8)[bl]{\(1\)}}


\put(38,18){\makebox(0,8)[bl]{\(0\)}}
\put(45,18){\makebox(0,8)[bl]{\(0\)}}
\put(52,18){\makebox(0,8)[bl]{\(0\)}}

\put(38,14){\makebox(0,8)[bl]{\(0\)}}
\put(45,14){\makebox(0,8)[bl]{\(0\)}}
\put(52,14){\makebox(0,8)[bl]{\(0\)}}

\put(38,10){\makebox(0,8)[bl]{\(1\)}}
\put(45,10){\makebox(0,8)[bl]{\(0\)}}
\put(52,10){\makebox(0,8)[bl]{\(1\)}}

\put(38,06){\makebox(0,8)[bl]{\(0\)}}
\put(45,06){\makebox(0,8)[bl]{\(0\)}}
\put(52,06){\makebox(0,8)[bl]{\(0\)}}

\put(38,02){\makebox(0,8)[bl]{\(1\)}}
\put(45,02){\makebox(0,8)[bl]{\(0\)}}
\put(52,02){\makebox(0,8)[bl]{\(0\)}}


\put(07,43){\makebox(0,8)[bl]{\(X^{i}_{1}\)}}
\put(13,43){\makebox(0,8)[bl]{\(X^{i}_{2}\)}}
\put(19,43){\makebox(0,8)[bl]{\(X^{i}_{3}\)}}
\put(25,43){\makebox(0,8)[bl]{\(X^{i}_{4}\)}}
\put(31,43){\makebox(0,8)[bl]{\(X^{i}_{5}\)}}
\put(37,43){\makebox(0,8)[bl]{\(X^{m}_{1}\)}}
\put(44,43){\makebox(0,8)[bl]{\(X^{m}_{2}\)}}
\put(51,43){\makebox(0,8)[bl]{\(X^{m}_{3}\)}}

\end{picture}
\end{center}


  Here for each system part
 (from the corresponding morphological class) the best design
 alternatives (as an ideal)
 are selected (e.g., by expert judgment).
 In the illustrative example (Fig. 3),
 the ideal design alternatives are:
 \(X^{1}_{1}\), \(X^{i}_{3}\), and \(X^{m}_{3}\).
 Thus, the ideal point (i.e., solution) is:~
\(S_{0} = X^{1}_{1} \star ... \star X^{i}_{3} \star ... \star
 X^{m}_{3}\).
 Unfortunately, this solution \(S_{o}\) is inadmissible (by
 compatibility).
%
 Admissible solutions are the following:
~\(S_{1} = X^{1}_{2} \star ... \star X^{i}_{3} \star ... \star
 X^{m}_{1}\)
 ~ and ~
~\(S_{2} = X^{1}_{5} \star ... \star X^{i}_{3} \star ... \star
 X^{m}_{3}\).

 Let ~\(\rho ( S', S''  )\)~ be a proximity
 (e.g., by elements)
 for two composite design
 alternatives ~\(S', S'' \in  \{ S \}\).
 Then it is reasonable to search for the following solution
 ~\(S^{*} \in \{ S^{a} \}  \subseteq \{S\} \)~
 (\(\{S^{a}\} \)~ is a set of admissible solutions):~
%
%
  \( S^{*} = Arg ~ \min_{ S \in \{S^{a}\} }   ~  \rho ( S , S_{o}  )  \).
 Clearly, in the illustrative example solution ~
 \(S_{2} = X^{1}_{5} \star ... \star X^{i}_{3} \star ... \star
 X^{m}_{3}\)~
 is more close to ideal solution ~\(S_{o}\)~
 (i.e., ~\(  \rho (S_{2}, S_{o}) \preceq   \rho (S_{1}, S_{o}) \)).
 Generally, various versions of proximity
 (as real functions, vectors, etc.)
 can by examined
 (e.g., \cite{ayr69}, \cite{dubov86}).
%


\subsection{Pareto-Based Morphological Approach}

 An integrated method (MA and multicriteria decision making, an
enumerative method) was suggested as follows (e.g., [4]):

 {\it Stage 1.} Usage of basic MA to get a set of admissible compositions.

 {\it Stage 2.} Generation of criteria for evaluation of the admissible
 compositions.

 {\it Stage 3.} Evaluation of admissible compositions upon criteria and
 selection of Pareto-efficient solutions.

 Fig. 4 illustrates Pareto-based MA. Concurrently, binary compatibility
 estimates are
 depicted in Table 4.
  Here admissible solutions are the following:
~\(S_{1} = X^{1}_{2} \star ... \star X^{i}_{3} \star ... \star
 X^{m}_{1}\),
%
%
~\(S_{2} = X^{1}_{5} \star ... \star X^{i}_{3} \star ... \star
 X^{m}_{3}\),
 and
~\(S_{3} = X^{1}_{5} \star ... \star X^{i}_{5} \star ... \star
 X^{m}_{3}\).
 Further, the solutions have to be evaluated upon  criteria and
 Pareto-efficient solution(s) will be selected.
%
 It is important to note that the estimate vector for each DA
 can contain estimates of compatibility as well.
%
 Pareto-based morphological approach was used by several students during the
 author's course (instead of HMMD)
 (\cite{lev06a}, \cite{lev09edu}, \cite{lev10}).

\begin{center}
\begin{picture}(80,51)
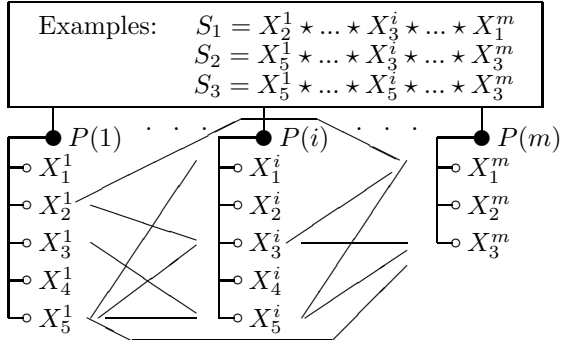
\put(07,00){\makebox(0,0)[bl] {Fig. 4. Illustration for
Pareto-based MA}}

\put(4,5){\makebox(0,8)[bl]{\(X^{1}_{5}\)}}
\put(4,10){\makebox(0,8)[bl]{\(X^{1}_{4}\)}}
\put(4,15){\makebox(0,8)[bl]{\(X^{1}_{3}\)}}
\put(4,20){\makebox(0,8)[bl]{\(X^{1}_{2}\)}}
\put(4,25){\makebox(0,8)[bl]{\(X^{1}_{1}\)}}

\put(6,31){\line(-1,0){6}}

\put(0,31){\line(0,-1){24}}

\put(0,27){\line(1,0){2}} \put(2.5,27){\circle{1}}
\put(0,22){\line(1,0){2}} \put(2.5,22){\circle{1}}
\put(0,17){\line(1,0){2}} \put(2.5,17){\circle{1}}
\put(0,12){\line(1,0){2}} \put(2.5,12){\circle{1}}
\put(0,7){\line(1,0){2}} \put(2.5,7){\circle{1}}


\put(09,22.5){\line(2,1){22}}

\put(31,33.5){\line(1,0){10}}

\put(41,33.5){\line(2,-1){11}}

\put(11,7){\line(2,3){14}}

\put(13,7){\line(1,0){12}}

\put(11,22){\line(3,-1){14}}

\put(11,17){\line(3,-2){14}}

\put(12,7){\line(4,3){13}}

\put(10.5,7){\line(2,-1){6}} \put(16.5,4){\line(1,0){26.5}}
\put(43,4){\line(1,1){10}}


\put(32,5){\makebox(0,8)[bl]{\(X^{i}_{5}\)}}
\put(32,10){\makebox(0,8)[bl]{\(X^{i}_{4}\)}}
\put(32,15){\makebox(0,8)[bl]{\(X^{i}_{3}\)}}
\put(32,20){\makebox(0,8)[bl]{\(X^{i}_{2}\)}}
\put(32,25){\makebox(0,8)[bl]{\(X^{i}_{1}\)}}

\put(34,31){\line(-1,0){6}}

\put(28,31){\line(0,-1){24}}

\put(28,27){\line(1,0){2}} \put(30.5,27){\circle{1}}
\put(28,22){\line(1,0){2}} \put(30.5,22){\circle{1}}
\put(28,17){\line(1,0){2}} \put(30.5,17){\circle{1}}
\put(28,12){\line(1,0){2}} \put(30.5,12){\circle{1}}
\put(28,7){\line(1,0){2}} \put(30.5,7){\circle{1}}



\put(39,7){\line(2,3){14}}

\put(39.5,7){\line(3,2){13.5}}

\put(39,17){\line(1,0){14}}

\put(37,17){\line(3,2){14}}


\put(61,15){\makebox(0,8)[bl]{\(X^{m}_{3}\)}}
\put(61,20){\makebox(0,8)[bl]{\(X^{m}_{2}\)}}
\put(61,25){\makebox(0,8)[bl]{\(X^{m}_{1}\)}}

\put(63,31){\line(-1,0){6}}

\put(57,31){\line(0,-1){14}}

\put(57,27){\line(1,0){2}} \put(59.5,27){\circle{1}}
\put(57,22){\line(1,0){2}} \put(59.5,22){\circle{1}}
\put(57,17){\line(1,0){2}} \put(59.5,17){\circle{1}}



\put(6,31){\circle*{2}} \put(34,31){\circle*{2}}
\put(63,31){\circle*{2}}

\put(06,31){\line(0,1){4}} \put(34,31){\line(0,1){4}}
\put(63,31){\line(0,1){4}}

\put(08,29){\makebox(0,8)[bl]{\(P(1)\) }}
\put(36,29){\makebox(0,8)[bl]{\(P(i)\) }}
\put(65,29){\makebox(0,8)[bl]{\(P(m)\) }}


\put(17,32){\makebox(0,8)[bl]{ .~ .~ .}}
\put(45,32){\makebox(0,8)[bl]{ .~ .~ .}}


\put(00,35){\line(1,0){71}} \put(00,49){\line(1,0){71}}
\put(00,35){\line(0,1){14}} \put(71,35){\line(0,1){14}}




\put(04,44){\makebox(0,8)[bl]{Examples:  ~~~\(S_{1}=X^{1}_{2}\star
... \star X^{i}_{3} \star ... \star X^{m}_{1}\)}}

\put(20,40){\makebox(0,8)[bl]{ ~~~\(S_{2}=X^{1}_{5}\star ... \star
X^{i}_{3} \star ... \star X^{m}_{3}\)}}

\put(20,36){\makebox(0,8)[bl]{ ~~~\(S_{3}=X^{1}_{5}\star ... \star
X^{i}_{5} \star ... \star X^{m}_{3}\)}}

\end{picture}
%
\begin{picture}(57,55)
\put(05,50){\makebox(0,0)[bl]{Table 4. Binary compatibility}}



\put(00,00){\line(1,0){57}} \put(00,42){\line(1,0){57}}
\put(00,48){\line(1,0){57}}

\put(00,0){\line(0,1){48}} \put(06,0){\line(0,1){48}}
\put(57,0){\line(0,1){48}}


\put(12,42){\line(0,1){6}} \put(18,42){\line(0,1){6}}
\put(24,42){\line(0,1){6}} \put(30,42){\line(0,1){6}}
\put(36,42){\line(0,1){6}} \put(43,42){\line(0,1){6}}
\put(50,42){\line(0,1){6}}

\put(0.7,37){\makebox(0,8)[bl]{\(X^{1}_{1}\)}}
\put(0.7,33){\makebox(0,8)[bl]{\(X^{1}_{2}\)}}
\put(0.7,29){\makebox(0,8)[bl]{\(X^{1}_{3}\)}}
\put(0.7,25){\makebox(0,8)[bl]{\(X^{1}_{4}\)}}
\put(0.7,21){\makebox(0,8)[bl]{\(X^{1}_{5}\)}}
\put(0.7,17){\makebox(0,8)[bl]{\(X^{i}_{1}\)}}
\put(0.7,13){\makebox(0,8)[bl]{\(X^{i}_{2}\)}}
\put(0.7,09){\makebox(0,8)[bl]{\(X^{i}_{3}\)}}
\put(0.7,05){\makebox(0,8)[bl]{\(X^{i}_{4}\)}}
\put(0.7,01){\makebox(0,8)[bl]{\(X^{i}_{5}\)}}


\put(08,38){\makebox(0,8)[bl]{\(0\)}}
\put(14,38){\makebox(0,8)[bl]{\(0\)}}
\put(20,38){\makebox(0,8)[bl]{\(0\)}}
\put(26,38){\makebox(0,8)[bl]{\(0\)}}
\put(32,38){\makebox(0,8)[bl]{\(0\)}}

\put(38,38){\makebox(0,8)[bl]{\(0\)}}
\put(45,38){\makebox(0,8)[bl]{\(0\)}}
\put(52,38){\makebox(0,8)[bl]{\(0\)}}


\put(08,34){\makebox(0,8)[bl]{\(0\)}}
\put(14,34){\makebox(0,8)[bl]{\(0\)}}
\put(20,34){\makebox(0,8)[bl]{\(1\)}}
\put(26,34){\makebox(0,8)[bl]{\(0\)}}
\put(32,34){\makebox(0,8)[bl]{\(0\)}}

\put(38,34){\makebox(0,8)[bl]{\(1\)}}
\put(45,34){\makebox(0,8)[bl]{\(0\)}}
\put(52,34){\makebox(0,8)[bl]{\(0\)}}


\put(08,30){\makebox(0,8)[bl]{\(0\)}}
\put(14,30){\makebox(0,8)[bl]{\(0\)}}
\put(20,30){\makebox(0,8)[bl]{\(0\)}}
\put(26,30){\makebox(0,8)[bl]{\(0\)}}
\put(32,30){\makebox(0,8)[bl]{\(1\)}}

\put(38,30){\makebox(0,8)[bl]{\(0\)}}
\put(45,30){\makebox(0,8)[bl]{\(0\)}}
\put(52,30){\makebox(0,8)[bl]{\(0\)}}


\put(08,26){\makebox(0,8)[bl]{\(0\)}}
\put(14,26){\makebox(0,8)[bl]{\(0\)}}
\put(20,26){\makebox(0,8)[bl]{\(0\)}}
\put(26,26){\makebox(0,8)[bl]{\(0\)}}
\put(32,26){\makebox(0,8)[bl]{\(0\)}}

\put(38,26){\makebox(0,8)[bl]{\(0\)}}
\put(45,26){\makebox(0,8)[bl]{\(0\)}}
\put(52,26){\makebox(0,8)[bl]{\(0\)}}


\put(08,22){\makebox(0,8)[bl]{\(1\)}}
\put(14,22){\makebox(0,8)[bl]{\(0\)}}
\put(20,22){\makebox(0,8)[bl]{\(1\)}}
\put(26,22){\makebox(0,8)[bl]{\(0\)}}
\put(32,22){\makebox(0,8)[bl]{\(1\)}}

\put(38,22){\makebox(0,8)[bl]{\(0\)}}
\put(45,22){\makebox(0,8)[bl]{\(0\)}}
\put(52,22){\makebox(0,8)[bl]{\(1\)}}


\put(38,18){\makebox(0,8)[bl]{\(0\)}}
\put(45,18){\makebox(0,8)[bl]{\(0\)}}
\put(52,18){\makebox(0,8)[bl]{\(0\)}}

\put(38,14){\makebox(0,8)[bl]{\(0\)}}
\put(45,14){\makebox(0,8)[bl]{\(0\)}}
\put(52,14){\makebox(0,8)[bl]{\(0\)}}

\put(38,10){\makebox(0,8)[bl]{\(1\)}}
\put(45,10){\makebox(0,8)[bl]{\(0\)}}
\put(52,10){\makebox(0,8)[bl]{\(1\)}}

\put(38,06){\makebox(0,8)[bl]{\(0\)}}
\put(45,06){\makebox(0,8)[bl]{\(0\)}}
\put(52,06){\makebox(0,8)[bl]{\(0\)}}

\put(38,02){\makebox(0,8)[bl]{\(1\)}}
\put(45,02){\makebox(0,8)[bl]{\(0\)}}
\put(52,02){\makebox(0,8)[bl]{\(1\)}}


\put(07,43){\makebox(0,8)[bl]{\(X^{i}_{1}\)}}
\put(13,43){\makebox(0,8)[bl]{\(X^{i}_{2}\)}}
\put(19,43){\makebox(0,8)[bl]{\(X^{i}_{3}\)}}
\put(25,43){\makebox(0,8)[bl]{\(X^{i}_{4}\)}}
\put(31,43){\makebox(0,8)[bl]{\(X^{i}_{5}\)}}
\put(37,43){\makebox(0,8)[bl]{\(X^{m}_{1}\)}}
\put(44,43){\makebox(0,8)[bl]{\(X^{m}_{2}\)}}
\put(51,43){\makebox(0,8)[bl]{\(X^{m}_{3}\)}}

\end{picture}
\end{center}


\subsection{Linear Programming}

  In \cite{kol86}, MA is reduced to linear programming.
  Here constraints imposed on the solution are reduced to a set of
  inequalities of Boolean variables
  and
  quality criterion for the solution
  as an additive function
   is used.
%
%
 In this case, well-known methods for linear programming problems
 can be used.


\subsection{Multiple Choice Problem}

 The basic knapsack problem is
 (e.g.,
 \cite{gar79},
 \cite{keller04},
 \cite{mar90}):
 \[\max\sum_{i=1}^{m} c_{i} x_{i}
 ~~~s.t.~\sum_{i=1}^{m} a_{i} x_{i} \leq b,
 ~x_{i} \in \{ 0,1\}, ~ i=\overline{1,m},\]
%
 ¨ where \(x_{i}=1\) if item \(i\) is selected,
 \(c_{i}\) is a value ("utility") for item \(i\), and
 \(a_{i}\) is a weight (or resource required).
 Often nonnegative coefficients are assumed.
 The problem is NP-hard (\cite{gar79}, \cite{mar90})
 and can be solved by enumerative methods
 (e.g., Branch-and-Bound, dynamic programming),
 approximate schemes with a limited
 relative error (e.g., \cite{keller04}, \cite{mar90}).
 In  multiple choice problem
(e.g., \cite{mar90}),
 the items
 are divided into groups
 and we select element(s) from each group
 while taking into account a total resource constraint (or constraints).
 Here each element has two indices: ~\((i,j)\),
 where \(i\) corresponds to number of group and
 \(j\) corresponds to number of item in the group.
 In the case of multicriteria description,
 each element (i.e., \((i,j)\)) has vector profit:
  ~\( \overline{c_{i,j}}    = (  c^{1}_{i,j}, ..., c^{\xi}_{i,j}, ... , c^{r}_{i,j} ) \)~
  and multicriteria multiple choice problem is
    \cite{levsaf10}:
 \[\max\sum_{i=1}^{m} \sum_{j=1}^{q_{i}} c^{\xi}_{ij} x_{ij}, ~~~ \forall \xi =
 \overline{1,r}
%
%
 ~~~~~~~s.t.~\sum_{i=1}^{m} \sum_{j=1}^{q_{i}} a_{ij} x_{ij} \leq
 b,
 ~~\sum_{j=1}^{q_{i}} x_{ij}=1 ~\forall i=\overline{1,m},
 ~~x_{ij} \in \{0,1\}.\]
%
 For this problem formulation,
 it is reasonable to search for
 Pareto-efficient solutions.
 This design approach was used for design and redesign/improvement of
 applied systems
 (software, hardware,
 communication)
 (\cite{lev11},
 \cite{levsaf10},
  \cite{sousa06}).
 Here
 the following solving schemes can be used
  \cite{levsaf10}:
 (i) enumerative algorithm based on dynamic programming,
 (ii) heuristic based on preliminary multicriteria ranking of elements
 to get their priorities
 and
 step-by-step packing the knapsack (i.e., greedy approach),
%
 (iii) multicriteria ranking of elements to get their ordinal
 priorities and usage of approximate solving scheme
 (as for knapsack)
 based on discrete space of system excellence
 (as later in HMMD).
%

\subsection{Quadratic Assignment Problem}


%
 Assignment/allocation problems are widely used in many domains
 (e.g.,  \cite{cela98},
 \cite{gar79},
  \cite{pardalos94}).
%
%
%
 Simple  assignment problem involves nonnegative correspondence matrix
 ~\( \Upsilon = ||c_{i,j}||\) ~(\(i=\overline{1,n}\),
 \(j=\overline{1,n}\))~
 where \(c_{i,j}\) is a profit ('utility') to assign
 element \(i\) to position \(j\).
 The problem is (e.g., \cite{gar79}):


     {\it Find assignment} ~\(\pi =(\pi(1),...\pi(i),...,\pi(n))\)~
     {\it of elements}
    \(i\) (\(i=\overline{1,n}\))
      {\it to positions}
      ~\(\pi(i)\)
      {\it which}
     {\it corresponds to a total effectiveness:}~
     \(\sum_{i=1}^{n} c_{i,\pi(i)} \rightarrow \max\).



%
%
%
 More complicated well-known model as
  quadratic assignment problem (QAP)
 includes interconnection between elements of different groups
 (each group corresponds to a certain position)
 (e.g., \cite{cela98}, \cite{lev98}, \cite{pardalos94}).
%
%
%
 Let a nonnegative value \( d (i,j_{1},k,j_{2}) \)
 be a profit of compatibility between item
 \(j_{1}\) in group \(J_{i}\) and item \(j_{2}\) in group \(J_{k}\).
 Also, this value of compatibility is added to the objective
 function.
 QAP may be considered as a version of MA.
 Thus,
 QAP can be formulated as follows:
%
%
 \[ \max  \sum_{i=1}^{m} \sum_{j=1}^{q_{i}}¨c_{i,j}x_{i,j}+
  \sum_{l < k} \sum_{j_{1}=1}^{q_{l}}
 \sum_{j_{2}=1}^{q_{k}} ¨d(l,j_{1},k,j_{2})~x_{l,j_{1}}
 ¨x_{k,j_{2}},
  ~~ l=\overline{1,m}, ~~k=\overline{1,m}; \]
 \[ s.t.~~ \sum_{i=1}^{m} \sum_{j=1}^{q_{i}} ~a_{i,j} ~x_{i,j} \leq
 b,
 ~~ \sum_{j=1}^{q_{i}} x_{i,j} \leq 1 ~\forall i=\overline{1,m}, ~~
  x_{i,j} \in \{0,1\}.\]
%
%
%
%
%
 QAP is NP-hard.
 Enumerative methods (e.g., branch-and-bound) or heuristics
 (e.g., greedy algorithms, tabu search, genetic algorithms)
 are usually used for the problem.
%
%
%
%
 In  the case of
  multicriteria assignment problem,
  the objective function is
 transformed into a vector function, i.e.,
%
 ~\(c_{i,j} ~\Rightarrow
 \overline{c_{i,j}} = (c^{1}_{i,j},...,c^{\xi
 }_{i,j},...,c^{r}_{i,j})\)
%
%
 and the vector objective function is, for example:

 \[ (~ \sum_{i=1}^{m} \sum_{j=1}^{n} c^{1}_{i,j} x_{i,j}, ... ,  \sum_{i=1}^{m} \sum_{j=1}^{n} c^{\xi}_{i,j} x_{i,j}, ... ,
  \sum_{i=1}^{m} \sum_{j=1}^{n} c^{r}_{i,j} x_{i,j} ~) .\]
%


%
%
%
%
%
%
%
 Here Pareto-efficient solutions are usually searched for.
%
%
%
%
%
%
%
%
%


\subsection{Hierarchical Morphological Multicriteria Design}

 A basic description of
 Hierarchical Morphological Multicriteria Design (HMMD)
 is contained in
 (\cite{lev98},
   \cite{lev06},
    \cite{lev09}).
%
%
%
 The assumptions of HMMD are the following:
 ~(a) a tree-like structure of the system;
 ~(b) a composite estimate for system quality
     that integrates components (subsystems, parts) qualities and
     qualities of interconnections
      (hereinafter referred as 'IC')
     across subsystems;
 ~(c) monotonic criteria for the system and its components;
 and
 ~(d) quality of system components and IC are evaluated
   on the basis of coordinated ordinal scales.
 The designations are:
  ~(1) design alternatives (DAs) for
  nodes of the model;
  ~(2) priorities of DAs (\(r=\overline{1,k}\);
      \(1\) corresponds to the best level);
  ~(3) ordinal compatibility estimates for each pair of DAs
  (\(w=\overline{0,l}\); \(l\) corresponds to the best level).
 The basic phases of HMMD are:

  {\it Phase 1.} Design of the tree-like system model (a preliminary
  phase).

  {\it Phase 2.} Generating DAs for leaf nodes of the system model.

  {\it Phase 3.} Hierarchical selection and composing of DAs into composite
    DAs for the corresponding higher level of the system
    hierarchy (morphological clique problem).

  {\it Phase 4.} Analysis and improvement of the resultant composite DAs
  (decisions).

 Let ~\(S\) be a system consisting of ~\(m\) parts (components):
 ~\(P(1),...,P(i),...,P(m)\) ~(Fig. 1).
 A set of DAs
 is generated for each system part above.
 The problem is:

 {\it Find composite design alternative}
 ~ \(S=S(1)\star ...\star S(i)\star ...\star S(m)\)~~
 {\it of}~ DAs {\it (one representative design alternative}
 ~\(S(i)\)~
 {\it for each system component/part}
 ~\(P(i)\), ~\(i=\overline{1,m}\))~
 {\it with non-zero} ~IC~ {\it estimates between the selected design alternatives.}

 A discrete space of the system quality is based on the
 following vector (Fig. 5):
 ~\(N(S)=(w(S);n(S))\),
 ~where \(w(S)\) is the minimum of pairwise compatibility
 between DAs which correspond to different system components
 (i.e.,
 \(~\forall ~P_{j_{1}}\) and \( P_{j_{2}}\),
 \(1 \leq j_{1} \neq j_{2} \leq m\))
 in \(S\),
 ~\(n(S)=(n_{1},...,n_{r},...n_{k})\),
 ~where ~\(n_{r}\) is the number of DAs of the \(r\)th quality in ~\(S\)
 ~(\(\sum^{k}_{r=1} n_{r} = m\)).
 Here composite solutions (composite DAs) are searched for
 which are nondominated by ~\(N(S)\)~ (i.e., Pareto-efficient solutions)
 (Fig. 5).

 In (\cite{lev98}, \cite{lev06}),
 the described combinatorial problem is called
 {\it morphological clique problem},
 this problem is NP-hard
 (because a more simple its subproblem is NP hard \cite{knu92}).
   Generally,
   the following layers of system excellence can be considered
   (e.g., \cite{lev98}):
  ~(i) ideal point;
  ~(ii) Pareto-efficient points;
  ~(iii) a neighborhood of Pareto-efficient  DAs
 (e.g., a composite decision of this set can be
 transformed into a Pareto-efficient point on the basis of an
 improvement action(s)).
 Clearly, the compatibility component of vector ~\(N(S)\)
 can be considered on the basis of a poset-like scale too
 (as \(n(S)\) ).
 In this case, the discrete space of
 system excellence will be an analogical lattice
 (\cite{levf01}, \cite{lev06}).
 Fig. 6 and Fig. 7 illustrate HMMD
 (by a numerical example for a system consisting of three parts
 ~\(S = X \star Y \star Z\)).
 Priorities of DAs are shown in Fig. 6 in parentheses and are
 depicted in Fig. 7.
 Table 5 contains compatibility estimates
 (they are pointed out in Fig. 7 too).
 In the example, composite decisions are
 (Pareto-efficient  solutions) (Fig. 5, Fig. 6, Fig. 7, Fig. 8):
%
%
 ~\(S_{1}=X_{2}\star Y_{2}\star Z_{2}\), ~\(N(S_{1}) = (1;2,1,0)\);
 ~\(S_{2}=X_{1}\star Y_{2}\star Z_{2}\), ~\(N(S_{2}) = (2;1,2,0)\);
 ~\(S_{3}=X_{1}\star Y_{1}\star Z_{3}\), ~\(N(S_{3}) = (3;0,2,1)\).
%

 HMMD method was used to design various modular systems
 (e.g., packaged software, communication networks, security system,
 web-hosting system, car, telemetric system,
 test inputs in system testing,
 concrete technology, immunoassay technology,
 management for smart home, sensor node)
 (\cite{lev98},  \cite{lev05},
  \cite{lev06},
  \cite{lev08a},
   \cite{lev11},
   \cite{levnis01},
  \cite{levfir05},
  \cite{levlast06},
   \cite{levkhod07},
 \cite{levleus09},
   \cite{levshar09},
 \cite{levfim10},
  \cite{levsaf10}).
 In addition, HMMD was used in modular system improvement
 processes
 (\cite{lev98},
  \cite{lev05},
  \cite{lev06},
  \cite{lev10ir},
   \cite{levdan05},
  \cite{levand11}).

\begin{center}
\begin{picture}(114,82)
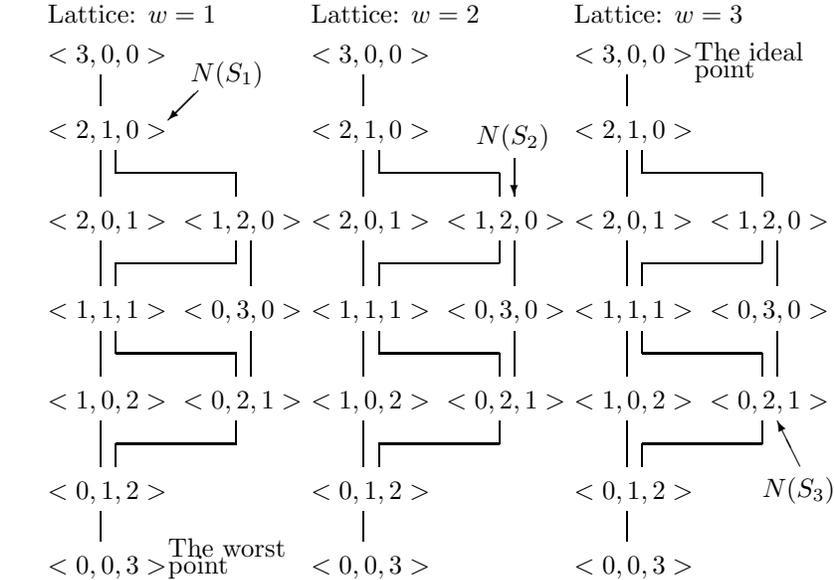
\put(0,0){\makebox(0,0)[bl] {Fig. 5. Space of system quality (3
system parts, 3 levels of element quality)}}


\put(05,78){\makebox(0,0)[bl]{Lattice: \(w=1\) }}

\put(05,72){\makebox(0,0)[bl]{\(<3,0,0>\) }}

\put(12,67){\line(0,1){4}}
\put(05,62){\makebox(0,0)[bl]{\(<2,1,0>\)}}

\put(12,55){\line(0,1){6}}
\put(05,50){\makebox(0,0)[bl]{\(<2,0,1>\) }}

\put(12,43){\line(0,1){6}}
\put(05,38){\makebox(0,0)[bl]{\(<1,1,1>\) }}

\put(12,31){\line(0,1){6}}
\put(05,26){\makebox(0,0)[bl]{\(<1,0,2>\) }}





\put(12,19){\line(0,1){6}}
\put(05,14){\makebox(0,0)[bl]{\(<0,1,2>\) }}

\put(12,9){\line(0,1){4}}

\put(05,04){\makebox(0,0)[bl]{\(<0,0,3>\) }}

\put(21,07){\makebox(0,0)[bl]{The worst}}
\put(21,04){\makebox(0,0)[bl]{point}}

\put(14,58){\line(0,1){3}} \put(30,58){\line(-1,0){16}}
\put(30,55){\line(0,1){3}}

\put(23,50){\makebox(0,0)[bl]{\(<1,2,0>\) }}

\put(30,49){\line(0,-1){3}} \put(30,46){\line(-1,0){16}}
\put(14,46){\line(0,-1){3}}
\put(32,43){\line(0,1){6}}
\put(23,38){\makebox(0,0)[bl]{\(<0,3,0>\) }}

\put(14,34){\line(0,1){3}} \put(30,34){\line(-1,0){16}}
\put(30,31){\line(0,1){3}}

\put(32,31){\line(0,1){6}}
\put(23,26){\makebox(0,0)[bl]{\(<0,2,1>\) }}

\put(30,25){\line(0,-1){3}} \put(30,22){\line(-1,0){16}}
\put(14,22){\line(0,-1){3}}


\put(24,69.4){\makebox(0,0)[bl]{\(N(S_{1})\)}}
\put(25,69){\vector(-1,-1){4}}


\put(40,78){\makebox(0,0)[bl]{Lattice: \(w=2\) }}

\put(40,72){\makebox(0,0)[bl]{\(<3,0,0>\) }}

\put(47,67){\line(0,1){4}}
\put(40,62){\makebox(0,0)[bl]{\(<2,1,0>\)}}

\put(47,55){\line(0,1){6}}
\put(40,50){\makebox(0,0)[bl]{\(<2,0,1>\) }}

\put(47,43){\line(0,1){6}}
\put(40,38){\makebox(0,0)[bl]{\(<1,1,1>\) }}

\put(47,31){\line(0,1){6}}
\put(40,26){\makebox(0,0)[bl]{\(<1,0,2>\) }}

\put(62,61){\makebox(0,0)[bl]{\(N(S_{2})\)}}
\put(67,60){\vector(0,-1){5}}


\put(47,19){\line(0,1){6}}
\put(40,14){\makebox(0,0)[bl]{\(<0,1,2>\) }}

\put(47,9){\line(0,1){4}}

\put(40,04){\makebox(0,0)[bl]{\(<0,0,3>\) }}

\put(49,58){\line(0,1){3}} \put(65,58){\line(-1,0){16}}
\put(65,55){\line(0,1){3}}

\put(58,50){\makebox(0,0)[bl]{\(<1,2,0>\) }}

\put(65,49){\line(0,-1){3}} \put(65,46){\line(-1,0){16}}
\put(49,46){\line(0,-1){3}}
\put(67,43){\line(0,1){6}}
\put(58,38){\makebox(0,0)[bl]{\(<0,3,0>\) }}

\put(49,34){\line(0,1){3}} \put(65,34){\line(-1,0){16}}
\put(65,31){\line(0,1){3}}

\put(67,31){\line(0,1){6}}
\put(58,26){\makebox(0,0)[bl]{\(<0,2,1>\) }}

\put(65,25){\line(0,-1){3}} \put(65,22){\line(-1,0){16}}
\put(49,22){\line(0,-1){3}}



\put(75,78){\makebox(0,0)[bl]{Lattice: \(w=3\) }}

\put(91,73){\makebox(0,0)[bl]{The ideal}}
\put(91,70){\makebox(0,0)[bl]{point}}

\put(75,72){\makebox(0,0)[bl]{\(<3,0,0>\) }}

\put(82,67){\line(0,1){4}}
\put(75,62){\makebox(0,0)[bl]{\(<2,1,0>\)}}

\put(82,55){\line(0,1){6}}
\put(75,50){\makebox(0,0)[bl]{\(<2,0,1>\) }}

\put(82,43){\line(0,1){6}}
\put(75,38){\makebox(0,0)[bl]{\(<1,1,1>\) }}

\put(82,31){\line(0,1){6}}
\put(75,26){\makebox(0,0)[bl]{\(<1,0,2>\) }}


\put(82,19){\line(0,1){6}}
\put(75,14){\makebox(0,0)[bl]{\(<0,1,2>\) }}

\put(82,9){\line(0,1){4}}

\put(75,04){\makebox(0,0)[bl]{\(<0,0,3>\) }}

\put(84,58){\line(0,1){3}} \put(100,58){\line(-1,0){16}}
\put(100,55){\line(0,1){3}}

\put(93,50){\makebox(0,0)[bl]{\(<1,2,0>\) }}

\put(100,49){\line(0,-1){3}} \put(100,46){\line(-1,0){16}}
\put(84,46){\line(0,-1){3}}
\put(102,43){\line(0,1){6}}
\put(93,38){\makebox(0,0)[bl]{\(<0,3,0>\) }}

\put(84,34){\line(0,1){3}} \put(100,34){\line(-1,0){16}}
\put(100,31){\line(0,1){3}}

\put(102,31){\line(0,1){6}}
\put(93,26){\makebox(0,0)[bl]{\(<0,2,1>\) }}

\put(100,25){\line(0,-1){3}} \put(100,22){\line(-1,0){16}}
\put(84,22){\line(0,-1){3}}


\put(100,14){\makebox(0,0)[bl]{\(N(S_{3})\)}}
\put(105,19){\vector(-1,2){3}}


\end{picture}
\end{center}

\begin{center}
\begin{picture}(55,41)
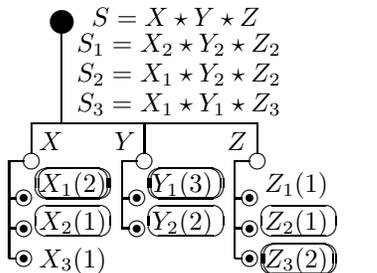
\put(0,0){\makebox(0,0)[bl] {Fig. 6. Example of composition}}

\put(4,4){\makebox(0,8)[bl]{\(X_{3}(1)\)}}
\put(4,09){\makebox(0,8)[bl]{\(X_{2}(1)\)}}
\put(4,14){\makebox(0,8)[bl]{\(X_{1}(2)\)}}

\put(8.5,16){\oval(10,4)} \put(8.5,16){\oval(9.2,4)}

\put(8.5,11){\oval(10,4)}

\put(19,09){\makebox(0,8)[bl]{\(Y_{2}(2)\)}}
\put(19,14){\makebox(0,8)[bl]{\(Y_{1}(3)\)}}

\put(23.5,16){\oval(10,4)} \put(23.5,16){\oval(9.2,4)}

\put(23.5,11){\oval(10,4)}

\put(34,4){\makebox(0,8)[bl]{\(Z_{3}(2)\)}}
\put(34,09){\makebox(0,8)[bl]{\(Z_{2}(1)\)}}
\put(34,14){\makebox(0,8)[bl]{\(Z_{1}(1)\)}}

\put(38.5,11){\oval(10,4)}

\put(38.5,06){\oval(10,4)} \put(38.5,06){\oval(9.2,4)}

\put(3,19){\circle{2}} \put(18,19){\circle{2}}
\put(33,19){\circle{2}}

\put(0,19){\line(1,0){02}} \put(15,19){\line(1,0){02}}
\put(30,19){\line(1,0){02}}

\put(0,19){\line(0,-1){13}} \put(15,19){\line(0,-1){09}}
\put(30,19){\line(0,-1){13}}

\put(30,14){\line(1,0){01}} \put(30,10){\line(1,0){01}}
\put(30,6){\line(1,0){01}}

\put(32,14){\circle{2}} \put(32,14){\circle*{1}}
\put(32,10){\circle{2}} \put(32,10){\circle*{1}}
\put(32,6){\circle{2}} \put(32,6){\circle*{1}}

\put(15,10){\line(1,0){01}} \put(15,14){\line(1,0){01}}

\put(17,10){\circle{2}} \put(17,10){\circle*{1}}

\put(17,14){\circle{2}}

\put(17,14){\circle*{1}}

\put(0,6){\line(1,0){01}} \put(0,10){\line(1,0){01}}
\put(0,14){\line(1,0){01}}

\put(2,10){\circle{2}} \put(2,14){\circle{2}}
\put(2,10){\circle*{1}} \put(2,14){\circle*{1}}
\put(2,6){\circle{2}} \put(2,6){\circle*{1}}

\put(3,24){\line(0,-1){04}} \put(18,24){\line(0,-1){04}}
\put(33,24){\line(0,-1){04}}

\put(3,24){\line(1,0){30}}

\put(7,24){\line(0,1){14}} \put(7,37.5){\circle*{3}}

\put(04,20.5){\makebox(0,8)[bl]{\(X\) }}
\put(14,20.5){\makebox(0,8)[bl]{\(Y\) }}
\put(29,20.5){\makebox(0,8)[bl]{\(Z\) }}

\put(11,37){\makebox(0,8)[bl]{\(S = X \star Y \star Z \) }}

\put(9,33){\makebox(0,8)[bl]{\(S_{1}=X_{2}\star Y_{2}\star
Z_{2}\)}}

\put(9,29){\makebox(0,8)[bl]{\(S_{2}=X_{1}\star Y_{2}\star
Z_{2}\)}}

\put(9,25){\makebox(0,8)[bl]{\(S_{3}=X_{1}\star Y_{1}\star
Z_{3}\)}}

\end{picture}
%
\begin{picture}(55,36)
\put(0,0){\makebox(0,0)[bl] {Fig. 7. Concentric presentation}}

\put(1,08){\line(0,1){6}} \put(2,08){\line(0,1){6}}

\put(01,08){\line(1,0){18}} \put(01,14){\line(1,0){18}}

\put(3,10){\makebox(0,0)[bl]{\(Z_{3}\)}} \put(8,08){\line(0,1){6}}
\put(9,10){\makebox(0,0)[bl]{\(Z_{2}\)}}

\put(11,11){\oval(5,8)}

\put(5,11){\oval(5,8)} \put(5,11){\oval(4.2,8)}

\put(14,10){\makebox(0,0)[bl]{\(Z_{1}\)}}
\put(19,08){\line(0,1){6}}

\put(25,08){\line(1,0){14}} \put(25,14){\line(1,0){14}}
\put(25,08){\line(0,1){6}}
\put(27,08){\line(0,1){6}}
\put(28,10){\makebox(0,0)[bl]{\(Y_{2}\)}}
\put(33,08){\line(0,1){6}}
\put(34,10){\makebox(0,0)[bl]{\(Y_{1}\)}}
\put(39,08){\line(0,1){6}}

\put(30,11){\oval(5,8)}

\put(36,11){\oval(5,8)} \put(36,11){\oval(4.2,8)}

\put(19,18){\line(0,1){18}} \put(25,18){\line(0,1){18}}
\put(19,18){\line(1,0){6}}
\put(20,19.5){\makebox(0,0)[bl]{\(X_{3}\)}}
\put(20,24){\makebox(0,0)[bl]{\(X_{2}\)}}

\put(22,25.5){\oval(8,4)}

\put(22,31.5){\oval(8,4)} \put(22,31.5){\oval(7.2,4)}

 \put(19,28){\line(1,0){6}}
\put(20,30){\makebox(0,0)[bl]{\(X_{1}\)}}
\put(19,34){\line(1,0){6}} \put(19,36){\line(1,0){6}}


\put(10,08){\line(0,-1){3}} \put(10,5){\line(1,0){20}}
\put(30,5){\line(0,1){3}}

\put(31,5){\makebox(0,0)[bl]{\(2\)}}


\put(10,14){\line(0,1){17}} \put(10,31){\line(1,0){09}}
\put(11,27){\makebox(0,0)[bl]{\(2\)}}

\put(12,14){\line(0,1){11}} \put(12,25){\line(1,0){07}}
\put(13,20){\makebox(0,0)[bl]{\(1\)}}

\put(30,14){\line(0,1){11}} \put(30,25){\line(-1,0){05}}
\put(27,20){\makebox(0,0)[bl]{\(3\)}}

\put(31,14){\line(0,1){17}} \put(31,31){\line(-1,0){07}}
\put(32,27){\makebox(0,0)[bl]{\(2\)}}


\put(36,14){\line(0,1){18}} \put(36,32){\line(-1,0){11}}
\put(37,24){\makebox(0,0)[bl]{\(3\)}}

\put(04,08){\line(0,-1){4}} \put(04,4){\line(1,0){32}}
\put(36,4){\line(0,1){4}}

\put(37,4){\makebox(0,0)[bl]{\(3\)}}

\put(04,14){\line(0,1){18}} \put(04,32){\line(1,0){14}}
\put(05,25){\makebox(0,0)[bl]{\(3\)}}

\end{picture}
%
\begin{picture}(36,35)

\put(0,30){\makebox(0,0)[bl]{Table 5. Compatibility}}

\put(00,00){\line(1,0){36}} \put(00,22){\line(1,0){36}}
\put(00,28){\line(1,0){36}}

\put(00,0){\line(0,1){28}} \put(06,0){\line(0,1){28}}
\put(36,0){\line(0,1){28}}


\put(12,22){\line(0,1){6}} \put(18,22){\line(0,1){6}}
\put(24,22){\line(0,1){6}} \put(30,22){\line(0,1){6}}

\put(01,18){\makebox(0,8)[bl]{\(X_{1}\)}}
\put(01,14){\makebox(0,8)[bl]{\(X_{2}\)}}
\put(01,10){\makebox(0,8)[bl]{\(X_{3}\)}}
\put(01,06){\makebox(0,8)[bl]{\(Y_{1}\)}}
\put(01,02){\makebox(0,8)[bl]{\(Y_{2}\)}}



\put(08,18){\makebox(0,8)[bl]{\(3\)}}
\put(14,18){\makebox(0,8)[bl]{\(2\)}}
\put(20,18){\makebox(0,8)[bl]{\(0\)}}
\put(26,18){\makebox(0,8)[bl]{\(2\)}}
\put(32,18){\makebox(0,8)[bl]{\(3\)}}


\put(08,14){\makebox(0,8)[bl]{\(0\)}}
\put(14,14){\makebox(0,8)[bl]{\(3\)}}
\put(20,14){\makebox(0,8)[bl]{\(0\)}}
\put(26,14){\makebox(0,8)[bl]{\(1\)}}
\put(32,14){\makebox(0,8)[bl]{\(0\)}}


\put(08,10){\makebox(0,8)[bl]{\(0\)}}
\put(14,10){\makebox(0,8)[bl]{\(0\)}}
\put(20,10){\makebox(0,8)[bl]{\(0\)}}
\put(26,10){\makebox(0,8)[bl]{\(0\)}}
\put(32,10){\makebox(0,8)[bl]{\(1\)}}


\put(20,06){\makebox(0,8)[bl]{\(0\)}}
\put(26,06){\makebox(0,8)[bl]{\(0\)}}
\put(32,06){\makebox(0,8)[bl]{\(3\)}}

\put(20,02){\makebox(0,8)[bl]{\(0\)}}
\put(26,02){\makebox(0,8)[bl]{\(2\)}}
\put(32,02){\makebox(0,8)[bl]{\(0\)}}


\put(07,24){\makebox(0,8)[bl]{\(Y_{1}\)}}
\put(13,24){\makebox(0,8)[bl]{\(Y_{2}\)}}
\put(19,24){\makebox(0,8)[bl]{\(Z_{1}\)}}
\put(25,24){\makebox(0,8)[bl]{\(Z_{2}\)}}
\put(31,24){\makebox(0,8)[bl]{\(Z_{3}\)}}

\end{picture}
\end{center}

\begin{center}
\begin{picture}(62,63)
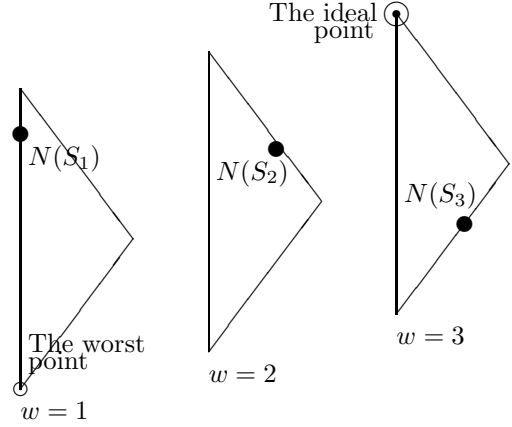
\put(01,0){\makebox(0,0)[bl]{Fig. 8. Illustration for space of
quality}}

\put(0,010){\line(0,1){40}} \put(0,010){\line(3,4){15}}
\put(0,050){\line(3,-4){15}}

\put(25,015){\line(0,1){40}} \put(25,015){\line(3,4){15}}
\put(25,055){\line(3,-4){15}}

\put(50,020){\line(0,1){40}} \put(50,020){\line(3,4){15}}
\put(50,060){\line(3,-4){15}}



\put(0,44){\circle*{2}}
\put(1,39){\makebox(0,0)[bl]{\(N(S_{1})\)}}

\put(34,42){\circle*{2}}
\put(26,37){\makebox(0,0)[bl]{\(N(S_{2})\)}}

\put(59,32){\circle*{2}}
\put(51,34){\makebox(0,0)[bl]{\(N(S_{3})\)}}

\put(50,60){\circle*{1}} \put(50,60){\circle{3}}


\put(33,59){\makebox(0,0)[bl]{The ideal}}
\put(39,56){\makebox(0,0)[bl]{point}}


\put(0,6){\makebox(0,0)[bl]{\(w=1\)}}
\put(25,11){\makebox(0,0)[bl]{\(w=2\)}}
\put(50,16){\makebox(0,0)[bl]{\(w=3\)}}

\put(01,15){\makebox(0,0)[bl]{The worst}}
\put(01,12){\makebox(0,0)[bl]{point}}

\put(0,10){\circle*{0.5}} \put(0,10){\circle{1.6}}

\end{picture}
\end{center}


\subsection{Morphological Multicriteria Design (Uncertainty)}

 The version of HMMD under uncertainty has been suggested in
 \cite{lev98}.
 Here a brief description of the approach is presented
 (by a simplified example).
%
%
%
 Fuzzy estimates for DAs or/and IC are considered.
 The following designations are used:
 ~(1)  \(i\) is an index corresponding to the design alternative;
 ~(2) \(\mu^{r}(i) \) is a membership function of priority
 \(r(i) \in \{ 1,2,3 \}\),
 we consider the following set: \({\mu^{r}_{l}(i), ~l=1,...,3}\)
 ~(\(l\) corresponds to value of priority);
 and
 ~(3) \(\mu^{w}( i_{1}, i_{2} ) \) is a membership function of compatibility
 \(w( i_{1}, i_{2} )\) between alternatives \(i_{1}\) and \(i_{2}\),
 we use the following set: ~\( \{ \mu^{w}_{k}( i_{1}, i_{2} ), ~k=0,...,3 \}\)
 ~(\(k\) corresponds to value of pairwise compatibility).
 Thus,
 let \({ \overline{ \mu^{w} (i_{1},i_{2}) } }\) be the following vector
 (i.e., pairwise compatibility vector):
 ~~\((\mu^{w}_{3}(i_{1},i_{2}), \mu^{w}_{2}(i_{1},i_{2}), \mu^{w}_{1}(i_{1},i_{2}),
 \mu^{w}_{0}(i_{1},i_{2}))\).

 Now let \(r^{\alpha } (i)\) and \(w^{\alpha } (i_{1},i_{2})\)
 be aggregated estimates for design alternative \(i\),
 and for a pair of design alternatives \((i_{1},i_{2})\)
 accordingly.

 Here a basic system example is depicted in Fig. 9.
 Table 6 contains normalized fuzzy priorities
 \(\{\mu^{r}_{l} (i)\}\) and realistic aggregated priorities
 \(\{r_{\alpha } (i)\}\).
 Table 7  contain normalized fuzzy compatibility
 \(\{\mu^{w}_{k} (i_{1},i_{2})\}\), and Table 8 presents
 realistic aggregated compatibility
  \(\{w^{\alpha } (i_{1},i_{2})\}\).

\begin{center}
\begin{picture}(55,27)

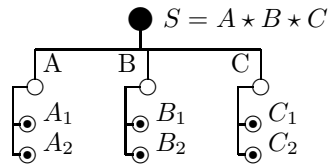
\put(0,0){\makebox(0,0)[bl] {Fig. 9. Example of system}}

\put(4,6){\makebox(0,8)[bl]{\(A_{2}\)}}
\put(4,10){\makebox(0,8)[bl]{\(A_{1}\)}}

\put(19,6){\makebox(0,8)[bl]{\(B_{2}\)}}
\put(19,10){\makebox(0,8)[bl]{\(B_{1}\)}}

\put(34,6){\makebox(0,8)[bl]{\(C_{2}\)}}
\put(34,10){\makebox(0,8)[bl]{\(C_{1}\)}}

\put(3,15){\circle{2}} \put(18,15){\circle{2}}
\put(33,15){\circle{2}}

\put(0,15){\line(1,0){02}} \put(15,15){\line(1,0){02}}
\put(30,15){\line(1,0){02}}

\put(0,15){\line(0,-1){9}} \put(15,15){\line(0,-1){9}}
\put(30,15){\line(0,-1){9}}

\put(30,10){\line(1,0){01}} \put(30,6){\line(1,0){01}}

\put(32,10){\circle{2}} \put(32,10){\circle*{1}}
\put(32,6){\circle{2}} \put(32,6){\circle*{1}}

\put(15,6){\line(1,0){01}} \put(15,10){\line(1,0){01}}

\put(17,6){\circle{2}} \put(17,6){\circle*{1}}

\put(17,10){\circle{2}} \put(17,10){\circle*{1}}

\put(0,6){\line(1,0){01}} \put(0,10){\line(1,0){01}}

\put(2,6){\circle{2}} \put(2,10){\circle{2}}
\put(2,6){\circle*{1}} \put(2,10){\circle*{1}}

\put(3,20){\line(0,-1){04}} \put(18,20){\line(0,-1){04}}
\put(33,20){\line(0,-1){04}}

\put(3,20){\line(1,0){30}} \put(17,20){\line(0,1){4}}
\put(17,24){\circle*{3}}

\put(4,17){\makebox(0,8)[bl]{A}} \put(14,17){\makebox(0,8)[bl]{B}}
\put(29,17){\makebox(0,8)[bl]{C}}

\put(20,23){\makebox(0,8)[bl]{\(S = A\star B\star C\)}}

\end{picture}
%
\begin{picture}(55,41)

\put(6,34){\makebox(0,0)[bl]{Table 6. Fuzzy priorities}}

\put(00,00){\line(1,0){51}}

\put(00,026){\line(1,0){51}} \put(00,032){\line(1,0){51}}

\put(00,0){\line(0,1){032}}

\put(11,0){\line(0,1){032}}

\put(21,0){\line(0,1){032}}

\put(31,0){\line(0,1){032}}

\put(41,0){\line(0,1){032}}

\put(51,0){\line(0,1){032}}


\put(01,02){\makebox(0,8)[bl]{\(C_{2}\)}}
\put(13,02){\makebox(0,8)[bl]{\(0.00\)}}
\put(23,02){\makebox(0,8)[bl]{\(1.00\)}}
\put(33,02){\makebox(0,8)[bl]{\(0.00\)}}
\put(45,02){\makebox(0,8)[bl]{\(2\)}}

\put(01,06){\makebox(0,8)[bl]{\(C_{1}\)}}
\put(13,06){\makebox(0,8)[bl]{\(1.00\)}}
\put(23,06){\makebox(0,8)[bl]{\(0.00\)}}
\put(33,06){\makebox(0,8)[bl]{\(0.00\)}}
\put(45,06){\makebox(0,8)[bl]{\(1\)}}

\put(01,10){\makebox(0,8)[bl]{\(B_{2}\)}}
\put(13,10){\makebox(0,8)[bl]{\(0.85\)}}
\put(23,10){\makebox(0,8)[bl]{\(0.15\)}}
\put(33,10){\makebox(0,8)[bl]{\(0.00\)}}
\put(45,10){\makebox(0,8)[bl]{\(1\)}}

\put(01,14){\makebox(0,8)[bl]{\(B_{1}\)}}
\put(13,14){\makebox(0,8)[bl]{\(0.15\)}}
\put(23,14){\makebox(0,8)[bl]{\(0.65\)}}
\put(33,14){\makebox(0,8)[bl]{\(0.20\)}}
\put(45,14){\makebox(0,8)[bl]{\(2\)}}

\put(01,18){\makebox(0,8)[bl]{\(A_{2}\)}}
\put(13,18){\makebox(0,8)[bl]{\(0.00\)}}
\put(23,18){\makebox(0,8)[bl]{\(0.05\)}}
\put(33,18){\makebox(0,8)[bl]{\(0.95\)}}
\put(45,18){\makebox(0,8)[bl]{\(3\)}}

\put(01,22){\makebox(0,8)[bl]{\(A_{1}\)}}
\put(13,22){\makebox(0,8)[bl]{\(1.00\)}}
\put(23,22){\makebox(0,8)[bl]{\(0.00\)}}
\put(33,22){\makebox(0,8)[bl]{\(0.00\)}}
\put(45,22){\makebox(0,8)[bl]{\(1\)}}

\put(01,28){\makebox(0,8)[bl]{DAs \(i\)}}

\put(12,28){\makebox(0,8)[bl]{\( \mu^{r}_{1}(i)  \)}}

\put(22,28){\makebox(0,8)[bl]{\( \mu^{r}_{2}(i)  \)}}

\put(32,28){\makebox(0,8)[bl]{\( \mu^{r}_{3}(i)  \)}}

\put(42,28){\makebox(0,8)[bl]{\( r^{\alpha }(i)  \)}}

\end{picture}
\end{center}

 Thus,
 the following relationships over the set of DAs
 can be obtained (by fuzzy priorities):
%
 (a) \(A_{1} >  A_{2}\);
 (b) \(B_{1} > B_{2}\); and
 (c) \(C_{1}\) and  \( C_{2}\) are non-comparable.
%
%
%
 Now it is reasonable to examine the following cases:

 {\it Case 1:} deterministic (aggregated) estimates of priorities
 for DAs \(\{r^{\alpha} (i) \}\)
 and deterministic (aggregated) estimates of compatibility for
 IC
 \(\{w^{\alpha} (i_{1},i_{2}) \}\) (a basic case).

{\it Case 2:} estimates of DAs are aggregated (deterministic)
 \(\{r^{\alpha } (i) \}\),
 and estimates of compatibility are fuzzy
 \(\{\mu^{w}_{k} (i_{1},i_{2}) \}\), \(\forall (i_{1},i_{2})\).

 {\it Case 3:} estimates of DAs are fuzzy
  \(\{\mu^{r}_{l} (i) \}\) \(\forall i\)
 and estimates of IC are aggregated (deterministic)
 \(\{w^{\alpha} (i_{1},i_{2}) \}\), \(\forall (i_{1},i_{2})\).

 {\it Case 4:} estimates of DAs are fuzzy
  \(\{\mu^{r}_{l} (i) \}\) \(\forall i\)
   and estimates of compatibility are fuzzy
 \(\{\mu^{w}_{k} (i_{1},i_{2}) \}\), \(\forall (i_{1},i_{2})\).

\begin{center}
\begin{picture}(110,31)

\put(32,26){\makebox(0,0)[bl]{Table 7. Fuzzy compatibility}}



\put(00,00){\line(1,0){110}}

\put(00,18){\line(1,0){110}} \put(00,24){\line(1,0){110}}

\put(00,0){\line(0,1){24}}

\put(10,0){\line(0,1){24}}

\put(110,0){\line(0,1){24}}


\put(35,18){\line(0,1){6}}

\put(60,18){\line(0,1){6}}

\put(85,18){\line(0,1){6}}


\put(01,14){\makebox(0,8)[bl]{\(A_{1}\)}}
\put(01,10){\makebox(0,8)[bl]{\(A_{2}\)}}
\put(01,06){\makebox(0,8)[bl]{\(B_{1}\)}}
\put(01,02){\makebox(0,8)[bl]{\(B_{2}\)}}


\put(61,02){\makebox(0,8)[bl]{\(0.0;0.4;0.5;0.1\)}}
\put(86,02){\makebox(0,8)[bl]{\(0.2;0.5;0.3;0.0\)}}


\put(61,06){\makebox(0,8)[bl]{\(0.6;0.3;0.1;0.0\)}}
\put(86,06){\makebox(0,8)[bl]{\(0.0;0.5;0.5;0.0\)}}

\put(11,10){\makebox(0,8)[bl]{\(0.1;0.2;0.4;0.3\)}}
\put(36,10){\makebox(0,8)[bl]{\(0.7;0.3;0.0;0.0\)}}
\put(61,10){\makebox(0,8)[bl]{\(0.4;0.4;0.2;0.0\)}}
\put(86,10){\makebox(0,8)[bl]{\(0.0;0.7;0.3;0.0\)}}


\put(11,14){\makebox(0,8)[bl]{\(0.5;0.2;0.3;0.0\)}}
\put(36,14){\makebox(0,8)[bl]{\(0.0;0.3;0.4;0.3\)}}
\put(61,14){\makebox(0,8)[bl]{\(0.0;0.4;0.5;0.1\)}}
\put(86,14){\makebox(0,8)[bl]{\(0.2;0.4;0.3;0.1\)}}

\put(21,20){\makebox(0,8)[bl]{\(B_{1}\)}}

\put(46,20){\makebox(0,8)[bl]{\(B_{2}\)}}

\put(71,20){\makebox(0,8)[bl]{\(C_{1}\)}}

\put(96,20){\makebox(0,8)[bl]{\(C_{2}\)}}

\end{picture}
\end{center}

\begin{center}
\begin{picture}(50,31)

\put(0,26){\makebox(0,0)[bl]{Table 8. Aggregated
 compatibility}}

\put(00,00){\line(1,0){50}}

\put(00,18){\line(1,0){50}} \put(00,24){\line(1,0){50}}

\put(00,0){\line(0,1){24}}

\put(10,0){\line(0,1){24}}

\put(50,0){\line(0,1){24}}


\put(20,18){\line(0,1){6}} \put(30,18){\line(0,1){6}}
\put(40,18){\line(0,1){6}}


\put(01,14){\makebox(0,8)[bl]{\(A_{1}\)}}
\put(01,10){\makebox(0,8)[bl]{\(A_{2}\)}}
\put(01,06){\makebox(0,8)[bl]{\(B_{1}\)}}
\put(01,02){\makebox(0,8)[bl]{\(B_{2}\)}}

\put(34,02){\makebox(0,8)[bl]{\(1\)}}
\put(44,02){\makebox(0,8)[bl]{\(2\)}}

\put(34,06){\makebox(0,8)[bl]{\(3\)}}
\put(44,06){\makebox(0,8)[bl]{\(1\)}}

\put(14,10){\makebox(0,8)[bl]{\(1\)}}
\put(24,10){\makebox(0,8)[bl]{\(3\)}}
\put(34,10){\makebox(0,8)[bl]{\(2\)}}
\put(44,10){\makebox(0,8)[bl]{\(2\)}}

\put(14,14){\makebox(0,8)[bl]{\(3\)}}
\put(24,14){\makebox(0,8)[bl]{\(1\)}}
\put(34,14){\makebox(0,8)[bl]{\(1\)}}
\put(44,14){\makebox(0,8)[bl]{\(2\)}}

\put(12,20){\makebox(0,8)[bl]{\(B_{1}\)}}

\put(22,20){\makebox(0,8)[bl]{\(B_{2}\)}}

\put(32,20){\makebox(0,8)[bl]{\(C_{1}\)}}

\put(42,20){\makebox(0,8)[bl]{\(C_{2}\)}}

\end{picture}
\end{center}

 Fig. 10 illustrates the cases above
 (top index of composite DAs corresponds to the case).
 Clearly, that main solving method is based on two stages:
 (1) generation of feasible composite decisions; and
 (2) selection of Pareto-efficient decisions.
 Unfortunately, it is reasonable to point out the following two
 significant features of our synthesis problem with fuzzy
 estimates:
 (a) complexity of corresponding combinatorial problems  is
 increasing because a number of analyzed composite decisions is
 more than in deterministic case; and
 (b) it is necessary to construct a preference rule to select the
 best fuzzy decision(s).


\begin{center}
\begin{picture}(88,102)
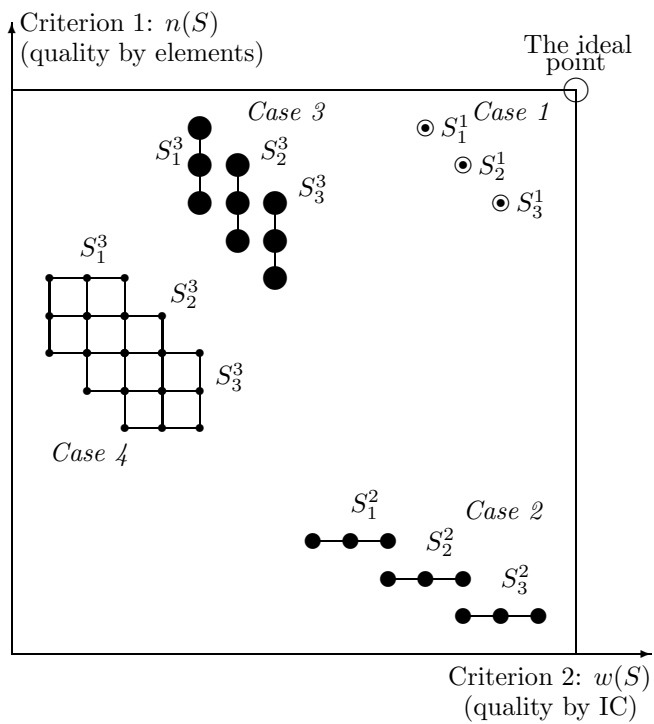
\put(0,0){\makebox(0,0)[bl]{Fig. 10. Illustrative space for
composite DAs (from \cite{lev98})}}

\put(01,97){\makebox(0,8)[bl]{Criterion 1: \(n(S)\) }}
\put(01,93){\makebox(0,8)[bl]{(quality by elements)}}

\put(58,10){\makebox(0,8)[bl]{Criterion 2: \(w(S)\)}}
\put(60,06){\makebox(0,8)[bl]{(quality by IC)}}


\put(0,15){\vector(1,0){85}}

\put(0,15){\vector(0,1){84}}

\put(0,90){\line(1,0){75}}

\put(75,15){\line(0,1){75}}


\put(68,95){\makebox(0,8)[bl]{The ideal}}
\put(71,92){\makebox(0,8)[bl]{point}}

\put(75,90){\circle{3}}


\put(61,86){\makebox(0,8)[bl]{{\it Case 1}}}

\put(55,85){\circle{2}} \put(55,85){\circle*{1}}
\put(57,83){\makebox(0,8)[bl]{\(S^{1}_{1}\)}}

\put(60,80){\circle{2}} \put(60,80){\circle*{1}}
\put(62,78){\makebox(0,8)[bl]{\(S^{1}_{2}\)}}

\put(65,75){\circle{2}} \put(65,75){\circle*{1}}
\put(67,73){\makebox(0,8)[bl]{\(S^{1}_{3}\)}}


\put(60,33){\makebox(0,8)[bl]{{\it Case 2}}}

\put(40,30){\circle*{2}} \put(45,30){\circle*{2}}
\put(50,30){\circle*{2}}

\put(40,30){\line(1,0){10}}

\put(45,33){\makebox(0,8)[bl]{\(S^{2}_{1}\)}}
\put(50,25){\circle*{2}} \put(55,25){\circle*{2}}
\put(60,25){\circle*{2}}

\put(50,25){\line(1,0){10}}

\put(55,28){\makebox(0,8)[bl]{\(S^{2}_{2}\)}}
\put(60,20){\circle*{2}} \put(65,20){\circle*{2}}
\put(70,20){\circle*{2}}

\put(60,20){\line(1,0){10}}

\put(65,23){\makebox(0,8)[bl]{\(S^{2}_{3}\)}}


\put(31,86){\makebox(0,8)[bl]{{\it Case 3}}}

\put(25,85){\circle*{3}} \put(25,80){\circle*{3}}
\put(25,75){\circle*{3}}

\put(25,85){\line(0,-1){10}}

\put(19,80){\makebox(0,8)[bl]{\(S^{3}_{1}\)}}

\put(30,80){\circle*{3}} \put(30,75){\circle*{3}}
\put(30,70){\circle*{3}}


\put(30,70){\line(0,1){10}}

\put(33,80){\makebox(0,8)[bl]{\(S^{3}_{2}\)}}
\put(35,75){\circle*{3}} \put(35,70){\circle*{3}}
\put(35,65){\circle*{3}}

\put(35,65){\line(0,1){10}}

\put(38,75){\makebox(0,8)[bl]{\(S^{3}_{3}\)}}


\put(05,40){\makebox(0,8)[bl]{{\it Case 4}}}

\put(05,65){\circle*{1}} \put(05,60){\circle*{1}}
\put(05,55){\circle*{1}}

\put(10,65){\circle*{1}} \put(10,60){\circle*{1}}
\put(10,55){\circle*{1}}

\put(15,65){\circle*{1}} \put(15,60){\circle*{1}}
\put(15,55){\circle*{1}}

\put(05,55){\line(0,1){10}} \put(10,55){\line(0,1){10}}
\put(15,55){\line(0,1){10}}

\put(05,65){\line(1,0){10}} \put(05,60){\line(1,0){10}}
\put(05,55){\line(1,0){10}}

\put(09,67){\makebox(0,8)[bl]{\(S^{3}_{1}\)}}


\put(10,60){\circle*{1}} \put(10,55){\circle*{1}}
\put(10,50){\circle*{1}}

\put(15,60){\circle*{1}} \put(15,55){\circle*{1}}
\put(15,50){\circle*{1}}

\put(20,60){\circle*{1}} \put(20,55){\circle*{1}}
\put(20,50){\circle*{1}}

\put(10,50){\line(0,1){10}} \put(15,50){\line(0,1){10}}
\put(20,50){\line(0,1){10}}

\put(10,60){\line(1,0){10}} \put(10,55){\line(1,0){10}}
\put(10,50){\line(1,0){10}}

\put(21,61){\makebox(0,8)[bl]{\(S^{3}_{2}\)}}


\put(15,55){\circle*{1}} \put(15,50){\circle*{1}}
\put(15,45){\circle*{1}}

\put(20,55){\circle*{1}} \put(20,50){\circle*{1}}
\put(20,45){\circle*{1}}

\put(25,55){\circle*{1}} \put(25,50){\circle*{1}}
\put(25,45){\circle*{1}}

\put(15,45){\line(0,1){10}} \put(20,45){\line(0,1){10}}
\put(25,45){\line(0,1){10}}

\put(15,55){\line(1,0){10}} \put(15,50){\line(1,0){10}}
\put(15,45){\line(1,0){10}}

\put(27,50){\makebox(0,8)[bl]{\(S^{3}_{3}\)}}


\end{picture}
\end{center}


\section{Design Examples for GSM Network}

 In recent two decades, the significance of GSM network has been  increased
 (e.g.,
 \cite{chen08},
  \cite{gh00}, \cite{haine99},
 \cite{lin97},  \cite{mehr97}, \cite{rah93},
 \cite{tisal97}).
 Thus,
 there exists a need of the design and maintenance of
 this kind of communication systems.
 Here a numerical example for design of GSM network
 (a modification of an example from \cite{levvis07})
  is used to illustrate and to compare
 several MA-based methods:
 basic MA,
 method of closeness to ideal point,
 Pareto-based MA,
 multiple choice problem, and
 HMMD.

 \subsection{Initial Example}

  The general tree-like simplified model of GSM network is as follows
 (Fig. 11, the developers of DAs are pointed out in parentheses):

~~

 {\bf 0.} GSM network ~\(S = A \star B \).

 {\bf 1.} Switching SubSystem SSS ~( \( A = M \star L \)).

 {\it 1.1.} Mobile Switching Center/Visitors Location Register MSC/VLR
 ~\( M:\)~
 \(M_{1}\) (Motorola),
 \(M_{2}\) (Alcatel),
 \(M_{3}\) (Huawei),
 \(M_{4}\) (Siemens), and
 \(M_{5}\) (Ericsson).

 {\it 1.2.} Home Location Register/Authentification Center HLR/AC
 ~\( L:\)~
 \(L_{1}\) (Motorola),
 \(L_{2}\) (Ericsson),
 \(L_{3}\) (Alcatel), and
 \(L_{4}\) (Huawei).

 {\bf 2.} Base Station SubSystem BSS
 ~(\( B = V \star U \star T \)).

 {\it 2.1.} Base Station Controller BSC
 ~\( V:\)~
 \(V_{1}\) (Motorola),
 \(V_{2}\) (Ericsson),
 \(V_{3}\) (Alcatel),
 \(V_{4}\) (Huawei),
 \(V_{5}\) (Nokia), and
 \(V_{6}\) (Siemens).

 {\it 2.2.} Base Transceiver Station BTS
 ~\( U:\)~
  \(U_{1}\) (Motorola),
  \(U_{2}\) (Ericsson),
  \(U_{3}\) (Alcatel),
  \(U_{4}\) (Huawei), and
  \(U_{5}\) (Nokia).

 {\it 2.3.} Transceivers TRx
 ~\( T:\)~
 \(T_{1}\) (Alcatel),
 \(T_{2}\) (Ericsson),
 \(T_{3}\) (Motorola),
 \(T_{4}\) (Huawei), and
 \(T_{5}\) (Siemens).

~~

 Note an initial set of possible composite decisions contained
 \(3000\) combinations
 (\( 5 \times 4 \times 6 \times  5 \times 5 \)).

\begin{center}
\begin{picture}(137,39)

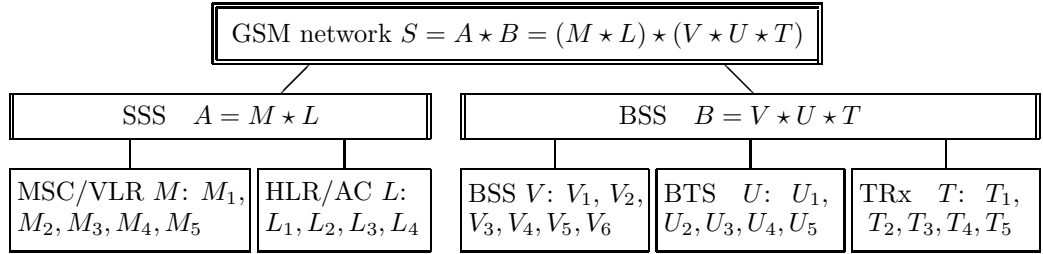
\put(26,0){\makebox(0,0)[bl]{Fig. 11. General simplified structure
of GSM network}}


\put(29.5,32){\makebox(0,0)[bl]{GSM  network  \(S = A \star B =
 (M \star L) \star ( V \star U \star T) \)}}

\put(27,30){\line(1,0){81}} \put(27,38){\line(1,0){81}}
\put(27,30){\line(0,1){08}} \put(108,30){\line(0,1){08}}

\put(27.5,30.5){\line(1,0){80}} \put(27.5,37.5){\line(1,0){80}}
\put(27.5,30.5){\line(0,1){07}} \put(107.5,30){\line(0,1){07}}


\put(40,30){\line(-1,-1){4}}

\put(81,22){\makebox(0,0)[bl]{BSS ~ \(B = V \star U \star T \)}}

\put(60,20){\line(1,0){77}} \put(60,26){\line(1,0){77}}
\put(60,20){\line(0,1){06}} \put(137,20){\line(0,1){06}}
\put(60.5,20){\line(0,1){06}} \put(136.5,20){\line(0,1){06}}


\put(95,30){\line(1,-1){4}}

\put(15,22){\makebox(0,0)[bl]{SSS ~ \(A = M \star L \)}}

\put(00,20){\line(1,0){56}} \put(00,26){\line(1,0){56}}
\put(00,20){\line(0,1){06}} \put(56,20){\line(0,1){06}}
\put(00.5,20){\line(0,1){06}} \put(55.5,20){\line(0,1){06}}


\put(124.5,16){\line(0,1){4}}

\put(113,11){\makebox(0,0)[bl]{TRx  ~ \(T \):~ \(T_{1}\), }}
\put(113,07){\makebox(0,0)[bl]{~\(T_{2}, T_{3},  T_{4}, T_{5}\)}}

\put(112,05){\line(1,0){25}} \put(112,16){\line(1,0){25}}
\put(112,05){\line(0,1){11}} \put(137,05){\line(0,1){11}}


\put(98.5,16){\line(0,1){4}}

\put(87,11){\makebox(0,0)[bl]{BTS  ~ \(U \):~ \(U_{1}\), }}
\put(87,07){\makebox(0,0)[bl]{\(U_{2}, U_{3},  U_{4}, U_{5}\)}}

\put(86,05){\line(1,0){25}} \put(86,16){\line(1,0){25}}
\put(86,05){\line(0,1){11}} \put(111,05){\line(0,1){11}}


\put(72.5,16){\line(0,1){4}}

\put(61,11){\makebox(0,0)[bl]{BSS  \(V \): \(V_{1}\), \(V_{2}\),
}}

\put(61,07){\makebox(0,0)[bl]{\(V_{3},  V_{4}, V_{5}, V_{6}\)}}

\put(60,05){\line(1,0){25}} \put(60,16){\line(1,0){25}}
\put(60,05){\line(0,1){11}} \put(85,05){\line(0,1){11}}


\put(44.5,16){\line(0,1){4}}

\put(34,11){\makebox(0,0)[bl]{HLR/AC  \(L \):}}
\put(34,07){\makebox(0,0)[bl]{\(L_{1}, L_{2}, L_{3}, L_{4}\)}}

\put(33,05){\line(1,0){23}} \put(33,16){\line(1,0){23}}
\put(33,05){\line(0,1){11}} \put(56,05){\line(0,1){11}}


\put(16,16){\line(0,1){4}}

\put(01,11){\makebox(0,0)[bl]{MSC/VLR  \(M \): \(M_{1}\),}}
\put(01,07){\makebox(0,0)[bl]{\(M_{2}, M_{3}, M_{4}, M_{5} \)}}

\put(00,05){\line(1,0){32}} \put(00,16){\line(1,0){32}}
\put(00,05){\line(0,1){11}} \put(32,05){\line(0,1){11}}

\end{picture}
\end{center}

 Let us consider criteria for system components as follows
 (weights of criteria are pointed out in parentheses):

 {\bf 1.} \(M\):~
  maximal number of datapathes
  (1000 pathes) ~(\(C_{m1}\),
  0.2);
  maximal capacity VLR (100000 subscribers) ~(\(C_{m2}\), 0.2);
  price index ( 100000/price(USD) ) ~(\(C_{m3}\), 0.2);
  power consumption ( 1/power consumption (kWt) ) ~(\(C_{m4}\), 0.2); and
  number of communication and signaling interfaces ~(\(C_{m5}\), 0.2).

 {\bf 2.}  \(L\):~
  maximal number of subscribers (100000 subscribers)
  ~(\(C_{l1}\), 0.25);
  volume of service provided ~(\(C_{l2}\), 0.25);
  reliability (scale \([1,...,10]\)) (\(C_{l3}\), 0.25); and
  integratability
 (scale \([1,...,10]\))
   (\(C_{l4}\), 0.25 ).

 {\bf 3.}  \(V\):~
  price index ( 100000/cost (USD) ) ~(\(C_{v1}\), 0.25);
 maximal number of BTS ~(\(C_{v2}\), 0.25);
  handover quality ~(\(C_{v3}\), 0.25); and
 throughput
   ~(\(C_{v4}\), 0.25).

 {\bf 4.}  \(U\):~
  maximal number of TRx ~(\(C_{u1}\), 0.25);
  capacity  ~(\(C_{u2}\), 0.25);
  price index  ( 100000/cost(USD) ) ~(\(C_{u3}\), 0.25); and
  reliability (scale \([1,...,10]\)) ~(\(C_{u4}\), 0.25).

 {\bf 5.}  \(T\):~
 maximum power-carrying capacity  ~(\(C_{t1}\), 0.3);
   throughput
    ~(\(C_{t2}\), 0.2);
%
  price index ( 100000/ cost(USD) ) ~(\(C_{t3}\), 0.25); and
   reliability (scale \([1,...,10]\)) ~(\(C_{t4}\), 0.25).

%
  Tables 9, 10, 11, 12, and 13 contain estimates of DAs
  upon criteria above  (data from catalogues, expert judgment)
   and their resultant priorities
   (the priorities are based on multicriteria ranking by
  an outranking technique
   \cite{roy96}).
%
 Compatibility estimates
 are contained in Tables 14 and 15
  (expert judgment).

\begin{center}
\begin{picture}(64,40)
\put(12,36){\makebox(0,0)[bl]{Table 9. Estimates for \(M\)}}

\put(00,0){\line(1,0){64}} \put(00,22){\line(1,0){64}}
\put(10,28){\line(1,0){40}} \put(00,34){\line(1,0){64}}

\put(00,0){\line(0,1){34}} \put(10,00){\line(0,1){34}}
\put(50,0){\line(0,1){34}} \put(64,0){\line(0,1){34}}

\put(18,22){\line(0,1){6}} \put(26,22){\line(0,1){6}}
\put(34,22){\line(0,1){6}} \put(42,22){\line(0,1){6}}

\put(01,30){\makebox(0,0)[bl]{DAs}}
\put(23,30){\makebox(0,0)[bl]{Criteria}}

\put(51,30){\makebox(0,0)[bl]{Priority}}
\put(56,27){\makebox(0,0)[bl]{\(r\)}}


\put(11,24){\makebox(0,0)[bl]{\(C_{m1}\)}}
\put(19,24){\makebox(0,0)[bl]{\(C_{m2}\)}}
\put(27,24){\makebox(0,0)[bl]{\(C_{m3}\)}}
\put(35,24){\makebox(0,0)[bl]{\(C_{m4}\)}}
\put(43,24){\makebox(0,0)[bl]{\(C_{m5}\)}}


\put(01,18){\makebox(0,0)[bl]{\(M_{1}\)}}
\put(01,14){\makebox(0,0)[bl]{\(M_{2}\)}}
\put(01,10){\makebox(0,0)[bl]{\(M_{3}\)}}
\put(01,06){\makebox(0,0)[bl]{\(M_{4}\)}}
\put(01,02){\makebox(0,0)[bl]{\(M_{5}\)}}

\put(13,18){\makebox(0,0)[bl]{\(3.7\)}}
\put(21,18){\makebox(0,0)[bl]{\(8.6\)}}
\put(29,18){\makebox(0,0)[bl]{\(6\)}}
\put(37,18){\makebox(0,0)[bl]{\(5.1\)}}
\put(45,18){\makebox(0,0)[bl]{\(4\)}}

\put(56,18){\makebox(0,0)[bl]{\(2\)}}


\put(13,14){\makebox(0,0)[bl]{\(4.0\)}}
\put(21,14){\makebox(0,0)[bl]{\(11\)}}
\put(29,14){\makebox(0,0)[bl]{\(8\)}}
\put(37,14){\makebox(0,0)[bl]{\(7\)}}
\put(45,14){\makebox(0,0)[bl]{\(5\)}}

\put(56,14){\makebox(0,0)[bl]{\(3\)}}


\put(13,10){\makebox(0,0)[bl]{\(4.1\)}}
\put(21,10){\makebox(0,0)[bl]{\(10\)}}
\put(29,10){\makebox(0,0)[bl]{\(9\)}}
\put(37,10){\makebox(0,0)[bl]{\(7\)}}
\put(45,10){\makebox(0,0)[bl]{\(4\)}}

\put(56,10){\makebox(0,0)[bl]{\(3\)}}


\put(13,06){\makebox(0,0)[bl]{\(3.2\)}}
\put(21,06){\makebox(0,0)[bl]{\(7\)}}
\put(29,06){\makebox(0,0)[bl]{\(5\)}}
\put(37,06){\makebox(0,0)[bl]{\(6\)}}
\put(45,06){\makebox(0,0)[bl]{\(3\)}}

\put(56,06){\makebox(0,0)[bl]{\(1\)}}


\put(13,02){\makebox(0,0)[bl]{\(3.5\)}}
\put(21,02){\makebox(0,0)[bl]{\(8.7\)}}
\put(29,02){\makebox(0,0)[bl]{\(6.2\)}}
\put(37,02){\makebox(0,0)[bl]{\(5\)}}
\put(45,02){\makebox(0,0)[bl]{\(4\)}}

\put(56,02){\makebox(0,0)[bl]{\(2\)}}


\end{picture}
\end{center}

\begin{center}
\begin{picture}(59,40)

\put(09,32){\makebox(0,0)[bl]{Table 10. Estimates for \(L\)}}

\put(00,0){\line(1,0){56}} \put(00,18){\line(1,0){56}}
\put(10,24){\line(1,0){32}} \put(00,30){\line(1,0){56}}

\put(00,0){\line(0,1){30}} \put(10,00){\line(0,1){30}}
\put(42,0){\line(0,1){30}} \put(56,0){\line(0,1){30}}

\put(18,18){\line(0,1){6}} \put(26,18){\line(0,1){6}}
\put(34,18){\line(0,1){6}}

\put(01,26){\makebox(0,0)[bl]{DAs}}
\put(20,26){\makebox(0,0)[bl]{Criteria}}
\put(43,26){\makebox(0,0)[bl]{Priority}}
\put(48,23){\makebox(0,0)[bl]{\(r\)}}

\put(11,20){\makebox(0,0)[bl]{\(C_{l1}\)}}
\put(19,20){\makebox(0,0)[bl]{\(C_{l2}\)}}
\put(27,20){\makebox(0,0)[bl]{\(C_{l3}\)}}
\put(35,20){\makebox(0,0)[bl]{\(C_{l4}\)}}


\put(01,14){\makebox(0,0)[bl]{\(L_{1}\)}}
\put(01,10){\makebox(0,0)[bl]{\(L_{2}\)}}
\put(01,06){\makebox(0,0)[bl]{\(L_{3}\)}}
\put(01,02){\makebox(0,0)[bl]{\(L_{4}\)}}


\put(13,14){\makebox(0,0)[bl]{\(9\)}}
\put(21,14){\makebox(0,0)[bl]{\(7\)}}
\put(29,14){\makebox(0,0)[bl]{\(7\)}}
\put(37,14){\makebox(0,0)[bl]{\(8\)}}

\put(48,14){\makebox(0,0)[bl]{\(1\)}}


\put(13,10){\makebox(0,0)[bl]{\(10\)}}
\put(21,10){\makebox(0,0)[bl]{\(4\)}}
\put(29,10){\makebox(0,0)[bl]{\(9\)}}
\put(37,10){\makebox(0,0)[bl]{\(8\)}}

\put(48,10){\makebox(0,0)[bl]{\(1\)}}


\put(13,06){\makebox(0,0)[bl]{\(12\)}}
\put(21,06){\makebox(0,0)[bl]{\(8\)}}
\put(29,06){\makebox(0,0)[bl]{\(10\)}}
\put(37,06){\makebox(0,0)[bl]{\(10\)}}

\put(48,06){\makebox(0,0)[bl]{\(2\)}}


\put(13,02){\makebox(0,0)[bl]{\(9\)}}
\put(21,02){\makebox(0,0)[bl]{\(5\)}}
\put(29,02){\makebox(0,0)[bl]{\(8\)}}
\put(37,02){\makebox(0,0)[bl]{\(8\)}}

\put(48,02){\makebox(0,0)[bl]{\(1\)}}


\end{picture}
%
\begin{picture}(56,44)
\put(09,40){\makebox(0,0)[bl]{Table 11. Estimates for \(V\)}}

\put(00,0){\line(1,0){56}} \put(00,26){\line(1,0){56}}
\put(10,32){\line(1,0){32}} \put(00,38){\line(1,0){56}}

\put(00,0){\line(0,1){38}} \put(10,00){\line(0,1){38}}
\put(42,0){\line(0,1){38}} \put(56,0){\line(0,1){38}}

\put(18,26){\line(0,1){6}} \put(26,26){\line(0,1){6}}
\put(34,26){\line(0,1){6}}

\put(01,34){\makebox(0,0)[bl]{DAs}}
\put(20,34){\makebox(0,0)[bl]{Criteria}}
\put(43,34){\makebox(0,0)[bl]{Priority}}
\put(48,31){\makebox(0,0)[bl]{\(r\)}}

\put(11,28){\makebox(0,0)[bl]{\(C_{v1}\)}}
\put(19,28){\makebox(0,0)[bl]{\(C_{v2}\)}}
\put(27,28){\makebox(0,0)[bl]{\(C_{v3}\)}}
\put(35,28){\makebox(0,0)[bl]{\(C_{v4}\)}}


\put(01,22){\makebox(0,0)[bl]{\(V_{1}\)}}
\put(01,18){\makebox(0,0)[bl]{\(V_{2}\)}}
\put(01,14){\makebox(0,0)[bl]{\(V_{3}\)}}
\put(01,10){\makebox(0,0)[bl]{\(V_{4}\)}}
\put(01,06){\makebox(0,0)[bl]{\(V_{5}\)}}
\put(01,02){\makebox(0,0)[bl]{\(V_{6}\)}}


\put(13,22){\makebox(0,0)[bl]{\(6\)}}
\put(21,22){\makebox(0,0)[bl]{\(4\)}}
\put(29,22){\makebox(0,0)[bl]{\(3\)}}
\put(37,22){\makebox(0,0)[bl]{\(4\)}}

\put(48,22){\makebox(0,0)[bl]{\(1\)}}


\put(13,18){\makebox(0,0)[bl]{\(7\)}}
\put(21,18){\makebox(0,0)[bl]{\(5\)}}
\put(29,18){\makebox(0,0)[bl]{\(7\)}}
\put(37,18){\makebox(0,0)[bl]{\(7\)}}

\put(48,18){\makebox(0,0)[bl]{\(2\)}}


\put(13,14){\makebox(0,0)[bl]{\(9\)}}
\put(21,14){\makebox(0,0)[bl]{\(7\)}}
\put(29,14){\makebox(0,0)[bl]{\(10\)}}
\put(37,14){\makebox(0,0)[bl]{\(7\)}}

\put(48,14){\makebox(0,0)[bl]{\(3\)}}


\put(13,10){\makebox(0,0)[bl]{\(7\)}}
\put(21,10){\makebox(0,0)[bl]{\(5\)}}
\put(29,10){\makebox(0,0)[bl]{\(8\)}}
\put(37,10){\makebox(0,0)[bl]{\(6\)}}

\put(48,10){\makebox(0,0)[bl]{\(2\)}}


\put(13,06){\makebox(0,0)[bl]{\(6\)}}
\put(21,06){\makebox(0,0)[bl]{\(3\)}}
\put(29,06){\makebox(0,0)[bl]{\(4\)}}
\put(37,06){\makebox(0,0)[bl]{\(4\)}}

\put(48,06){\makebox(0,0)[bl]{\(1\)}}


\put(13,02){\makebox(0,0)[bl]{\(10\)}}
\put(21,02){\makebox(0,0)[bl]{\(6\)}}
\put(29,02){\makebox(0,0)[bl]{\(9\)}}
\put(37,02){\makebox(0,0)[bl]{\(7\)}}

\put(48,02){\makebox(0,0)[bl]{\(3\)}}


\end{picture}
\end{center}

\begin{center}
\begin{picture}(59,40)

\put(09,36){\makebox(0,0)[bl]{Table 12. Estimates for \(U\)}}

\put(00,0){\line(1,0){56}} \put(00,22){\line(1,0){56}}
\put(10,28){\line(1,0){32}} \put(00,34){\line(1,0){56}}

\put(00,0){\line(0,1){34}} \put(10,00){\line(0,1){34}}
\put(42,0){\line(0,1){34}} \put(56,0){\line(0,1){34}}

\put(18,22){\line(0,1){6}} \put(26,22){\line(0,1){6}}
\put(34,22){\line(0,1){6}}

\put(01,30){\makebox(0,0)[bl]{DAs}}
\put(20,30){\makebox(0,0)[bl]{Criteria}}
\put(43,30){\makebox(0,0)[bl]{Priority}}
\put(48,27){\makebox(0,0)[bl]{\(r\)}}

\put(11,24){\makebox(0,0)[bl]{\(C_{u1}\)}}
\put(19,24){\makebox(0,0)[bl]{\(C_{u2}\)}}
\put(27,24){\makebox(0,0)[bl]{\(C_{u3}\)}}
\put(35,24){\makebox(0,0)[bl]{\(C_{u4}\)}}


\put(01,18){\makebox(0,0)[bl]{\(U_{1}\)}}
\put(01,14){\makebox(0,0)[bl]{\(U_{2}\)}}
\put(01,10){\makebox(0,0)[bl]{\(U_{3}\)}}
\put(01,06){\makebox(0,0)[bl]{\(U_{4}\)}}
\put(01,02){\makebox(0,0)[bl]{\(U_{5}\)}}

\put(13,18){\makebox(0,0)[bl]{\(2\)}}
\put(21,18){\makebox(0,0)[bl]{\(7\)}}
\put(29,18){\makebox(0,0)[bl]{\(5\)}}
\put(37,18){\makebox(0,0)[bl]{\(8\)}}

\put(48,18){\makebox(0,0)[bl]{\(1\)}}


\put(13,14){\makebox(0,0)[bl]{\(4\)}}
\put(21,14){\makebox(0,0)[bl]{\(10\)}}
\put(29,14){\makebox(0,0)[bl]{\(6\)}}
\put(37,14){\makebox(0,0)[bl]{\(10\)}}

\put(48,14){\makebox(0,0)[bl]{\(3\)}}


\put(13,10){\makebox(0,0)[bl]{\(3\)}}
\put(21,10){\makebox(0,0)[bl]{\(9\)}}
\put(29,10){\makebox(0,0)[bl]{\(6\)}}
\put(37,10){\makebox(0,0)[bl]{\(10\)}}

\put(48,10){\makebox(0,0)[bl]{\(2\)}}


\put(13,06){\makebox(0,0)[bl]{\(3\)}}
\put(21,06){\makebox(0,0)[bl]{\(6\)}}
\put(29,06){\makebox(0,0)[bl]{\(3\)}}
\put(37,06){\makebox(0,0)[bl]{\(7\)}}

\put(48,06){\makebox(0,0)[bl]{\(1\)}}


\put(13,02){\makebox(0,0)[bl]{\(3\)}}
\put(21,02){\makebox(0,0)[bl]{\(10\)}}
\put(29,02){\makebox(0,0)[bl]{\(6\)}}
\put(37,02){\makebox(0,0)[bl]{\(9\)}}

\put(48,02){\makebox(0,0)[bl]{\(2\)}}


\end{picture}
%
\begin{picture}(56,40)
\put(09,36){\makebox(0,0)[bl]{Table 13. Estimates for \(T\)}}

\put(00,0){\line(1,0){56}} \put(00,22){\line(1,0){56}}
\put(10,28){\line(1,0){32}} \put(00,34){\line(1,0){56}}

\put(00,0){\line(0,1){34}} \put(10,00){\line(0,1){34}}
\put(42,0){\line(0,1){34}} \put(56,0){\line(0,1){34}}

\put(18,22){\line(0,1){6}} \put(26,22){\line(0,1){6}}
\put(34,22){\line(0,1){6}}

\put(01,30){\makebox(0,0)[bl]{DAs}}
\put(20,30){\makebox(0,0)[bl]{Criteria}}
\put(43,30){\makebox(0,0)[bl]{Priority}}
\put(48,27){\makebox(0,0)[bl]{\(r\)}}

\put(11,24){\makebox(0,0)[bl]{\(C_{t1}\)}}
\put(19,24){\makebox(0,0)[bl]{\(C_{t2}\)}}
\put(27,24){\makebox(0,0)[bl]{\(C_{t3}\)}}
\put(35,24){\makebox(0,0)[bl]{\(C_{t4}\)}}


\put(01,18){\makebox(0,0)[bl]{\(T_{1}\)}}
\put(01,14){\makebox(0,0)[bl]{\(T_{2}\)}}
\put(01,10){\makebox(0,0)[bl]{\(T_{3}\)}}
\put(01,06){\makebox(0,0)[bl]{\(T_{4}\)}}
\put(01,02){\makebox(0,0)[bl]{\(T_{5}\)}}

\put(13,18){\makebox(0,0)[bl]{\(9\)}}
\put(21,18){\makebox(0,0)[bl]{\(7\)}}
\put(29,18){\makebox(0,0)[bl]{\(10\)}}
\put(37,18){\makebox(0,0)[bl]{\(7\)}}

\put(48,18){\makebox(0,0)[bl]{\(3\)}}


\put(13,14){\makebox(0,0)[bl]{\(6\)}}
\put(21,14){\makebox(0,0)[bl]{\(4\)}}
\put(29,14){\makebox(0,0)[bl]{\(3\)}}
\put(37,14){\makebox(0,0)[bl]{\(4\)}}

\put(48,14){\makebox(0,0)[bl]{\(1\)}}


\put(13,10){\makebox(0,0)[bl]{\(7\)}}
\put(21,10){\makebox(0,0)[bl]{\(5\)}}
\put(29,10){\makebox(0,0)[bl]{\(7\)}}
\put(37,10){\makebox(0,0)[bl]{\(7\)}}

\put(48,10){\makebox(0,0)[bl]{\(2\)}}


\put(13,06){\makebox(0,0)[bl]{\(7\)}}
\put(21,06){\makebox(0,0)[bl]{\(5\)}}
\put(29,06){\makebox(0,0)[bl]{\(8\)}}
\put(37,06){\makebox(0,0)[bl]{\(6\)}}

\put(48,06){\makebox(0,0)[bl]{\(2\)}}


\put(13,02){\makebox(0,0)[bl]{\(6\)}}
\put(21,02){\makebox(0,0)[bl]{\(3\)}}
\put(29,02){\makebox(0,0)[bl]{\(4\)}}
\put(37,02){\makebox(0,0)[bl]{\(4\)}}

\put(48,02){\makebox(0,0)[bl]{\(1\)}}


\end{picture}
\end{center}

\begin{center}
\begin{picture}(40,58)
\put(0,30){\makebox(0,0)[bl]{Table 14. Compatibility}}

\put(00,0){\line(1,0){27}} \put(00,22){\line(1,0){27}}
\put(00,28){\line(1,0){27}}

\put(00,0){\line(0,1){28}} \put(07,0){\line(0,1){28}}
\put(27,0){\line(0,1){28}}

\put(01,18){\makebox(0,0)[bl]{\(M_{1}\)}}
\put(01,14){\makebox(0,0)[bl]{\(M_{2}\)}}
\put(01,10){\makebox(0,0)[bl]{\(M_{3}\)}}
\put(01,06){\makebox(0,0)[bl]{\(M_{4}\)}}
\put(01,02){\makebox(0,0)[bl]{\(M_{5}\)}}

\put(12,22){\line(0,1){6}} \put(17,22){\line(0,1){6}}
\put(22,22){\line(0,1){6}}




\put(07.4,24){\makebox(0,0)[bl]{\(L_{1}\)}}
\put(12.4,24){\makebox(0,0)[bl]{\(L_{2}\)}}
\put(17.4,24){\makebox(0,0)[bl]{\(L_{3}\)}}
\put(22.4,24){\makebox(0,0)[bl]{\(L_{4}\)}}


\put(09,18){\makebox(0,0)[bl]{\(3\)}}
\put(14,18){\makebox(0,0)[bl]{\(2\)}}
\put(19,18){\makebox(0,0)[bl]{\(0\)}}
\put(24,18){\makebox(0,0)[bl]{\(3\)}}

\put(09,14){\makebox(0,0)[bl]{\(2\)}}
\put(14,14){\makebox(0,0)[bl]{\(3\)}}
\put(19,14){\makebox(0,0)[bl]{\(2\)}}
\put(24,14){\makebox(0,0)[bl]{\(0\)}}

\put(09,10){\makebox(0,0)[bl]{\(0\)}}
\put(14,10){\makebox(0,0)[bl]{\(2\)}}
\put(19,10){\makebox(0,0)[bl]{\(3\)}}
\put(24,10){\makebox(0,0)[bl]{\(2\)}}

\put(09,6){\makebox(0,0)[bl]{\(2\)}}
\put(14,6){\makebox(0,0)[bl]{\(3\)}}
\put(19,6){\makebox(0,0)[bl]{\(3\)}}
\put(24,6){\makebox(0,0)[bl]{\(3\)}}

\put(09,2){\makebox(0,0)[bl]{\(3\)}}
\put(14,2){\makebox(0,0)[bl]{\(3\)}}
\put(19,2){\makebox(0,0)[bl]{\(0\)}}
\put(24,2){\makebox(0,0)[bl]{\(3\)}}

\end{picture}
%
\begin{picture}(57,58)
\put(10,54){\makebox(0,0)[bl]{Table 15. Compatibility}}

\put(00,00){\line(1,0){57}} \put(00,46){\line(1,0){57}}
\put(00,52){\line(1,0){57}}

\put(00,0){\line(0,1){52}} \put(07,0){\line(0,1){52}}
\put(57,0){\line(0,1){52}}

\put(01,42){\makebox(0,0)[bl]{\(V_{1}\)}}
\put(01,38){\makebox(0,0)[bl]{\(V_{2}\)}}
\put(01,34){\makebox(0,0)[bl]{\(V_{3}\)}}
\put(01,30){\makebox(0,0)[bl]{\(V_{4}\)}}
\put(01,26){\makebox(0,0)[bl]{\(V_{5}\)}}
\put(01,22){\makebox(0,0)[bl]{\(V_{6}\)}}
\put(01,18){\makebox(0,0)[bl]{\(U_{1}\)}}
\put(01,14){\makebox(0,0)[bl]{\(U_{2}\)}}
\put(01,10){\makebox(0,0)[bl]{\(U_{3}\)}}
\put(01,06){\makebox(0,0)[bl]{\(U_{4}\)}}
\put(01,02){\makebox(0,0)[bl]{\(U_{5}\)}}

\put(12,46){\line(0,1){6}} \put(17,46){\line(0,1){6}}
\put(22,46){\line(0,1){6}} \put(27,46){\line(0,1){6}}
\put(32,46){\line(0,1){6}} \put(37,46){\line(0,1){6}}
\put(42,46){\line(0,1){6}} \put(47,46){\line(0,1){6}}
\put(52,46){\line(0,1){6}}



\put(07.4,48){\makebox(0,0)[bl]{\(U_{1}\)}}
\put(12.4,48){\makebox(0,0)[bl]{\(U_{2}\)}}
\put(17.4,48){\makebox(0,0)[bl]{\(U_{3}\)}}
\put(22.4,48){\makebox(0,0)[bl]{\(U_{4}\)}}
\put(27.4,48){\makebox(0,0)[bl]{\(U_{5}\)}}
\put(32.4,48){\makebox(0,0)[bl]{\(T_{1}\)}}
\put(37.4,48){\makebox(0,0)[bl]{\(T_{2}\)}}
\put(42.4,48){\makebox(0,0)[bl]{\(T_{3}\)}}
\put(47.4,48){\makebox(0,0)[bl]{\(T_{4}\)}}
\put(52.4,48){\makebox(0,0)[bl]{\(T_{5}\)}}

\put(09,42){\makebox(0,0)[bl]{\(2\)}}
\put(14,42){\makebox(0,0)[bl]{\(2\)}}
\put(19,42){\makebox(0,0)[bl]{\(2\)}}
\put(24,42){\makebox(0,0)[bl]{\(2\)}}
\put(29,42){\makebox(0,0)[bl]{\(3\)}}
\put(34,42){\makebox(0,0)[bl]{\(3\)}}
\put(39,42){\makebox(0,0)[bl]{\(2\)}}
\put(44,42){\makebox(0,0)[bl]{\(2\)}}
\put(49,42){\makebox(0,0)[bl]{\(2\)}}
\put(54,42){\makebox(0,0)[bl]{\(2\)}}

\put(09,38){\makebox(0,0)[bl]{\(3\)}}
\put(14,38){\makebox(0,0)[bl]{\(3\)}}
\put(19,38){\makebox(0,0)[bl]{\(3\)}}
\put(24,38){\makebox(0,0)[bl]{\(2\)}}
\put(29,38){\makebox(0,0)[bl]{\(0\)}}
\put(34,38){\makebox(0,0)[bl]{\(0\)}}
\put(39,38){\makebox(0,0)[bl]{\(3\)}}
\put(44,38){\makebox(0,0)[bl]{\(0\)}}
\put(49,38){\makebox(0,0)[bl]{\(3\)}}
\put(54,38){\makebox(0,0)[bl]{\(2\)}}

\put(09,34){\makebox(0,0)[bl]{\(3\)}}
\put(14,34){\makebox(0,0)[bl]{\(3\)}}
\put(19,34){\makebox(0,0)[bl]{\(3\)}}
\put(24,34){\makebox(0,0)[bl]{\(2\)}}
\put(29,34){\makebox(0,0)[bl]{\(0\)}}
\put(34,34){\makebox(0,0)[bl]{\(0\)}}
\put(39,34){\makebox(0,0)[bl]{\(3\)}}
\put(44,34){\makebox(0,0)[bl]{\(0\)}}
\put(49,34){\makebox(0,0)[bl]{\(3\)}}
\put(54,34){\makebox(0,0)[bl]{\(2\)}}

\put(09,30){\makebox(0,0)[bl]{\(3\)}}
\put(14,30){\makebox(0,0)[bl]{\(2\)}}
\put(19,30){\makebox(0,0)[bl]{\(0\)}}
\put(24,30){\makebox(0,0)[bl]{\(2\)}}
\put(29,30){\makebox(0,0)[bl]{\(3\)}}
\put(34,30){\makebox(0,0)[bl]{\(0\)}}
\put(39,30){\makebox(0,0)[bl]{\(2\)}}
\put(44,30){\makebox(0,0)[bl]{\(0\)}}
\put(49,30){\makebox(0,0)[bl]{\(2\)}}
\put(54,30){\makebox(0,0)[bl]{\(2\)}}

\put(09,26){\makebox(0,0)[bl]{\(3\)}}
\put(14,26){\makebox(0,0)[bl]{\(0\)}}
\put(19,26){\makebox(0,0)[bl]{\(0\)}}
\put(24,26){\makebox(0,0)[bl]{\(2\)}}
\put(29,26){\makebox(0,0)[bl]{\(0\)}}
\put(34,26){\makebox(0,0)[bl]{\(2\)}}
\put(39,26){\makebox(0,0)[bl]{\(2\)}}
\put(44,26){\makebox(0,0)[bl]{\(0\)}}
\put(49,26){\makebox(0,0)[bl]{\(2\)}}
\put(54,26){\makebox(0,0)[bl]{\(2\)}}

\put(09,22){\makebox(0,0)[bl]{\(0\)}}
\put(14,22){\makebox(0,0)[bl]{\(3\)}}
\put(19,22){\makebox(0,0)[bl]{\(2\)}}
\put(24,22){\makebox(0,0)[bl]{\(3\)}}
\put(29,22){\makebox(0,0)[bl]{\(2\)}}
\put(34,22){\makebox(0,0)[bl]{\(3\)}}
\put(39,22){\makebox(0,0)[bl]{\(0\)}}
\put(44,22){\makebox(0,0)[bl]{\(2\)}}
\put(49,22){\makebox(0,0)[bl]{\(2\)}}
\put(54,22){\makebox(0,0)[bl]{\(0\)}}


\put(34,18){\makebox(0,0)[bl]{\(2\)}}
\put(39,18){\makebox(0,0)[bl]{\(0\)}}
\put(44,18){\makebox(0,0)[bl]{\(0\)}}
\put(49,18){\makebox(0,0)[bl]{\(2\)}}
\put(54,18){\makebox(0,0)[bl]{\(3\)}}

\put(34,14){\makebox(0,0)[bl]{\(0\)}}
\put(39,14){\makebox(0,0)[bl]{\(2\)}}
\put(44,14){\makebox(0,0)[bl]{\(0\)}}
\put(49,14){\makebox(0,0)[bl]{\(3\)}}
\put(54,14){\makebox(0,0)[bl]{\(0\)}}

\put(34,10){\makebox(0,0)[bl]{\(0\)}}
\put(39,10){\makebox(0,0)[bl]{\(2\)}}
\put(44,10){\makebox(0,0)[bl]{\(0\)}}
\put(49,10){\makebox(0,0)[bl]{\(3\)}}
\put(54,10){\makebox(0,0)[bl]{\(0\)}}

\put(34,6){\makebox(0,0)[bl]{\(0\)}}
\put(39,6){\makebox(0,0)[bl]{\(3\)}}
\put(44,6){\makebox(0,0)[bl]{\(3\)}}
\put(49,6){\makebox(0,0)[bl]{\(0\)}}
\put(54,6){\makebox(0,0)[bl]{\(0\)}}

\put(34,2){\makebox(0,0)[bl]{\(3\)}}
\put(39,2){\makebox(0,0)[bl]{\(0\)}}
\put(44,2){\makebox(0,0)[bl]{\(2\)}}
\put(49,2){\makebox(0,0)[bl]{\(2\)}}
\put(54,2){\makebox(0,0)[bl]{\(0\)}}

\end{picture}
\end{center}

\subsection{Morphological Analysis}

 In the case of basic MA,
 binary compatibility estimates are used.
 To decrease the dimension of the considered numerical example,
 the following version of MA is examined.
 Let us consider more strong requirements to compatibility:
 (Tables 16 and 17):
 (i) new compatibility estimate equals \(1\)
  if the old estimate was equal  \(3\),
 (ii) new compatibility estimate equals \(1\)
  if the old estimate was equal \(0\) or  \(1\) or \(2\).
 Clearly, here we can get some negative results, for example:
 (a) admissible solutions are absent,
 (b) some sufficiently good solutions
 (e.g., solutions with one/two compatibility estimate at the only admissible/good
 levels
 as \(1\) or \(2\))
 will be lost.
 As a result,
 the following admissible DAs can be analyzed:

 (1) nine  DAs for \(A\):~
 \(A_{1} = M_{1} \star L_{1}\),
 \(A_{2} = M_{1} \star L_{4}\),
 \(A_{3} = M_{2} \star L_{2}\),
 \(A_{4} = M_{3} \star L_{3}\),
 \(A_{5} = M_{4} \star L_{2}\),
 \(A_{6} = M_{4} \star L_{3}\),
 \(A_{7} = M_{5} \star L_{1}\),
 \(A_{8} = M_{5} \star L_{2}\), and
 \(A_{9} = M_{5} \star L_{4}\);

 (2) five DAs for \(B\):~
 \(B_{1} = V_{1} \star U_{5} \star T_{1}\),
 \(B_{2} = V_{2} \star U_{2} \star T_{4}\),
 \(B_{3} = V_{2} \star U_{3} \star T_{4}\),
 \(B_{4} = V_{3} \star U_{2} \star T_{4}\), and
 \(B_{5} = V_{3} \star U_{3} \star T_{4}\);

 and the resultant composite DAs are:~
 \(S_{1} = A_{1} \star B_{1}\),
 \(S_{2} = A_{2} \star B_{1}\),
 \(S_{3} = A_{3} \star B_{1}\),
 \(S_{4} = A_{4} \star B_{1}\),
 \(S_{5} = A_{5} \star B_{1}\),
 \(S_{6} = A_{6} \star B_{1}\),
 \(S_{7} = A_{7} \star B_{1}\),
 \(S_{8} = A_{8} \star B_{1}\),
 \(S_{9} = A_{9} \star B_{1}\);
 \(S_{10} = A_{1} \star B_{2}\),
 \(S_{11} = A_{2} \star B_{2}\),
 \(S_{12} = A_{3} \star B_{2}\),
 \(S_{13} = A_{4} \star B_{2}\),
 \(S_{14} = A_{5} \star B_{2}\),
 \(S_{15} = A_{6} \star B_{2}\),
 \(S_{16} = A_{7} \star B_{2}\),
 \(S_{17} = A_{8} \star B_{2}\),
 \(S_{18} = A_{9} \star B_{2}\);
  \(S_{19} = A_{1} \star B_{3}\),
 \(S_{20} = A_{2} \star B_{3}\),
 \(S_{21} = A_{3} \star B_{3}\),
 \(S_{22} = A_{4} \star B_{3}\),
 \(S_{23} = A_{5} \star B_{3}\),
 \(S_{24} = A_{6} \star B_{3}\),
 \(S_{25} = A_{7} \star B_{3}\),
 \(S_{26} = A_{8} \star B_{3}\),
 \(S_{27} = A_{9} \star B_{3}\);
  \(S_{28} = A_{1} \star B_{4}\),
 \(S_{29} = A_{2} \star B_{4}\),
 \(S_{30} = A_{3} \star B_{4}\),
 \(S_{31} = A_{4} \star B_{4}\),
 \(S_{32} = A_{5} \star B_{4}\),
 \(S_{33} = A_{6} \star B_{4}\),
 \(S_{34} = A_{7} \star B_{4}\),
 \(S_{35} = A_{8} \star B_{4}\),
 \(S_{36} = A_{9} \star B_{4}\);
  \(S_{37} = A_{1} \star B_{5}\),
 \(S_{38} = A_{2} \star B_{5}\),
 \(S_{39} = A_{3} \star B_{5}\),
 \(S_{40} = A_{4} \star B_{5}\),
 \(S_{41} = A_{5} \star B_{5}\),
 \(S_{42} = A_{6} \star B_{5}\),
 \(S_{43} = A_{7} \star B_{5}\),
 \(S_{44} = A_{8} \star B_{5}\), and
 \(S_{45} = A_{9} \star B_{5}\).

 Finally, the next step has to consist in selection of the best
 solution.

\begin{center}
\begin{picture}(40,58)
\put(0,30){\makebox(0,0)[bl]{Table 16. Compatibility}}

\put(00,0){\line(1,0){27}} \put(00,22){\line(1,0){27}}
\put(00,28){\line(1,0){27}}

\put(00,0){\line(0,1){28}} \put(07,0){\line(0,1){28}}
\put(27,0){\line(0,1){28}}

\put(01,18){\makebox(0,0)[bl]{\(M_{1}\)}}
\put(01,14){\makebox(0,0)[bl]{\(M_{2}\)}}
\put(01,10){\makebox(0,0)[bl]{\(M_{3}\)}}
\put(01,06){\makebox(0,0)[bl]{\(M_{4}\)}}
\put(01,02){\makebox(0,0)[bl]{\(M_{5}\)}}

\put(12,22){\line(0,1){6}} \put(17,22){\line(0,1){6}}
\put(22,22){\line(0,1){6}}




\put(07.4,24){\makebox(0,0)[bl]{\(L_{1}\)}}
\put(12.4,24){\makebox(0,0)[bl]{\(L_{2}\)}}
\put(17.4,24){\makebox(0,0)[bl]{\(L_{3}\)}}
\put(22.4,24){\makebox(0,0)[bl]{\(L_{4}\)}}


\put(09,18){\makebox(0,0)[bl]{\(1\)}}
\put(14,18){\makebox(0,0)[bl]{\(0\)}}
\put(19,18){\makebox(0,0)[bl]{\(0\)}}
\put(24,18){\makebox(0,0)[bl]{\(1\)}}

\put(09,14){\makebox(0,0)[bl]{\(0\)}}
\put(14,14){\makebox(0,0)[bl]{\(1\)}}
\put(19,14){\makebox(0,0)[bl]{\(0\)}}
\put(24,14){\makebox(0,0)[bl]{\(0\)}}

\put(09,10){\makebox(0,0)[bl]{\(0\)}}
\put(14,10){\makebox(0,0)[bl]{\(0\)}}
\put(19,10){\makebox(0,0)[bl]{\(1\)}}
\put(24,10){\makebox(0,0)[bl]{\(0\)}}

\put(09,6){\makebox(0,0)[bl]{\(0\)}}
\put(14,6){\makebox(0,0)[bl]{\(1\)}}
\put(19,6){\makebox(0,0)[bl]{\(1\)}}
\put(24,6){\makebox(0,0)[bl]{\(1\)}}

\put(09,2){\makebox(0,0)[bl]{\(1\)}}
\put(14,2){\makebox(0,0)[bl]{\(1\)}}
\put(19,2){\makebox(0,0)[bl]{\(0\)}}
\put(24,2){\makebox(0,0)[bl]{\(1\)}}

\end{picture}
%
\begin{picture}(57,58)
\put(10,54){\makebox(0,0)[bl]{Table 17. Compatibility}}

\put(00,00){\line(1,0){57}} \put(00,46){\line(1,0){57}}
\put(00,52){\line(1,0){57}}

\put(00,0){\line(0,1){52}} \put(07,0){\line(0,1){52}}
\put(57,0){\line(0,1){52}}

\put(01,42){\makebox(0,0)[bl]{\(V_{1}\)}}
\put(01,38){\makebox(0,0)[bl]{\(V_{2}\)}}
\put(01,34){\makebox(0,0)[bl]{\(V_{3}\)}}
\put(01,30){\makebox(0,0)[bl]{\(V_{4}\)}}
\put(01,26){\makebox(0,0)[bl]{\(V_{5}\)}}
\put(01,22){\makebox(0,0)[bl]{\(V_{6}\)}}
\put(01,18){\makebox(0,0)[bl]{\(U_{1}\)}}
\put(01,14){\makebox(0,0)[bl]{\(U_{2}\)}}
\put(01,10){\makebox(0,0)[bl]{\(U_{3}\)}}
\put(01,06){\makebox(0,0)[bl]{\(U_{4}\)}}
\put(01,02){\makebox(0,0)[bl]{\(U_{5}\)}}

\put(12,46){\line(0,1){6}} \put(17,46){\line(0,1){6}}
\put(22,46){\line(0,1){6}} \put(27,46){\line(0,1){6}}
\put(32,46){\line(0,1){6}} \put(37,46){\line(0,1){6}}
\put(42,46){\line(0,1){6}} \put(47,46){\line(0,1){6}}
\put(52,46){\line(0,1){6}}



\put(07.4,48){\makebox(0,0)[bl]{\(U_{1}\)}}
\put(12.4,48){\makebox(0,0)[bl]{\(U_{2}\)}}
\put(17.4,48){\makebox(0,0)[bl]{\(U_{3}\)}}
\put(22.4,48){\makebox(0,0)[bl]{\(U_{4}\)}}
\put(27.4,48){\makebox(0,0)[bl]{\(U_{5}\)}}
\put(32.4,48){\makebox(0,0)[bl]{\(T_{1}\)}}
\put(37.4,48){\makebox(0,0)[bl]{\(T_{2}\)}}
\put(42.4,48){\makebox(0,0)[bl]{\(T_{3}\)}}
\put(47.4,48){\makebox(0,0)[bl]{\(T_{4}\)}}
\put(52.4,48){\makebox(0,0)[bl]{\(T_{5}\)}}

\put(09,42){\makebox(0,0)[bl]{\(0\)}}
\put(14,42){\makebox(0,0)[bl]{\(0\)}}
\put(19,42){\makebox(0,0)[bl]{\(0\)}}
\put(24,42){\makebox(0,0)[bl]{\(0\)}}
\put(29,42){\makebox(0,0)[bl]{\(1\)}}
\put(34,42){\makebox(0,0)[bl]{\(1\)}}
\put(39,42){\makebox(0,0)[bl]{\(0\)}}
\put(44,42){\makebox(0,0)[bl]{\(0\)}}
\put(49,42){\makebox(0,0)[bl]{\(0\)}}
\put(54,42){\makebox(0,0)[bl]{\(0\)}}

\put(09,38){\makebox(0,0)[bl]{\(1\)}}
\put(14,38){\makebox(0,0)[bl]{\(1\)}}
\put(19,38){\makebox(0,0)[bl]{\(1\)}}
\put(24,38){\makebox(0,0)[bl]{\(0\)}}
\put(29,38){\makebox(0,0)[bl]{\(0\)}}
\put(34,38){\makebox(0,0)[bl]{\(0\)}}
\put(39,38){\makebox(0,0)[bl]{\(1\)}}
\put(44,38){\makebox(0,0)[bl]{\(0\)}}
\put(49,38){\makebox(0,0)[bl]{\(1\)}}
\put(54,38){\makebox(0,0)[bl]{\(0\)}}

\put(09,34){\makebox(0,0)[bl]{\(1\)}}
\put(14,34){\makebox(0,0)[bl]{\(1\)}}
\put(19,34){\makebox(0,0)[bl]{\(1\)}}
\put(24,34){\makebox(0,0)[bl]{\(0\)}}
\put(29,34){\makebox(0,0)[bl]{\(0\)}}
\put(34,34){\makebox(0,0)[bl]{\(0\)}}
\put(39,34){\makebox(0,0)[bl]{\(1\)}}
\put(44,34){\makebox(0,0)[bl]{\(0\)}}
\put(49,34){\makebox(0,0)[bl]{\(1\)}}
\put(54,34){\makebox(0,0)[bl]{\(0\)}}

\put(09,30){\makebox(0,0)[bl]{\(1\)}}
\put(14,30){\makebox(0,0)[bl]{\(0\)}}
\put(19,30){\makebox(0,0)[bl]{\(0\)}}
\put(24,30){\makebox(0,0)[bl]{\(0\)}}
\put(29,30){\makebox(0,0)[bl]{\(1\)}}
\put(34,30){\makebox(0,0)[bl]{\(0\)}}
\put(39,30){\makebox(0,0)[bl]{\(0\)}}
\put(44,30){\makebox(0,0)[bl]{\(0\)}}
\put(49,30){\makebox(0,0)[bl]{\(0\)}}
\put(54,30){\makebox(0,0)[bl]{\(0\)}}

\put(09,26){\makebox(0,0)[bl]{\(1\)}}
\put(14,26){\makebox(0,0)[bl]{\(0\)}}
\put(19,26){\makebox(0,0)[bl]{\(0\)}}
\put(24,26){\makebox(0,0)[bl]{\(0\)}}
\put(29,26){\makebox(0,0)[bl]{\(0\)}}
\put(34,26){\makebox(0,0)[bl]{\(0\)}}
\put(39,26){\makebox(0,0)[bl]{\(0\)}}
\put(44,26){\makebox(0,0)[bl]{\(0\)}}
\put(49,26){\makebox(0,0)[bl]{\(0\)}}
\put(54,26){\makebox(0,0)[bl]{\(0\)}}

\put(09,22){\makebox(0,0)[bl]{\(0\)}}
\put(14,22){\makebox(0,0)[bl]{\(1\)}}
\put(19,22){\makebox(0,0)[bl]{\(0\)}}
\put(24,22){\makebox(0,0)[bl]{\(1\)}}
\put(29,22){\makebox(0,0)[bl]{\(0\)}}
\put(34,22){\makebox(0,0)[bl]{\(1\)}}
\put(39,22){\makebox(0,0)[bl]{\(0\)}}
\put(44,22){\makebox(0,0)[bl]{\(0\)}}
\put(49,22){\makebox(0,0)[bl]{\(0\)}}
\put(54,22){\makebox(0,0)[bl]{\(0\)}}


\put(34,18){\makebox(0,0)[bl]{\(0\)}}
\put(39,18){\makebox(0,0)[bl]{\(0\)}}
\put(44,18){\makebox(0,0)[bl]{\(0\)}}
\put(49,18){\makebox(0,0)[bl]{\(0\)}}
\put(54,18){\makebox(0,0)[bl]{\(1\)}}

\put(34,14){\makebox(0,0)[bl]{\(0\)}}
\put(39,14){\makebox(0,0)[bl]{\(0\)}}
\put(44,14){\makebox(0,0)[bl]{\(0\)}}
\put(49,14){\makebox(0,0)[bl]{\(1\)}}
\put(54,14){\makebox(0,0)[bl]{\(0\)}}

\put(34,10){\makebox(0,0)[bl]{\(0\)}}
\put(39,10){\makebox(0,0)[bl]{\(0\)}}
\put(44,10){\makebox(0,0)[bl]{\(0\)}}
\put(49,10){\makebox(0,0)[bl]{\(1\)}}
\put(54,10){\makebox(0,0)[bl]{\(0\)}}

\put(34,6){\makebox(0,0)[bl]{\(0\)}}
\put(39,6){\makebox(0,0)[bl]{\(1\)}}
\put(44,6){\makebox(0,0)[bl]{\(1\)}}
\put(49,6){\makebox(0,0)[bl]{\(0\)}}
\put(54,6){\makebox(0,0)[bl]{\(0\)}}

\put(34,2){\makebox(0,0)[bl]{\(1\)}}
\put(39,2){\makebox(0,0)[bl]{\(0\)}}
\put(44,2){\makebox(0,0)[bl]{\(0\)}}
\put(49,2){\makebox(0,0)[bl]{\(0\)}}
\put(54,2){\makebox(0,0)[bl]{\(0\)}}

\end{picture}
\end{center}

\subsection{Method of Closeness to Ideal Point}

 Here the initial set of admissible solutions corresponds
 to the solution set that was obtained in previous case
  (i.e., basic MA).
 Evidently, this approach depended on the kind of the proximity
 between the ideal point
 ~( \(S^{I}\) )~
  and examined solutions.

 First of all, let us consider estimate vector for each admissible solution
 (basic estimates are contained in Tables 9, 10, 11, 12, and 13):
 \[ \overline{z} = ( z_{M} \bigcup z_{L} \bigcup z_{V} \bigcup z_{U} \bigcup z_{T} )
 =\]
  \[ ( z_{m1},z_{m2},z_{m3},z_{m4},z_{m5},z_{l1},z_{l2},z_{l3},z_{l4},z_{v1},z_{v2},z_{v3},z_{v4},
  z_{u1},z_{u2},z_{u3},z_{u4},z_{t1},z_{t2},z_{t3},z_{t4} ) .\]
 On the other hand, it may be reasonable to consider a simplified version of the estimate vector as follows:
  ~\( \widehat{z} = ( r_{M}, r_{L},  r_{V},  r_{U},  r_{T} )\),
 where ~\(r_{M},r_{L},r_{V},r_{U},r_{T}\)~ are the priorities of
 DAs which are obtained for local DAs
 (for \(M\), for \(L\), for \(V\), for \(U\), and for \(T\);
 Table 9, Table 10, Table 11, Table 12, Table 13).
 To simplify the considered example,
 the second case of the estimate
 vector is used.
 Thus,
 the resultant vector estimates
 (i.e.,  \( \{  \widehat{z} \} \)  )
 for examined 45 admissible solutions are contained in Tables 18 and 19.
 Evidently, it is reasonable to consider the estimate vector for the ideal solution as
 follows:~
  \( \widehat{z_{I}} = ( 1, 1, 1, 1, 1 )\).
  Now let us use a simplified proximity function
 between ideal solution \(I\) and design alternative DA
   as follows (i.e, metric like \(l^{2}\)):
 \[ \rho (I,DA)  =  \sqrt{ \sum_{k \in \{M,L,V,U,T \} }(z_{k}(I)-z_{k}(DA))^{2}  } .\]
 The resultant proximity is presented in Tables 18 and 19.
  Finally, the best composite design alternative  (by the minimal proximity) is:~
 \(S^{I}_{0} = S_{23} = A_{5}\star B_{3} = M_{3}\star L_{1} \star V_{1} \star U_{2} \star T_{3} \)
 (\(\rho = 1.7321\)).
 Several composite DAs are very close to the best one,
 for example:~

  \(S^{I}_{1} = S_{19} = A_{1}\star B_{3} = M_{1}\star L_{1} \star V_{2} \star U_{3} \star T_{4}\)
   (\(\rho = 2.0\)),
  \(S^{I}_{2} = S_{20} = A_{2}\star B_{3} = M_{1}\star L_{4} \star V_{2} \star U_{3} \star T_{4}\)
  (\(\rho = 2.0\)),
  \(S^{I}_{3} = S_{24} = A_{6}\star B_{3} = M_{4}\star L_{3} \star V_{2} \star U_{3} \star T_{4}\)
  (\(\rho = 2.0\)),
  \(S^{I}_{4} = S_{25} = A_{7}\star B_{3} = M_{5}\star L_{1} \star V_{2} \star U_{3} \star T_{4}\)
  (\(\rho = 2.0\)),
  \(S^{I}_{5} = S_{26} = A_{8}\star B_{3} = M_{5}\star L_{2} \star V_{3} \star U_{2} \star T_{4}\)
   (\(\rho = 2.0\)), and
  \(S^{I}_{6} = S_{27} = A_{9}\star B_{3} = M_{5}\star L_{4} \star V_{2} \star U_{3} \star T_{4}\)
   (\(\rho = 2.0\)).

 It may be reasonable to point out
 several prospective directions for
 the improvement of this method:

 (1) consideration of special types of proximity between solutions
 and the ideal point (e.g.,
 ordinal proximity,
 vector-like proximity \cite{lev98},
 etc.);

 (2) usage of special expert judgment interactive procedures for
  the assessment of the proximity;

 (3) consideration of a set of ideal points
 (the set can be generated by domain expert(s));
 and

 (4) design of special support visualization tools which will aid
  domain expert(s) in his/her (their) activity
  (i.e., generation of the ideal point and assessment of proximity).

 In addition let us list the basic approaches
 to generation of the ideal point(s):

 1. consideration of design alternative  with the estimate vector in which each
 component equals the best value of the design alternatives
 estimates (by the corresponding criterion, i.e., minimum or
 maximum);

 2. consideration of design alternative with the estimate vector in which each
 component equals the best value of the corresponding criterion scale
 (i.e., minimum or maximum);

 3. expert judgment based generation design alternative(s);

 4. projection of expert judgment based design alternatives into
 convex shell of the set of Pareto-efficient points; etc.

%
%

\begin{center}
\begin{picture}(71,111)
\put(1,107){\makebox(0,0)[bl]{Table 18. Estimates of admissible
solutions}}

\put(00,00){\line(1,0){70}} \put(00,95){\line(1,0){70}}
\put(00,105){\line(1,0){70}}

\put(00,0){\line(0,1){105}} \put(08,0){\line(0,1){105}}
\put(28,0){\line(0,1){105}} \put(49,0){\line(0,1){105}}
\put(70,0){\line(0,1){105}}


\put(0.6,101){\makebox(0,0)[bl]{DAs}}

\put(17,101){\makebox(0,0)[bl]{\(\widehat{z}\)}}

\put(28.6,101){\makebox(0,0)[bl]{Closeness to}}
\put(28.6,97){\makebox(0,0)[bl]{ideal point}}

\put(49.7,101){\makebox(0,0)[bl]{Membership}}
\put(49.7,97){\makebox(0,0)[bl]{of Pareto-set}}


\put(01,90){\makebox(0,0)[bl]{\(S_{1}\)}}
\put(01,86){\makebox(0,0)[bl]{\(S_{2}\)}}
\put(01,82){\makebox(0,0)[bl]{\(S_{3}\)}}
\put(01,78){\makebox(0,0)[bl]{\(S_{4}\)}}
\put(01,74){\makebox(0,0)[bl]{\(S_{5}\)}}
\put(01,70){\makebox(0,0)[bl]{\(S_{6}\)}}
\put(01,66){\makebox(0,0)[bl]{\(S_{7}\)}}
\put(01,62){\makebox(0,0)[bl]{\(S_{8}\)}}
\put(01,58){\makebox(0,0)[bl]{\(S_{9}\)}}
\put(01,54){\makebox(0,0)[bl]{\(S_{10}\)}}
\put(01,50){\makebox(0,0)[bl]{\(S_{11}\)}}
\put(01,46){\makebox(0,0)[bl]{\(S_{12}\)}}
\put(01,42){\makebox(0,0)[bl]{\(S_{13}\)}}
\put(01,38){\makebox(0,0)[bl]{\(S_{14}\)}}
\put(01,34){\makebox(0,0)[bl]{\(S_{15}\)}}
\put(01,30){\makebox(0,0)[bl]{\(S_{16}\)}}
\put(01,26){\makebox(0,0)[bl]{\(S_{17}\)}}
\put(01,22){\makebox(0,0)[bl]{\(S_{18}\)}}
\put(01,18){\makebox(0,0)[bl]{\(S_{19}\)}}
\put(01,14){\makebox(0,0)[bl]{\(S_{20}\)}}
\put(01,10){\makebox(0,0)[bl]{\(S_{21}\)}}
\put(01,06){\makebox(0,0)[bl]{\(S_{22}\)}}
\put(01,02){\makebox(0,0)[bl]{\(S_{23}\)}}

\put(09,89.5){\makebox(0,0)[bl]{\((2,1,1,2,3)\)}}
\put(33,90){\makebox(0,0)[bl]{\(2.4495\)}}
\put(57,90){\makebox(0,0)[bl]{No}}

\put(09,85.5){\makebox(0,0)[bl]{\((2,1,1,2,3)\)}}
\put(33,86){\makebox(0,0)[bl]{\(2.4495\)}}
\put(57,86){\makebox(0,0)[bl]{No}}

\put(09,81.5){\makebox(0,0)[bl]{\((3,1,1,2,3)\)}}
\put(33,82){\makebox(0,0)[bl]{\(3.0\)}}
\put(57,82){\makebox(0,0)[bl]{No}}

\put(09,77.5){\makebox(0,0)[bl]{\((3,2,1,2,3)\)}}
\put(33,78){\makebox(0,0)[bl]{\(3.1623\)}}
\put(57,78){\makebox(0,0)[bl]{No}}

\put(09,73.5){\makebox(0,0)[bl]{\((1,1,1,2,3)\)}}
\put(33,74){\makebox(0,0)[bl]{\(2.2361\)}}
\put(57,74){\makebox(0,0)[bl]{Yes}}

\put(09,69.5){\makebox(0,0)[bl]{\((1,2,1,2,3)\)}}
\put(33,70){\makebox(0,0)[bl]{\(2.4495\)}}
\put(57,70){\makebox(0,0)[bl]{No}}

\put(09,65.5){\makebox(0,0)[bl]{\((2,1,1,2,3)\)}}
\put(33,66){\makebox(0,0)[bl]{\(2.4495\)}}
\put(57,66){\makebox(0,0)[bl]{No}}

\put(09,61.5){\makebox(0,0)[bl]{\((2,1,1,2,3)\)}}
\put(33,62){\makebox(0,0)[bl]{\(2.4495\)}}
\put(57,62){\makebox(0,0)[bl]{No}}

\put(09,57.5){\makebox(0,0)[bl]{\((2,1,1,2,3)\)}}
\put(33,58){\makebox(0,0)[bl]{\(2.4495\)}}
\put(57,58){\makebox(0,0)[bl]{No}}

\put(09,53.5){\makebox(0,0)[bl]{\((2,1,2,3,2)\)}}
\put(33,54){\makebox(0,0)[bl]{\(2.6458\)}}
\put(57,54){\makebox(0,0)[bl]{No}}

\put(09,49.5){\makebox(0,0)[bl]{\((2,1,2,3,2)\)}}
\put(33,50){\makebox(0,0)[bl]{\(2.6458\)}}
\put(57,50){\makebox(0,0)[bl]{No}}

\put(09,45.5){\makebox(0,0)[bl]{\((3,1,2,3,2)\)}}
\put(33,46){\makebox(0,0)[bl]{\(3.1623\)}}
\put(57,46){\makebox(0,0)[bl]{No}}

\put(09,41.5){\makebox(0,0)[bl]{\((3,2,2,3,2)\)}}
\put(33,42){\makebox(0,0)[bl]{\(3.3166\)}}
\put(57,42){\makebox(0,0)[bl]{No}}

\put(09,37.5){\makebox(0,0)[bl]{\((1,1,2,3,2)\)}}
\put(33,38){\makebox(0,0)[bl]{\(2.4495\)}}
\put(57,38){\makebox(0,0)[bl]{No}}

\put(09,33.5){\makebox(0,0)[bl]{\((1,2,2,3,2)\)}}
\put(33,34){\makebox(0,0)[bl]{\(2.6458\)}}
\put(57,34){\makebox(0,0)[bl]{No}}

\put(09,29.5){\makebox(0,0)[bl]{\((2,1,2,3,2)\)}}
\put(33,30){\makebox(0,0)[bl]{\(2.6458\)}}
\put(57,30){\makebox(0,0)[bl]{No}}

\put(09,25.5){\makebox(0,0)[bl]{\((2,1,2,3,2)\)}}
\put(33,26){\makebox(0,0)[bl]{\(2.6458\)}}
\put(57,26){\makebox(0,0)[bl]{No}}

\put(09,21.5){\makebox(0,0)[bl]{\((2,1,2,3,2)\)}}
\put(33,22){\makebox(0,0)[bl]{\(2.6458\)}}
\put(57,22){\makebox(0,0)[bl]{No}}

\put(09,17.5){\makebox(0,0)[bl]{\((2,1,2,2,2)\)}}
\put(33,18){\makebox(0,0)[bl]{\(2.0\)}}
\put(57,18){\makebox(0,0)[bl]{No}}

\put(09,13.5){\makebox(0,0)[bl]{\((2,1,2,2,2)\)}}
\put(33,14){\makebox(0,0)[bl]{\(2.0\)}}
\put(57,14){\makebox(0,0)[bl]{No}}

\put(09,09.5){\makebox(0,0)[bl]{\((3,1,2,2,2)\)}}
\put(33,10){\makebox(0,0)[bl]{\(2.6458\)}}
\put(57,10){\makebox(0,0)[bl]{No}}

\put(09,05.5){\makebox(0,0)[bl]{\((3,2,2,2,2)\)}}
\put(33,06){\makebox(0,0)[bl]{\(2.8284\)}}
\put(57,06){\makebox(0,0)[bl]{No}}

\put(09,01.5){\makebox(0,0)[bl]{\((1,1,2,2,2)\)}}
\put(33,02){\makebox(0,0)[bl]{\(1.7321\)}}
\put(57,02){\makebox(0,0)[bl]{Yes}}

\end{picture}
%
\begin{picture}(71,108)
\put(1,103){\makebox(0,0)[bl]{Table 19. Estimates of admissible
solutions}}

\put(00,00){\line(1,0){70}} \put(00,91){\line(1,0){70}}
\put(00,101){\line(1,0){70}}

\put(00,0){\line(0,1){101}} \put(08,0){\line(0,1){101}}
\put(28,0){\line(0,1){101}} \put(49,0){\line(0,1){101}}
\put(70,0){\line(0,1){101}}


\put(0.6,97){\makebox(0,0)[bl]{DAs}}

\put(17,97){\makebox(0,0)[bl]{\(\widehat{z}\)}}

\put(28.6,97){\makebox(0,0)[bl]{Closeness to}}
\put(28.6,93){\makebox(0,0)[bl]{ideal point}}

\put(49.7,97){\makebox(0,0)[bl]{Membership}}
\put(49.7,93){\makebox(0,0)[bl]{of Pareto-set}}


\put(01,86){\makebox(0,0)[bl]{\(S_{24}\)}}
\put(01,82){\makebox(0,0)[bl]{\(S_{25}\)}}
\put(01,78){\makebox(0,0)[bl]{\(S_{26}\)}}
\put(01,74){\makebox(0,0)[bl]{\(S_{27}\)}}
\put(01,70){\makebox(0,0)[bl]{\(S_{28}\)}}
\put(01,66){\makebox(0,0)[bl]{\(S_{29}\)}}
\put(01,62){\makebox(0,0)[bl]{\(S_{30}\)}}
\put(01,58){\makebox(0,0)[bl]{\(S_{31}\)}}
\put(01,54){\makebox(0,0)[bl]{\(S_{32}\)}}
\put(01,50){\makebox(0,0)[bl]{\(S_{33}\)}}
\put(01,46){\makebox(0,0)[bl]{\(S_{34}\)}}
\put(01,42){\makebox(0,0)[bl]{\(S_{35}\)}}
\put(01,38){\makebox(0,0)[bl]{\(S_{36}\)}}
\put(01,34){\makebox(0,0)[bl]{\(S_{37}\)}}
\put(01,30){\makebox(0,0)[bl]{\(S_{38}\)}}
\put(01,26){\makebox(0,0)[bl]{\(S_{39}\)}}
\put(01,22){\makebox(0,0)[bl]{\(S_{40}\)}}
\put(01,18){\makebox(0,0)[bl]{\(S_{41}\)}}
\put(01,14){\makebox(0,0)[bl]{\(S_{42}\)}}
\put(01,10){\makebox(0,0)[bl]{\(S_{43}\)}}
\put(01,06){\makebox(0,0)[bl]{\(S_{44}\)}}
\put(01,02){\makebox(0,0)[bl]{\(S_{45}\)}}

\put(09,85.5){\makebox(0,0)[bl]{\((1,2,2,2,2)\)}}
\put(33,86){\makebox(0,0)[bl]{\(2.0\)}}
\put(57,86){\makebox(0,0)[bl]{No}}

\put(09,81.5){\makebox(0,0)[bl]{\((2,1,2,2,2)\)}}
\put(33,82){\makebox(0,0)[bl]{\(2.0\)}}
\put(57,82){\makebox(0,0)[bl]{No}}

\put(09,77.5){\makebox(0,0)[bl]{\((2,1,2,2,2)\)}}
\put(33,78){\makebox(0,0)[bl]{\(2.0\)}}
\put(57,78){\makebox(0,0)[bl]{No}}

\put(09,73.5){\makebox(0,0)[bl]{\((2,1,2,2,2)\)}}
\put(33,74){\makebox(0,0)[bl]{\(2.0\)}}
\put(57,74){\makebox(0,0)[bl]{No}}

\put(09,69.5){\makebox(0,0)[bl]{\((2,1,3,3,2)\)}}
\put(33,70){\makebox(0,0)[bl]{\(3.1623\)}}
\put(57,70){\makebox(0,0)[bl]{No}}

\put(09,65.5){\makebox(0,0)[bl]{\((2,1,3,3,2)\)}}
\put(33,66){\makebox(0,0)[bl]{\(3.1623\)}}
\put(57,66){\makebox(0,0)[bl]{No}}

\put(09,61.5){\makebox(0,0)[bl]{\((3,1,3,3,2)\)}}
\put(33,62){\makebox(0,0)[bl]{\(3.6056\)}}
\put(57,62){\makebox(0,0)[bl]{No}}

\put(09,57.5){\makebox(0,0)[bl]{\((3,2,3,3,2)\)}}
\put(33,58){\makebox(0,0)[bl]{\(3.7417\)}}
\put(57,58){\makebox(0,0)[bl]{No}}

\put(09,53.5){\makebox(0,0)[bl]{\((1,1,3,3,2)\)}}
\put(33,54){\makebox(0,0)[bl]{\(3.0\)}}
\put(57,54){\makebox(0,0)[bl]{No}}

\put(09,49.5){\makebox(0,0)[bl]{\((1,2,3,3,2)\)}}
\put(33,50){\makebox(0,0)[bl]{\(3.1623\)}}
\put(57,50){\makebox(0,0)[bl]{No}}

\put(09,45.5){\makebox(0,0)[bl]{\((2,1,3,3,2)\)}}
\put(33,46){\makebox(0,0)[bl]{\(3.1623\)}}
\put(57,46){\makebox(0,0)[bl]{No}}

\put(09,41.5){\makebox(0,0)[bl]{\((2,1,3,3,2)\)}}
\put(33,42){\makebox(0,0)[bl]{\(3.1623\)}}
\put(57,42){\makebox(0,0)[bl]{No}}

\put(09,37.5){\makebox(0,0)[bl]{\((2,1,3,3,2)\)}}
\put(33,38){\makebox(0,0)[bl]{\(3.1623\)}}
\put(57,38){\makebox(0,0)[bl]{No}}

\put(09,33.5){\makebox(0,0)[bl]{\((2,1,3,2,2)\)}}
\put(33,34){\makebox(0,0)[bl]{\(2.6458\)}}
\put(57,34){\makebox(0,0)[bl]{No}}

\put(09,29.5){\makebox(0,0)[bl]{\((2,1,3,2,2)\)}}
\put(33,30){\makebox(0,0)[bl]{\(2.6458\)}}
\put(57,30){\makebox(0,0)[bl]{No}}

\put(09,25.5){\makebox(0,0)[bl]{\((3,1,3,2,2)\)}}
\put(33,26){\makebox(0,0)[bl]{\(3.1623\)}}
\put(57,26){\makebox(0,0)[bl]{No}}

\put(09,21.5){\makebox(0,0)[bl]{\((3,2,3,2,2)\)}}
\put(33,22){\makebox(0,0)[bl]{\(3.3166\)}}
\put(57,22){\makebox(0,0)[bl]{No}}

\put(09,17.5){\makebox(0,0)[bl]{\((1,1,3,2,2)\)}}
\put(33,18){\makebox(0,0)[bl]{\(2.4495\)}}
\put(57,18){\makebox(0,0)[bl]{No}}

\put(09,13.5){\makebox(0,0)[bl]{\((1,2,3,2,2)\)}}
\put(33,14){\makebox(0,0)[bl]{\(2.6458\)}}
\put(57,14){\makebox(0,0)[bl]{No}}

\put(09,09.5){\makebox(0,0)[bl]{\((2,1,3,2,2)\)}}
\put(33,10){\makebox(0,0)[bl]{\(2.658\)}}
\put(57,10){\makebox(0,0)[bl]{No}}

\put(09,05.5){\makebox(0,0)[bl]{\((2,1,3,2,2)\)}}
\put(33,06){\makebox(0,0)[bl]{\(2.6458\)}}
\put(57,06){\makebox(0,0)[bl]{No}}

\put(09,01.5){\makebox(0,0)[bl]{\((2,1,3,2,2)\)}}
\put(33,02){\makebox(0,0)[bl]{\(2.6458\)}}
\put(57,02){\makebox(0,0)[bl]{No}}

\end{picture}
\end{center}

\subsection{Pareto-based Morphological Analysis}

  Here the initial set of admissible solutions corresponds
 to the previous case
 of basic MA.
 Two approaches can be used for mulricriteria assessment of
 admissible solutions:

 {\bf 1.} Basic method:
 selection of Pareto-efficient solutions over the set of
 admissible composite solutions on the basis of
  of usage
  of the initial set of criteria for assessment of
 each admissible composite DAs;

 {\bf 2.} Two-stage method:

 (i) assessment of initial components
 by the corresponding criteria and ranking of the alternative
 components to get an ordinal priority for each component,

 (ii) selection of Pareto-efficient solutions over the set of
 admissible composite solutions on the basis of
  of usage
  of the vector estimates which
 integrate priorities of solution components above.
 The results of the Pareto-based MA are presented in Tables 18 and
 19, i.e., the resultant (Pareto-efficient) DAs are:~
 (i) \(S^{P}_{1}  =S_{5} = A_{5}\star B_{1} = M_{4}\star L_{2} \star V_{1} \star U_{5} \star T_{1} \) ~and
 (ii) \(S^{P}_{2} = S_{23} = A_{5}\star B_{3} = M_{4}\star L_{2} \star V_{2} \star U_{3} \star T_{4}\).
%


\subsection{Multiple Choice Problem}

 Multiple choice problem with 5 groups of elements
 (i.e., for \(M\), \(L\), \(V\), \(U\), \(T\)) is examined
 (Fig. 12).
  Evidently, here it is reasonable to examine multicriteria
 multiple choice problem.
 In the example,
 a simplified problem solving approach is considered (Table 20):

 (i) a simple greedy heuristic based on element priorities is used;

 (ii) for each element (i.e., \(i,j\)) 'profit' is computed as follows:~
  \(c_{i,j} = 4 - r_{i,j}\);

 (iii) for each element  (i.e., \(i,j\))
 a required resource is computed as follows:~
  \(a_{i,j} = 11 - z_{i,j}\)~
  where
  \(z_{i,j}\) equals:
 (a) for \(M\): the estimate upon criterion \(C_{m3}\) (Table 9),
 (b) for \(L\):   \(1.0\),
 (c) for \(V\): the estimate upon criterion \(C_{v1}\) (Table 11),
 (d) for \(U\): the estimate upon criterion \(C_{u3}\) (Table 12), and
 (e) for \(T\): the estimate upon criterion \(C_{mt3}\) (Table 13).

\begin{center}
\begin{picture}(82,53)

\put(0,04){\makebox(0,0)[bl] {Fig. 12. Designed GSM network
(priorities of DAs}}

\put(14,0){\makebox(0,0)[bl]{are shown in parentheses)}}

\put(00,50){\circle*{3}}

\put(05,49){\makebox(0,0)[bl]{\(S=M \star L \star V \star U\star T
\)}}

\put(00,50){\line(0,-1){4}}

\put(00,45){\line(1,0){60}}


\put(00,45){\line(1,0){22}} \put(45,45){\line(1,0){33}}
\put(00,40){\line(0,1){10}} \put(22,40){\line(0,1){05}}
\put(45,40){\line(0,1){05}} \put(63,40){\line(0,1){05}}
\put(78,40){\line(0,1){05}} \put(00,40){\circle*{1}}

\put(22,40){\circle*{1}} \put(45,40){\circle*{1}}
\put(63,40){\circle*{1}} \put(78,40){\circle*{1}}

\put(02,40){\makebox(0,0)[bl]{\(M\)}}
\put(18,40){\makebox(0,0)[bl]{\(L\)}}
\put(47,40){\makebox(0,0)[bl]{\(V\)}}
\put(65,40){\makebox(0,0)[bl]{\(U\)}}
\put(80,40){\makebox(0,0)[bl]{\(T\)}}

\put(73,36){\makebox(0,0)[bl]{TRx}}
\put(74,32){\makebox(0,0)[bl]{\(T_{1}(3)\)}}
\put(74,28){\makebox(0,0)[bl]{\(T_{2}(1)\)}}
\put(74,24){\makebox(0,0)[bl]{\(T_{3}(2)\)}}
\put(74,20){\makebox(0,0)[bl]{\(T_{4}(2)\)}}
\put(74,16){\makebox(0,0)[bl]{\(T_{5}(1)\)}}

\put(59,36){\makebox(0,0)[bl]{BTS}}
\put(60,32){\makebox(0,0)[bl]{\(U_{1}(1)\)}}
\put(60,28){\makebox(0,0)[bl]{\(U_{2}(3)\)}}
\put(60,24){\makebox(0,0)[bl]{\(U_{3}(2)\)}}
\put(60,20){\makebox(0,0)[bl]{\(U_{4}(1)\)}}
\put(60,16){\makebox(0,0)[bl]{\(U_{5}(2)\)}}

\put(43,36){\makebox(0,0)[bl]{BSC}}
\put(44,32){\makebox(0,0)[bl]{\(V_{1}(1)\)}}
\put(44,28){\makebox(0,0)[bl]{\(V_{2}(2)\)}}
\put(44,24){\makebox(0,0)[bl]{\(V_{3}(3)\)}}
\put(44,20){\makebox(0,0)[bl]{\(V_{4}(2)\)}}
\put(44,16){\makebox(0,0)[bl]{\(V_{5}(1)\)}}
\put(44,12){\makebox(0,0)[bl]{\(V_{6}(3)\)}}

\put(17,35){\makebox(0,0)[bl]{HLR/}}
\put(17,32){\makebox(0,0)[bl]{AC}}
\put(18,28){\makebox(0,0)[bl]{\(L_{1}(1)\)}}
\put(18,24){\makebox(0,0)[bl]{\(L_{2}(1)\)}}
\put(18,20){\makebox(0,0)[bl]{\(L_{3}(2)\)}}
\put(18,16){\makebox(0,0)[bl]{\(L_{4}(1)\)}}

\put(00,35){\makebox(0,0)[bl]{MSC/ }}
\put(00,32){\makebox(0,0)[bl]{VLR}}
\put(00,28){\makebox(0,0)[bl]{\(M_{1}(2)\)}}
\put(00,24){\makebox(0,0)[bl]{\(M_{2}(3)\)}}
\put(00,20){\makebox(0,0)[bl]{\(M_{3}(3)\)}}
\put(00,16){\makebox(0,0)[bl]{\(M_{4}(1)\)}}
\put(00,12){\makebox(0,0)[bl]{\(M_{5}(2)\)}}

\end{picture}
\end{center}

\begin{center}
\begin{picture}(111,123)
\put(19,119){\makebox(0,0)[bl]{Table 20. Example for multiple
choice problem}}

\put(00,00){\line(1,0){110}} \put(00,103){\line(1,0){110}}
\put(00,117){\line(1,0){110}}

\put(00,0){\line(0,1){117}} \put(10,0){\line(0,1){117}}
\put(20,0){\line(0,1){117}} \put(35,0){\line(0,1){117}}
\put(55,0){\line(0,1){117}} \put(70,0){\line(0,1){117}}
\put(90,0){\line(0,1){117}} \put(110,0){\line(0,1){117}}


\put(01,113){\makebox(0,0)[bl]{No.}}
\put(01,109){\makebox(0,0)[bl]{\((i,j)\)}}

\put(11,113){\makebox(0,0)[bl]{DAs}}

\put(21,112.5){\makebox(0,0)[bl]{Priority}}
\put(26.5,108.5){\makebox(0,0)[bl]{\(r\)}}

\put(36,113){\makebox(0,0)[bl]{Resource }}
\put(36,109){\makebox(0,0)[bl]{requirement}}
\put(42,105){\makebox(0,0)[bl]{\(a_{i,j}\)}}

\put(57,112.5){\makebox(0,0)[bl]{\(c_{i,j}/a_{i,j}\)}}

\put(71,113){\makebox(0,0)[bl]{Selection}}
\put(71,109){\makebox(0,0)[bl]{(constraint:}}
\put(71,105){\makebox(0,0)[bl]{ \( \leq 14\))}}

\put(91,113){\makebox(0,0)[bl]{Selection}}
\put(91,109){\makebox(0,0)[bl]{(constraint:}}
\put(91,105){\makebox(0,0)[bl]{ \( \leq 15\))}}


\put(01,98){\makebox(0,0)[bl]{\((1,1)\)}}
\put(01,94){\makebox(0,0)[bl]{\((1,2)\)}}
\put(01,90){\makebox(0,0)[bl]{\((1,3)\)}}
\put(01,86){\makebox(0,0)[bl]{\((1,4)\)}}
\put(01,82){\makebox(0,0)[bl]{\((1,5)\)}}
\put(01,78){\makebox(0,0)[bl]{\((2,1)\)}}
\put(01,74){\makebox(0,0)[bl]{\((2,2)\)}}
\put(01,70){\makebox(0,0)[bl]{\((2,3)\)}}
\put(01,66){\makebox(0,0)[bl]{\((2,4)\)}}
\put(01,62){\makebox(0,0)[bl]{\((3,1)\)}}
\put(01,58){\makebox(0,0)[bl]{\((3,2)\)}}
\put(01,54){\makebox(0,0)[bl]{\((3,3)\)}}
\put(01,50){\makebox(0,0)[bl]{\((3,4)\)}}
\put(01,46){\makebox(0,0)[bl]{\((3,5)\)}}
\put(01,42){\makebox(0,0)[bl]{\((3,6)\)}}
\put(01,38){\makebox(0,0)[bl]{\((4,1)\)}}
\put(01,34){\makebox(0,0)[bl]{\((4,2)\)}}
\put(01,30){\makebox(0,0)[bl]{\((4,3)\)}}
\put(01,26){\makebox(0,0)[bl]{\((4,4)\)}}
\put(01,22){\makebox(0,0)[bl]{\((4,5)\)}}
\put(01,18){\makebox(0,0)[bl]{\((5,1)\)}}
\put(01,14){\makebox(0,0)[bl]{\((5,2)\)}}
\put(01,10){\makebox(0,0)[bl]{\((5,3)\)}}
\put(01,06){\makebox(0,0)[bl]{\((5,4)\)}}
\put(01,02){\makebox(0,0)[bl]{\((5,5)\)}}

\put(12,98){\makebox(0,0)[bl]{\(M_{1}\)}}
\put(12,94){\makebox(0,0)[bl]{\(M_{2}\)}}
\put(12,90){\makebox(0,0)[bl]{\(M_{3}\)}}
\put(12,86){\makebox(0,0)[bl]{\(M_{4}\)}}
\put(12,82){\makebox(0,0)[bl]{\(M_{5}\)}}
\put(12,78){\makebox(0,0)[bl]{\(L_{1}\)}}
\put(12,74){\makebox(0,0)[bl]{\(L_{2}\)}}
\put(12,70){\makebox(0,0)[bl]{\(L_{3}\)}}
\put(12,66){\makebox(0,0)[bl]{\(L_{4}\)}}
\put(12,62){\makebox(0,0)[bl]{\(V_{1}\)}}
\put(12,58){\makebox(0,0)[bl]{\(V_{2}\)}}
\put(12,54){\makebox(0,0)[bl]{\(V_{3}\)}}
\put(12,50){\makebox(0,0)[bl]{\(V_{4}\)}}
\put(12,46){\makebox(0,0)[bl]{\(V_{5}\)}}
\put(12,42){\makebox(0,0)[bl]{\(V_{6}\)}}
\put(12,38){\makebox(0,0)[bl]{\(U_{1}\)}}
\put(12,34){\makebox(0,0)[bl]{\(U_{2}\)}}
\put(12,30){\makebox(0,0)[bl]{\(U_{3}\)}}
\put(12,26){\makebox(0,0)[bl]{\(U_{4}\)}}
\put(12,22){\makebox(0,0)[bl]{\(U_{5}\)}}
\put(12,18){\makebox(0,0)[bl]{\(T_{1}\)}}
\put(12,14){\makebox(0,0)[bl]{\(T_{2}\)}}
\put(12,10){\makebox(0,0)[bl]{\(T_{3}\)}}
\put(12,06){\makebox(0,0)[bl]{\(T_{4}\)}}
\put(12,02){\makebox(0,0)[bl]{\(T_{5}\)}}

\put(26,98){\makebox(0,0)[bl]{\(2\)}}
\put(42,98){\makebox(0,0)[bl]{\(5.0\)}}
\put(59,98){\makebox(0,0)[bl]{\(0.4\)}}
\put(78,98){\makebox(0,0)[bl]{No}}
\put(98,98){\makebox(0,0)[bl]{No}}

\put(26,94){\makebox(0,0)[bl]{\(3\)}}
\put(42,94){\makebox(0,0)[bl]{\(3.0\)}}
\put(59,94){\makebox(0,0)[bl]{\(0.33\)}}
\put(78,94){\makebox(0,0)[bl]{No}}
\put(98,94){\makebox(0,0)[bl]{No}}

\put(26,90){\makebox(0,0)[bl]{\(3\)}}
\put(42,90){\makebox(0,0)[bl]{\(2.0\)}}
\put(59,90){\makebox(0,0)[bl]{\(0.5\)}}
\put(78,90){\makebox(0,0)[bl]{No}}
\put(98,90){\makebox(0,0)[bl]{No}}

\put(26,86){\makebox(0,0)[bl]{\(1\)}}
\put(42,86){\makebox(0,0)[bl]{\(6.0\)}}
\put(59,86){\makebox(0,0)[bl]{\(0.5\)}}
\put(78,86){\makebox(0,0)[bl]{Yes}}
\put(98,86){\makebox(0,0)[bl]{Yes}}

\put(26,82){\makebox(0,0)[bl]{\(2\)}}
\put(42,82){\makebox(0,0)[bl]{\(4.8\)}}
\put(59,82){\makebox(0,0)[bl]{\(0.38\)}}
\put(78,82){\makebox(0,0)[bl]{No}}
\put(98,82){\makebox(0,0)[bl]{No}}

\put(26,78){\makebox(0,0)[bl]{\(1\)}}
\put(42,78){\makebox(0,0)[bl]{\(1.0\)}}
\put(59,78){\makebox(0,0)[bl]{\(3.0\)}}
\put(78,78){\makebox(0,0)[bl]{Yes}}
\put(98,78){\makebox(0,0)[bl]{Yes}}

\put(26,74){\makebox(0,0)[bl]{\(1\)}}
\put(42,74){\makebox(0,0)[bl]{\(1.0\)}}
\put(59,74){\makebox(0,0)[bl]{\(3.0\)}}
\put(78,74){\makebox(0,0)[bl]{No}}
\put(98,74){\makebox(0,0)[bl]{No}}

\put(26,70){\makebox(0,0)[bl]{\(2\)}}
\put(42,70){\makebox(0,0)[bl]{\(1.0\)}}
\put(59,70){\makebox(0,0)[bl]{\(2.0\)}}
\put(78,70){\makebox(0,0)[bl]{No}}
\put(98,70){\makebox(0,0)[bl]{No}}

\put(26,66){\makebox(0,0)[bl]{\(1\)}}
\put(42,66){\makebox(0,0)[bl]{\(1.0\)}}
\put(59,66){\makebox(0,0)[bl]{\(3.0\)}}
\put(78,66){\makebox(0,0)[bl]{No}}
\put(98,66){\makebox(0,0)[bl]{No}}

\put(26,62){\makebox(0,0)[bl]{\(1\)}}
\put(42,62){\makebox(0,0)[bl]{\(5.0\)}}
\put(59,62){\makebox(0,0)[bl]{\(0.6\)}}
\put(78,62){\makebox(0,0)[bl]{No}}
\put(98,62){\makebox(0,0)[bl]{No}}

\put(26,58){\makebox(0,0)[bl]{\(2\)}}
\put(42,58){\makebox(0,0)[bl]{\(4.0\)}}
\put(59,58){\makebox(0,0)[bl]{\(0.5\)}}
\put(78,58){\makebox(0,0)[bl]{No}}
\put(98,58){\makebox(0,0)[bl]{No}}

\put(26,54){\makebox(0,0)[bl]{\(3\)}}
\put(42,54){\makebox(0,0)[bl]{\(2.0\)}}
\put(59,54){\makebox(0,0)[bl]{\(0.5\)}}
\put(78,54){\makebox(0,0)[bl]{No}}
\put(98,54){\makebox(0,0)[bl]{No}}

\put(26,50){\makebox(0,0)[bl]{\(2\)}}
\put(42,50){\makebox(0,0)[bl]{\(4.0\)}}
\put(59,50){\makebox(0,0)[bl]{\(0.5\)}}
\put(78,50){\makebox(0,0)[bl]{No}}
\put(98,50){\makebox(0,0)[bl]{No}}

\put(26,46){\makebox(0,0)[bl]{\(1\)}}
\put(42,46){\makebox(0,0)[bl]{\(5.0\)}}
\put(59,46){\makebox(0,0)[bl]{\(0.6\)}}
\put(78,46){\makebox(0,0)[bl]{No}}
\put(98,46){\makebox(0,0)[bl]{No}}

\put(26,42){\makebox(0,0)[bl]{\(3\)}}
\put(42,42){\makebox(0,0)[bl]{\(1.0\)}}
\put(59,42){\makebox(0,0)[bl]{\(1.0\)}}
\put(78,42){\makebox(0,0)[bl]{Yes}}
\put(98,42){\makebox(0,0)[bl]{Yes}}

\put(26,38){\makebox(0,0)[bl]{\(1\)}}
\put(42,38){\makebox(0,0)[bl]{\(6.0\)}}
\put(59,38){\makebox(0,0)[bl]{\(0.5\)}}
\put(78,38){\makebox(0,0)[bl]{No}}
\put(98,38){\makebox(0,0)[bl]{Yes}}

\put(26,34){\makebox(0,0)[bl]{\(3\)}}
\put(42,34){\makebox(0,0)[bl]{\(5.0\)}}
\put(59,34){\makebox(0,0)[bl]{\(0.2\)}}
\put(78,34){\makebox(0,0)[bl]{No}}
\put(98,34){\makebox(0,0)[bl]{No}}

\put(26,30){\makebox(0,0)[bl]{\(2\)}}
\put(42,30){\makebox(0,0)[bl]{\(5.0\)}}
\put(59,30){\makebox(0,0)[bl]{\(0.4\)}}
\put(78,30){\makebox(0,0)[bl]{Yes}}
\put(98,30){\makebox(0,0)[bl]{No}}

\put(26,26){\makebox(0,0)[bl]{\(3\)}}
\put(42,26){\makebox(0,0)[bl]{\(8.0\)}}
\put(59,26){\makebox(0,0)[bl]{\(0.39\)}}
\put(78,26){\makebox(0,0)[bl]{No}}
\put(98,26){\makebox(0,0)[bl]{No}}

\put(26,22){\makebox(0,0)[bl]{\(2\)}}
\put(42,22){\makebox(0,0)[bl]{\(5.0\)}}
\put(59,22){\makebox(0,0)[bl]{\(0.4\)}}
\put(78,22){\makebox(0,0)[bl]{No}}
\put(98,22){\makebox(0,0)[bl]{No}}

\put(26,18){\makebox(0,0)[bl]{\(3\)}}
\put(42,18){\makebox(0,0)[bl]{\(1.0\)}}
\put(59,18){\makebox(0,0)[bl]{\(1.0\)}}
\put(78,18){\makebox(0,0)[bl]{Yes}}
\put(98,18){\makebox(0,0)[bl]{Yes}}

\put(26,14){\makebox(0,0)[bl]{\(1\)}}
\put(42,14){\makebox(0,0)[bl]{\(8.0\)}}
\put(59,14){\makebox(0,0)[bl]{\(0.39\)}}
\put(78,14){\makebox(0,0)[bl]{No}}
\put(98,14){\makebox(0,0)[bl]{No}}

\put(26,10){\makebox(0,0)[bl]{\(2\)}}
\put(42,10){\makebox(0,0)[bl]{\(4.0\)}}
\put(59,10){\makebox(0,0)[bl]{\(0.5\)}}
\put(78,10){\makebox(0,0)[bl]{No}}
\put(98,10){\makebox(0,0)[bl]{No}}

\put(26,06){\makebox(0,0)[bl]{\(2\)}}
\put(42,06){\makebox(0,0)[bl]{\(3.0\)}}
\put(59,06){\makebox(0,0)[bl]{\(0.66\)}}
\put(78,06){\makebox(0,0)[bl]{No}}
\put(98,06){\makebox(0,0)[bl]{No}}

\put(26,02){\makebox(0,0)[bl]{\(1\)}}
\put(42,02){\makebox(0,0)[bl]{\(7.0\)}}
\put(59,02){\makebox(0,0)[bl]{\(0.42\)}}
\put(78,02){\makebox(0,0)[bl]{No}}
\put(98,02){\makebox(0,0)[bl]{No}}

\end{picture}
\end{center}

 Thus,
 the following simplified one-objective problem is considered:
 \[\max\sum_{i=1}^{5} \sum_{j=1}^{q_{i}} c_{ij} x_{ij} ~~~~
 s.t.~\sum_{i=1}^{5} \sum_{j=1}^{q_{i}} a_{ij} x_{ij} \leq
 b,
 ~~\sum_{j=1}^{q_{i}} x_{ij}=1 ~\forall i=\overline{1,5},  ~~x_{ij} \in \{0,1\},\]
 where \(q_{1} = 5\), \(q_{2} = 4\), \(q_{3} = 6\), \(q_{4} = 5\), \(q_{5} = 5\).
 After the usage of the greedy heuristic,
 the following composite DAs are obtained (Table 20):

 (1) resource constraint \(b=14\):~
  \( S^{C}_{1} = M_{4} \star L_{1} \star V_{6} \star U_{3} \star T_{1}\),

  (2) resource constraint \(b=15\):~
  \( S^{C}_{2} = M_{4} \star L_{1} \star V_{6} \star U_{1} \star
  T_{1}\).

\subsection{Hierarchical Morphological Design}

 A preliminary example for HMMD was presented in \cite{levvis07}
 (Fig. 13).
 For system part \(A\),
 we get the following Pareto-efficient composite DAs:
 (1) ~\(A_{1} = M_{4} \star L_{2} \), ~\(N(A_{1})= (3;2,0,0)\);
 (2) ~\(A_{2} = M_{4} \star L_{4} \), ~\(N(A_{2})= (3;2,0,0)\).
%
%
%
%
 For system part \(B\),
  we get the following Pareto-efficient  composite DAs:
 (1) ~\(B_{1} = V_{5} \star U_{1} \star T_{5}\), ~\(N(B_{1})=
 (2;3,0,0)\);
 (2) ~\(B_{2} = V_{5} \star U_{4} \star T_{2}\), ~\(N(B_{2})=
 (2;3,0,0)\);
 (3) ~\(B_{3} = V_{1} \star U_{5} \star T_{1}\), ~\(N(B_{3})=
 (3;1,1,1)\), and
 (4) ~\(B_{4} = V_{2} \star U_{3} \star T_{4}\), ~\(N(B_{4})=
 (3;0,3,0)\).
 Fig. 14 illustrates system quality for \(B\).


\begin{center}
\begin{picture}(82,108)

\put(0,04){\makebox(0,0)[bl] {Fig. 13. Designed GSM network
(priorities of DAs}}

\put(14,0){\makebox(0,0)[bl]{are shown in parentheses)}}

\put(00,105){\circle*{3}}

\put(05,104){\makebox(0,0)[bl]{~\(S=A\star B = (M\star L)\star
(V\star U\star T)\)}}

\put(03,100){\makebox(0,0)[bl] {\(S_{1}=A_{1}\star
B_{1}=(M_{4}\star L_{2})\star (V_{5}\star U_{1}\star T_{5})\)}}

\put(03,96){\makebox(0,0)[bl] {\(S_{2}=A_{1}\star
B_{2}=(M_{4}\star L_{2})\star (V_{5}\star U_{4}\star T_{2})\)}}

\put(03,92){\makebox(0,0)[bl] {\(S_{3}=A_{1}\star
B_{3}=(M_{4}\star L_{2})\star (V_{1}\star U_{5}\star T_{1})\)}}

\put(03,88){\makebox(0,0)[bl] {\(S_{4}=A_{2}\star
B_{1}=(M_{4}\star L_{4})\star (V_{5}\star U_{1}\star T_{5})\)}}

\put(03,84){\makebox(0,0)[bl] {\(S_{5}=A_{2}\star
B_{2}=(M_{4}\star L_{4})\star (V_{5}\star U_{4}\star T_{2})\)}}

\put(03,80){\makebox(0,0)[bl] {\(S_{6}=A_{2}\star
B_{3}=(M_{4}\star L_{4})\star (V_{1}\star U_{5}\star T_{1})\)}}

\put(03,76){\makebox(0,0)[bl] {\(S_{7}=A_{1}\star
B_{4}=(M_{4}\star L_{2})\star (V_{2}\star U_{3}\star T_{4})\)}}

\put(03,72){\makebox(0,0)[bl] {\(S_{8}=A_{2}\star
B_{4}=(M_{4}\star L_{4})\star (V_{2}\star U_{3}\star T_{4})\)}}

\put(02,67){\makebox(0,0)[bl]{SSS~ \(A = M\star L\)}}

\put(47,67){\makebox(0,0)[bl]{BSS~ \(B = V \star U \star T\)}}

\put(00,65){\line(0,1){32}} \put(00,65){\line(1,0){45}}
\put(45,60){\line(0,1){05}} \put(45,60){\circle*{2}}

\put(47,62){\makebox(0,0)[bl]{\(B_{1} = V_{5}\star U_{1}\star
T_{5}\)}}

\put(47,58){\makebox(0,0)[bl]{\(B_{2} = V_{5}\star U_{4}\star
T_{2}\)}}

\put(47,54){\makebox(0,0)[bl]{\(B_{3}=V_{1}\star U_{5}\star
T_{1}\)}}

\put(47,50){\makebox(0,0)[bl]{\(B_{4} = V_{2}\star U_{3}\star
 T_{4}\)}}

\put(45,40){\line(0,1){20}}

\put(00,60){\line(0,1){05}} \put(00,60){\circle*{2}}
\put(00,49){\line(0,1){55}}

\put(02,60){\makebox(0,0)[bl]{\(A_{1} = M_{4} \star L_{2}\)}}
\put(02,56){\makebox(0,0)[bl]{\(A_{2} = M_{4} \star L_{4}\)}}

\put(00,45){\line(1,0){22}} \put(45,45){\line(1,0){33}}
\put(00,40){\line(0,1){10}} \put(22,40){\line(0,1){05}}
\put(45,40){\line(0,1){05}} \put(63,40){\line(0,1){05}}
\put(78,40){\line(0,1){05}} \put(00,40){\circle*{1}}

\put(22,40){\circle*{1}} \put(45,40){\circle*{1}}
\put(63,40){\circle*{1}} \put(78,40){\circle*{1}}

\put(02,40){\makebox(0,0)[bl]{\(M\)}}
\put(18,40){\makebox(0,0)[bl]{\(L\)}}
\put(47,40){\makebox(0,0)[bl]{\(V\)}}
\put(65,40){\makebox(0,0)[bl]{\(U\)}}
\put(80,40){\makebox(0,0)[bl]{\(T\)}}

\put(73,36){\makebox(0,0)[bl]{TRx}}
\put(74,32){\makebox(0,0)[bl]{\(T_{1}(3)\)}}
\put(74,28){\makebox(0,0)[bl]{\(T_{2}(1)\)}}
\put(74,24){\makebox(0,0)[bl]{\(T_{3}(2)\)}}
\put(74,20){\makebox(0,0)[bl]{\(T_{4}(2)\)}}
\put(74,16){\makebox(0,0)[bl]{\(T_{5}(1)\)}}

\put(59,36){\makebox(0,0)[bl]{BTS}}
\put(60,32){\makebox(0,0)[bl]{\(U_{1}(1)\)}}
\put(60,28){\makebox(0,0)[bl]{\(U_{2}(3)\)}}
\put(60,24){\makebox(0,0)[bl]{\(U_{3}(2)\)}}
\put(60,20){\makebox(0,0)[bl]{\(U_{4}(1)\)}}
\put(60,16){\makebox(0,0)[bl]{\(U_{5}(2)\)}}

\put(43,36){\makebox(0,0)[bl]{BSC}}
\put(44,32){\makebox(0,0)[bl]{\(V_{1}(1)\)}}
\put(44,28){\makebox(0,0)[bl]{\(V_{2}(2)\)}}
\put(44,24){\makebox(0,0)[bl]{\(V_{3}(3)\)}}
\put(44,20){\makebox(0,0)[bl]{\(V_{4}(2)\)}}
\put(44,16){\makebox(0,0)[bl]{\(V_{5}(1)\)}}
\put(44,12){\makebox(0,0)[bl]{\(V_{6}(3)\)}}

\put(17,35){\makebox(0,0)[bl]{HLR/}}
\put(17,32){\makebox(0,0)[bl]{AC}}
\put(18,28){\makebox(0,0)[bl]{\(L_{1}(1)\)}}
\put(18,24){\makebox(0,0)[bl]{\(L_{2}(1)\)}}
\put(18,20){\makebox(0,0)[bl]{\(L_{3}(2)\)}}
\put(18,16){\makebox(0,0)[bl]{\(L_{4}(1)\)}}

\put(00,35){\makebox(0,0)[bl]{MSC/ }}
\put(00,32){\makebox(0,0)[bl]{VLR}}
\put(00,28){\makebox(0,0)[bl]{\(M_{1}(2)\)}}
\put(00,24){\makebox(0,0)[bl]{\(M_{2}(3)\)}}
\put(00,20){\makebox(0,0)[bl]{\(M_{3}(3)\)}}
\put(00,16){\makebox(0,0)[bl]{\(M_{4}(1)\)}}
\put(00,12){\makebox(0,0)[bl]{\(M_{5}(2)\)}}

\end{picture}
\end{center}

\begin{center}
\begin{picture}(80,62)
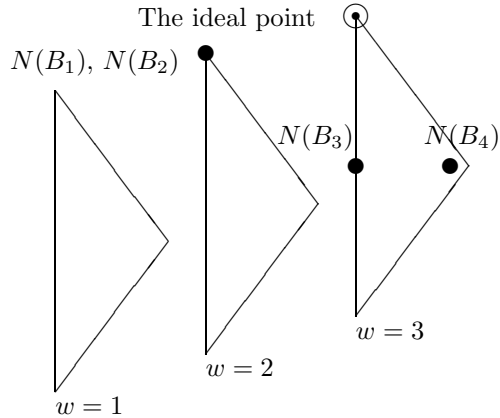
\put(10,0){\makebox(0,0)[bl]{Fig. 14. Space of system quality for
\(B\)}}



\put(10,010){\line(0,1){40}} \put(10,010){\line(3,4){15}}
\put(10,050){\line(3,-4){15}}

\put(30,015){\line(0,1){40}} \put(30,015){\line(3,4){15}}
\put(30,055){\line(3,-4){15}}

\put(50,020){\line(0,1){40}} \put(50,020){\line(3,4){15}}
\put(50,060){\line(3,-4){15}}

\put(30,55){\circle*{2}}
\put(4,52){\makebox(0,0)[bl]{\(N(B_{1})\), \(N(B_{2})\)}}

\put(50,40){\circle*{2}}
\put(39.5,42){\makebox(0,0)[bl]{\(N(B_{3})\)}}

\put(62.5,40){\circle*{2}}
\put(59,42){\makebox(0,0)[bl]{\(N(B_{4})\)}}

\put(50,60){\circle*{1}} \put(50,60){\circle{3}}

\put(21,58){\makebox(0,0)[bl]{The ideal point}}

\put(10,7){\makebox(0,0)[bl]{\(w=1\)}}
\put(30,12){\makebox(0,0)[bl]{\(w=2\)}}
\put(50,17){\makebox(0,0)[bl]{\(w=3\)}}
\end{picture}
\end{center}

 Now it is possible to combine
 the resultant composite DAs as follows (Fig. 13):

 (1) \(S^{H}_{1} = A_{1} \star B_{1}=(M_{4}\star L_{2})\star (V_{5}\star
U_{1}\star T_{5})\);
 (2) \(S^{H}_{2} = A_{1} \star B_{2}=(M_{4}\star L_{2})\star (V_{5}\star
U_{4}\star T_{2})\);

 (3) \(S^{H}_{3} = A_{1} \star B_{3}=(M_{4}\star L_{2})\star (V_{1}\star
U_{5}\star T_{1})\);
 (4) \(S^{H}_{4} = A_{2} \star B_{1}=(M_{4}\star L_{4})\star (V_{5}\star
U_{1}\star T_{5})\);

 (5) \(S^{H}_{5} = A_{2} \star B_{2}=(M_{4}\star L_{4})\star (V_{5}\star
U_{4}\star T_{2})\);
 (6) \(S^{H}_{6} = A_{2} \star B_{3}=(M_{4}\star L_{4})\star (V_{1}\star
 U_{5}\star T_{1})\);

 (7) \(S^{H}_{7} = A_{1} \star B_{3}=(M_{4}\star L_{2})\star (V_{2}\star
U_{3}\star T_{4})\); and
 (8) \(S^{H}_{8} = A_{2} \star B_{3}=(M_{4}\star L_{4})\star
(V_{2}\star U_{3}\star T_{4})\).

 Finally,
 it is reasonable to integrate quality vectors for components
 \(A\) and \(B\) to obtain the following
 quality vectors:~
 \(N(S^{H}_{1}) = (2;5,0,0)\),
 \(N(S^{H}_{2}) = (2;5,0,0)\),
 \(N(S^{H}_{3}) = (3;3,1,1)\),
 \(N(S^{H}_{4}) = (2;5,0,0)\),
 \(N(S^{H}_{5}) = (3;3,1,1)\), and
 \(N(S^{H}_{6}) = (3;3,1,1)\).
 \(N(S^{H}_{7}) = (3;2,3,0)\), and
 \(N(S^{H}_{8}) = (3;2,3,0)\).
 Further the obtained eight resultant composite decisions can be analyzed
 to select the best decision
 (e.g., additional multicriteria analysis, expert judgment).

\subsection{Brief Comparison and Discussion of Methods}

 Note,
 45 resultant solutions were obtained by basic MA.
 Table 21 integrates resultant composite solutions for four methods:
 (1) ideal point method (the best solution and six close solutions),
 (2) Pareto-based method (two solutions),
 (3) multiple choice problem (two solutions),
 (4) HMMD (eight solutions).
%

%
\begin{center}
\begin{picture}(114,101)
\put(20,97){\makebox(0,0)[bl]{Table 21. Integration of composite
solutions}}

\put(00,0){\line(1,0){114}} \put(00,35){\line(1,0){114}}
\put(00,46){\line(1,0){114}} \put(00,57){\line(1,0){114}}
\put(00,88){\line(1,0){114}} \put(00,95){\line(1,0){114}}

\put(00,0){\line(0,1){95}} \put(30,00){\line(0,1){95}}
\put(75,00){\line(0,1){95}} \put(114,00){\line(0,1){95}}


\put(01,90){\makebox(0,0)[bl]{Method}}
\put(31,90){\makebox(0,0)[bl]{Resultant composite DAs}}
\put(76,90){\makebox(0,0)[bl]{Quality vector (HMMD)}}


\put(01,83){\makebox(0,0)[bl]{1.Ideal-point}}
\put(04,79){\makebox(0,0)[bl]{method}}

\put(031,83){\makebox(0,0)[bl]{\(S^{I}_{0} = M_{4}\star L_{2}\star
V_{2}\star U_{3}\star T_{4}\)}}

\put(031,79){\makebox(0,0)[bl]{\(S^{I}_{1} = M_{1}\star L_{1}\star
V_{2}\star U_{3}\star T_{4}\)}}

\put(031,75){\makebox(0,0)[bl]{\(S^{I}_{2} = M_{1}\star L_{4}\star
V_{2}\star U_{3}\star T_{4}\)}}

\put(031,71){\makebox(0,0)[bl]{\(S^{I}_{3} = M_{4}\star L_{3}\star
V_{2}\star U_{3}\star T_{4}\)}}

\put(031,67){\makebox(0,0)[bl]{\(S^{I}_{4} = M_{5}\star L_{1}\star
V_{2}\star U_{3}\star T_{4}\)}}

\put(031,63){\makebox(0,0)[bl]{\(S^{I}_{5} = M_{5}\star L_{2}\star
V_{3}\star U_{2}\star T_{4}\)}}

\put(031,59){\makebox(0,0)[bl]{\(S^{I}_{6} = M_{5}\star L_{4}\star
V_{2}\star U_{3}\star T_{4}\)}}


\put(087,83){\makebox(0,0)[bl]{\((3;2,3,0)\)}}
\put(087,79){\makebox(0,0)[bl]{\((3;1,3,1)\)}}
\put(087,75){\makebox(0,0)[bl]{\((3;1,4,0)\)}}
\put(087,71){\makebox(0,0)[bl]{\((3;1,4,0)\)}}
\put(087,67){\makebox(0,0)[bl]{\((3;1,4,0)\)}}
\put(087,63){\makebox(0,0)[bl]{\((3;1,2,2)\)}}
\put(087,59){\makebox(0,0)[bl]{\((3;1,4,0)\)}}


\put(01,52){\makebox(0,0)[bl]{2.Pareto-based }}
\put(04,48){\makebox(0,0)[bl]{MA}}

\put(031,52){\makebox(0,0)[bl]{\(S^{P}_{1} = M_{4}\star L_{2}\star
V_{1}\star U_{5}\star T_{1}\)}}

\put(031,48){\makebox(0,0)[bl]{\(S^{P}_{2} = M_{4}\star L_{2}\star
V_{2}\star U_{3}\star T_{4}\)}}

\put(087,52){\makebox(0,0)[bl]{\((3;3,1,1)\)}}
\put(087,48){\makebox(0,0)[bl]{\((3;2,3,0)\)}}


\put(01,41){\makebox(0,0)[bl]{3.Multiple choice}}
\put(04,37){\makebox(0,0)[bl]{problem}}

\put(031,41){\makebox(0,0)[bl]{\(S^{C}_{1} = M_{4}\star L_{1}\star
V_{6}\star U_{3}\star T_{1}\)}}

\put(031,37){\makebox(0,0)[bl]{\(S^{C}_{2} = M_{4}\star L_{1}\star
V_{6}\star U_{1}\star T_{1}\)}}

\put(087,41){\makebox(0,0)[bl]{\((0;2,1,2)\)}}
\put(087,37){\makebox(0,0)[bl]{\((0;3,0,2)\)}}


\put(01,30){\makebox(0,0)[bl]{4.HMMD}}

\put(031,30){\makebox(0,0)[bl]{\(S^{H}_{1} = M_{4}\star L_{2}\star
V_{5}\star U_{1}\star T_{5}\)}}

\put(031,26){\makebox(0,0)[bl]{\(S^{H}_{2} = M_{4}\star L_{2}\star
V_{5}\star U_{4}\star T_{2}\)}}

\put(031,22){\makebox(0,0)[bl]{\(S^{H}_{3} = M_{4}\star L_{2}\star
V_{1}\star U_{5}\star T_{1}\)}}

\put(031,18){\makebox(0,0)[bl]{\(S^{H}_{4} = M_{4}\star L_{4}\star
V_{5}\star U_{1}\star T_{5}\)}}

\put(031,14){\makebox(0,0)[bl]{\(S^{H}_{5} = M_{4}\star L_{4}\star
V_{5}\star U_{4}\star T_{2}\)}}

\put(031,10){\makebox(0,0)[bl]{\(S^{H}_{6} = M_{4}\star L_{4}\star
V_{1}\star U_{5}\star T_{1}\)}}

\put(031,06){\makebox(0,0)[bl]{\(S^{H}_{7} = M_{4}\star L_{2}\star
V_{2}\star U_{3}\star T_{4}\)}}

\put(031,02){\makebox(0,0)[bl]{\(S^{H}_{8} = M_{4}\star L_{4}\star
V_{2}\star U_{3}\star T_{4}\)}}

\put(087,30){\makebox(0,0)[bl]{\((2;5,0,0)\)}}
\put(087,26){\makebox(0,0)[bl]{\((2;5,0,0)\)}}
\put(087,22){\makebox(0,0)[bl]{\((3;3,1,1)\)}}
\put(087,18){\makebox(0,0)[bl]{\((2;5,0,0)\)}}
\put(087,14){\makebox(0,0)[bl]{\((3;3,1,0)\)}}
\put(087,10){\makebox(0,0)[bl]{\((3;3,1,1)\)}}

\put(087,06){\makebox(0,0)[bl]{\((3;2,3,0)\)}}
\put(087,02){\makebox(0,0)[bl]{\((3;2,3,0)\)}}


\end{picture}
\end{center}

  Now let us discuss the obtained solutions (Table 21):

%
%

 {\it 1.} In the case of the first three methods
 (MA, ideal point method, and Pareto-based method),
 compatibility estimates
 at level \(3\) were used to combine solutions.
 Thus cardinality of the combinatorial space of admissible solutions
 was decreased (for the examples)
  and
 the resultant solution set does not involve solutions with
 compatibility estimates at level \(2\) and \(1\).
 In the other case, cardinality of the admissible solution set can be very high.
 High cardinality of the admissible solution set
 will lead to very high computational complexity
 (MA, ideal point method, Pareto-based method) and participation
 of domain expert(s) at the first method stage
 (i.e., generation of admissible solutions) will not be possible.

 {\it 2.} In the case of MA,
 a sufficiently large and rich set of admissible solutions was
 obtained: \(45\).
 Note, this solution set covers solutions sets
 for other methods (i.e., ideal-point method, Pareto-based method, HMMD).
 At the same time,
 the problem is:~
 {\it to analyze this large solution set}.

 {\it 3.} In the case of ideal point method, only solution \(S^{I}_{0}\)
 belongs to the set of Pareto-efficient solutions.
 The set of considered solutions
 \(\{ S^{I}_{1}, S^{I}_{2}, S^{I}_{3}, S^{I}_{4}, S^{I}_{5},  S^{I}_{6}
 \}\),
 which are close to the above-mentioned solution,
 is not sufficiently good by elements.
 At the same time, some good solutions are lost, for example:~
 \(S^{H}_{3}\),  \(S^{H}_{5}\),  \(S^{H}_{6}\), \(S^{H}_{8}\).

 {\it 4.} In the case of Pareto-based method,
 many good solutions are lost, for example:~
  \(S^{H}_{5}\),  \(S^{H}_{6}\),  \(S^{H}_{8}\), etc.

 {\it 5.} In the case of multiple choice problem,
 compatibility estimates are not examined.
 As a result,
 all obtained solutions are inadmissible.
 It can be reasonable to extend this kind of optimization models
 by additional logical constraints which will formalize the
 compatibility requirements.
 But it may lead to complicated models.

 {\it 6.} In the case of HMMD,
 the set of solutions is
 sufficiently rich and not very large at the same time
 (eight solutions).

 Table 22 contains an additional qualitative author's comparison of the examined
  methods.
 Here computational complexity is depended on
 enumerative computing and analysis of all admissible combinatorial solutions
  (i.e., admissible combinations).
  In the case of HMMD,
   the usage
  of hierarchical system structure decreases complexity of the
  computing process.
  In the case of Pareto-based MA,
  an analysis of Pareto-efficient solutions will required
  additional enumerative computing.
 Finally, column "Usefulness for expert(s)" (Table 22)
 corresponds to the following:
 (i) possibility to include the domain(s) expert(s) or/and decision
 maker(s) into the solving process
 (i.e., to include cognitive man-machine procedures into the design framework),
 (ii) understandability of
 the used design method
 to domain(s) expert(s) and/or decision maker(s).

\begin{center}
\begin{picture}(149,48)
\put(37,43){\makebox(0,0)[bl]{Table 22. Qualitative comparison of
used methods}}

\put(00,0){\line(1,0){149}} \put(00,27){\line(1,0){149}}
\put(00,41){\line(1,0){149}}

\put(00,0){\line(0,1){41}} \put(43,00){\line(0,1){41}}
\put(68,00){\line(0,1){41}} \put(90,00){\line(0,1){41}}
\put(131,0){\line(0,1){41}} \put(149,0){\line(0,1){41}}

\put(1,37){\makebox(0,0)[bl]{Method}}

\put(44,37){\makebox(0,0)[bl]{Computational}}
\put(44,33){\makebox(0,0)[bl]{complexity}}

\put(69,37){\makebox(0,0)[bl]{Taking into}}
\put(69,33){\makebox(0,0)[bl]{account }}
\put(69,29){\makebox(0,0)[bl]{compatibility}}

\put(91,37){\makebox(0,0)[bl]{Usefulness for selection}}
\put(91,33){\makebox(0,0)[bl]{of the best solutions}}

\put(132,37){\makebox(0,0)[bl]{Usefulness }}
\put(132,33){\makebox(0,0)[bl]{for }}
\put(132,29){\makebox(0,0)[bl]{expert(s)}}


\put(1,22){\makebox(0,0)[bl]{1.MA}}
\put(44,22){\makebox(0,0)[bl]{High}}
\put(69,22){\makebox(0,0)[bl]{Yes, binary}}
\put(91,22.5){\makebox(0,0)[bl]{Hard}}
\put(132,22.5){\makebox(0,0)[bl]{Hard}}


\put(1,18){\makebox(0,0)[bl]{2.Ideal-point method}}
\put(44,18){\makebox(0,0)[bl]{High}}
\put(69,18){\makebox(0,0)[bl]{Yes, binary}}
\put(91,18){\makebox(0,0)[bl]{Easy}}
\put(132,18.5){\makebox(0,0)[bl]{Good}}


\put(01,14){\makebox(0,0)[bl]{4.Pareto-based MA }}

\put(44,13.5){\makebox(0,0)[bl]{High}}

\put(69,14){\makebox(0,0)[bl]{Yes, binary}}

\put(91,14){\makebox(0,0)[bl]{Medium, analysis of }}
\put(91,10.5){\makebox(0,0)[bl]{Pareto-efficient solutions}}

\put(132,14.5){\makebox(0,0)[bl]{Good}}


\put(01,06){\makebox(0,0)[bl]{4.Multiple choice problem}}
\put(44,05.5){\makebox(0,0)[bl]{Low/Medium}}
\put(69,06){\makebox(0,0)[bl]{None}}
\put(91,06){\makebox(0,0)[bl]{Easy}}
\put(132,06.5){\makebox(0,0)[bl]{Medium}}


\put(01,02){\makebox(0,0)[bl]{5.HMMD}}
\put(44,01.5){\makebox(0,0)[bl]{Low/Medium}}
\put(69,02){\makebox(0,0)[bl]{Yes, ordinal }}
\put(91,02){\makebox(0,0)[bl]{Easy}}
\put(132,02.5){\makebox(0,0)[bl]{Good}}


\end{picture}
\end{center}

 Generally, the selection of the certain kind of the morphological method
 for a designed system has to be based on the following:
 (a) a type of the examined system class
 (structure, complexity of component interaction, etc.);
 (b) structure and complexity of the examined representative of
 the system class;
 (c) existence of an experienced design team;
 (d) possibility to implement some assessment procedures
 (for assessment of DAs and/or compatibility);
 (e) possibility to use computational recourses
  (e.g., computing environment, power software, computing personnel),
 and
 (f) possibility to use qualified domain(s) experts and/or decision makers.


\section{Towards Other Approaches}


 Generally,
 hierarchical design approaches are often based
 on a hierarchical model of the designed system
 and 'Bottom-Up' framework (Fig. 15).
%
%
%
 The list of some hierarchical design approaches,
 which are close to MA-based approaches and based on the framework
 above,
 is the following:
 (1) hierarchical design frameworks
   (e.g., \cite{kra79}, \cite{shupe87});
%
%
%
%
%
 (2) structural
 synthesis of technical systems
 based on MA, cluster analysis, and parametric optimization
       \cite{rakov96};
  (3) HTN (hierarchical task network) planning
  (e.g., \cite{erol96});
  (4) hierarchical decision making in design and manufacturing
   (e.g., \cite{bas89}, \cite{bas92}, \cite{ber87}, \cite{har92}, \cite{kup85});
 and
 (5) linguistic geometry approach
  (e.g., \cite{sti93b}).



 Here it is reasonable to point out some non-linear programming
 models which
 are targeted to modular system design as well.
 First, modular design of series and series-parallel information
 processing from the viewpoint of reliable software design
 while taking into account a total budget
 (i.e., multi-version software design)
 was investigated in  (\cite{ash92}, \cite{ash94}, \cite{ber93}).
 The authors
 suggested several generalizations of
 knapsack problem with non-linear objective function.
 Thus,
 the following kind of the optimization model for reliable modular software design
 can be examined (a basic case)
%
 \cite{ber93}:
 \[\max~\prod_{i=1}^{m} ~(1 -\prod_{j=1}^{q_{i}}  (1 - p_{ij} x_{ij} ) )
 ~~~~~s.t.~~\sum_{i=1}^{m} \sum_{j=1}^{q_{i}} d_{ij} x_{ij} \leq
 b,
 ~~\sum_{j=1}^{q_{i}} x_{ij}\geq 1 ~\forall i=\overline{1,m}, ~~x_{ij} \in \{0,1\},\]
 where \(p_{ij}\) is a reliability estimate of software module version
 \((i,j)\) (i.e., version \(j\) for module \(i\)),
 \(d_{ij}\) is a cost of software module version \((i,j)\).
 Fig. 16 illustrates the design problem above.
 Evidently, the obtained models are complicated ones and
 heuristics or enumerative techniques are used for the solving process
 (\cite{ash92}, \cite{ash94}, \cite{ber93}).
  In \cite{yam06}, the problems above are considered regarding
 the usage of multi-objective genetic algorithms.
 Second,
 design problems in chemical engineering
 systems require often
 examination of integer and continuous variables at the same time and, as a result,
 non-linear mixed integer programming models are formulated and used
  (e.g., \cite{flou95}, \cite{gros90}).

\begin{center}
\begin{picture}(127,74)
\put(35,0){\makebox(0,0)[bl] {Fig. 15. 'Bottom-Up' design scheme}}

\put(0,14){\line(1,0){28}} \put(0,5){\line(1,0){28}}
\put(0,5){\line(0,1){09}} \put(28,5){\line(0,1){09}}

\put(14,14){\vector(0,1){4}}

\put(1,10){\makebox(0,0)[bl]{Generation/}}
\put(1,06){\makebox(0,0)[bl]{definition of DAs}}


\put(26,15.5){\makebox(0,0)[bl]{.~.~.}}

\put(26,29.5){\makebox(0,0)[bl]{.~.~.}}

\put(0,27){\line(1,0){28}} \put(0,18){\line(1,0){28}}
\put(0,18){\line(0,1){9}} \put(28,18){\line(0,1){9}}

\put(14,27){\vector(1,1){5}}

\put(1,23){\makebox(0,0)[bl]{Selection of}}
\put(1,19){\makebox(0,0)[bl]{the best  DAs}}

\put(29,14){\line(1,0){28}} \put(29,5){\line(1,0){28}}
\put(29,5){\line(0,1){09}} \put(57,5){\line(0,1){09}}

\put(43,14){\vector(0,1){4}}

\put(30,10){\makebox(0,0)[bl]{Generation/}}
\put(30,06){\makebox(0,0)[bl]{definition of DAs}}


\put(29,27){\line(1,0){28}} \put(29,18){\line(1,0){28}}
\put(29,18){\line(0,1){9}} \put(57,18){\line(0,1){9}}

\put(43,27){\vector(-1,1){5}}

\put(29.5,23){\makebox(0,0)[bl]{Selection of}}
\put(29.5,19){\makebox(0,0)[bl]{the best DAs}}


\put(07.5,41){\line(1,0){43}} \put(07.5,32){\line(1,0){43}}
\put(07.5,32){\line(0,1){09}} \put(50.5,32){\line(0,1){09}}

\put(08.2,32){\line(0,1){09}} \put(49.8,32){\line(0,1){09}}

\put(28.5,41){\vector(4,3){6}}

\put(09,37){\makebox(0,0)[bl]{Composition/combination}}
\put(09,33.5){\makebox(0,0)[bl]{of the best DAs}}


\put(70,14){\line(1,0){28}} \put(70,5){\line(1,0){28}}
\put(70,5){\line(0,1){09}} \put(98,5){\line(0,1){09}}

\put(84,14){\vector(0,1){4}}

\put(71,10){\makebox(0,0)[bl]{Generation/}}
\put(71,06){\makebox(0,0)[bl]{definition of DAs}}


\put(96,15.5){\makebox(0,0)[bl]{.~.~.}}

\put(96,29.5){\makebox(0,0)[bl]{.~.~.}}

\put(70,27){\line(1,0){28}} \put(70,18){\line(1,0){28}}
\put(70,18){\line(0,1){9}} \put(98,18){\line(0,1){9}}

\put(84,27){\vector(1,1){5}}

\put(71,23){\makebox(0,0)[bl]{Selection of}}
\put(71,19){\makebox(0,0)[bl]{the best  DAs}}

\put(99,14){\line(1,0){28}} \put(99,5){\line(1,0){28}}
\put(99,5){\line(0,1){09}} \put(127,5){\line(0,1){09}}

\put(113,14){\vector(0,1){4}}
\put(100,10){\makebox(0,0)[bl]{Generation/}}
\put(100,06){\makebox(0,0)[bl]{definition of DAs}}


\put(99,27){\line(1,0){28}} \put(99,18){\line(1,0){28}}
\put(99,18){\line(0,1){9}} \put(127,18){\line(0,1){9}}

\put(113,27){\vector(-1,1){5}}

\put(99.5,23){\makebox(0,0)[bl]{Selection of}}
\put(99.5,19){\makebox(0,0)[bl]{the best DAs}}


\put(77.5,41){\line(1,0){43}} \put(77.5,32){\line(1,0){43}}
\put(77.5,32){\line(0,1){09}} \put(120.5,32){\line(0,1){09}}

\put(78.2,32){\line(0,1){09}} \put(119.8,32){\line(0,1){09}}

\put(98.5,41){\vector(-4,3){6}}

\put(79,37){\makebox(0,0)[bl]{Composition/combination}}
\put(79,33.5){\makebox(0,0)[bl]{of the best DAs}}

\put(43,50){\vector(2,1){8}} \put(82,50){\vector(-2,1){8}}
\put(62.5,50){\vector(0,1){4}}

\put(60,46){\makebox(0,0)[bl]{.~.~.}}

\put(40,73){\line(1,0){46}} \put(40,54){\line(1,0){46}}
\put(40,54){\line(0,1){19}} \put(86,54){\line(0,1){19}}

\put(41,72){\line(1,0){44}} \put(41,55){\line(1,0){44}}
\put(41,55){\line(0,1){17}} \put(85,55){\line(0,1){17}}

\put(41.5,68){\makebox(0,0)[bl]{1.Composition/combination}}
\put(44.5,64.5){\makebox(0,0)[bl]{of the best DAs}}
\put(41.5,60){\makebox(0,0)[bl]{2.Analysis of resultant}}
\put(44.5,56){\makebox(0,0)[bl]{composite DAs}}

\end{picture}
\end{center}

\begin{center}
\begin{picture}(131,36)

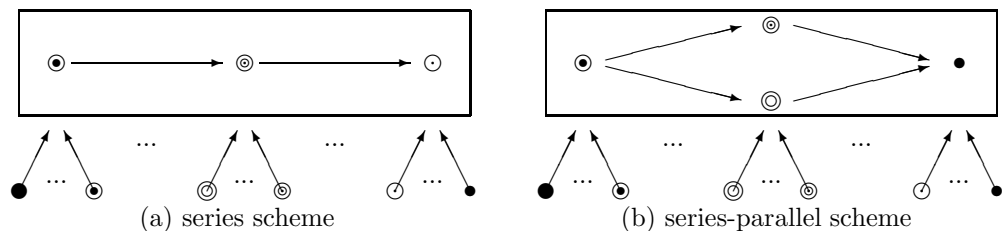
\put(20,00){\makebox(0,0)[bl] {Fig. 16. Modular design of series
or series/parallel system }}


\put(3.5,11){\makebox(0,8)[bl]{\(. . .\)}}

\put(0,10){\circle*{2}}

\put(10,10){\circle{2}} \put(10,10){\circle*{1}}

\put(00,10){\vector(1,2){4}} \put(10,10){\vector(-1,2){4}}


\put(15.5,16){\makebox(0,8)[bl]{\(. . .\)}}


\put(28.5,11){\makebox(0,8)[bl]{\(. . .\)}}

\put(25,10){\circle{2.5}} \put(25,10){\circle{1.4}}

\put(35,10){\circle{2.1}} \put(35,10){\circle{1.0}}
\put(35,10){\circle*{0.5}}

\put(25,10){\vector(1,2){4}} \put(35,10){\vector(-1,2){4}}


\put(40.5,16){\makebox(0,8)[bl]{\(. . .\)}}


\put(53.5,11){\makebox(0,8)[bl]{\(. . .\)}}

\put(50,10){\circle{2}} \put(50,10){\circle*{0.4}}

\put(60,10){\circle*{1.4}}

 \put(50,10){\vector(1,2){4}}
\put(60,10){\vector(-1,2){4}}


\put(00,20){\line(1,0){60}} \put(00,34){\line(1,0){60}}
\put(00,20){\line(0,1){14}} \put(60,20){\line(0,1){14}}


\put(5,27){\circle{2}} \put(05,27){\circle*{1}}
\put(7,27){\vector(1,0){20}}


\put(30,27){\circle{2.1}} \put(30,27){\circle{1.0}}
\put(30,27){\circle*{0.5}} \put(32,27){\vector(1,0){20}}

\put(55,27){\circle{2}} \put(55,27){\circle*{0.4}}


\put(16,4.5){\makebox(0,8)[bl]{(a) series scheme}}


\put(73.5,11){\makebox(0,8)[bl]{\(. . .\)}}

\put(70,10){\circle*{2}}

\put(80,10){\circle{2}} \put(80,10){\circle*{1}}

\put(70,10){\vector(1,2){4}} \put(80,10){\vector(-1,2){4}}


\put(85.5,16){\makebox(0,8)[bl]{\(. . .\)}}


\put(98.5,11){\makebox(0,8)[bl]{\(. . .\)}}

\put(95,10){\circle{2.5}} \put(95,10){\circle{1.4}}

\put(105,10){\circle{2.1}} \put(105,10){\circle{1.0}}
\put(105,10){\circle*{0.5}}

\put(95,10){\vector(1,2){4}} \put(105,10){\vector(-1,2){4}}


\put(110.5,16){\makebox(0,8)[bl]{\(. . .\)}}


\put(123.5,11){\makebox(0,8)[bl]{\(. . .\)}}

\put(120,10){\circle{2}} \put(120,10){\circle*{0.4}}

\put(130,10){\circle*{1.4}}

 \put(120,10){\vector(1,2){4}}
\put(130,10){\vector(-1,2){4}}


\put(70,20){\line(1,0){60}} \put(70,34){\line(1,0){60}}
\put(70,20){\line(0,1){14}} \put(130,20){\line(0,1){14}}


\put(75,27){\circle{2}} \put(75,27){\circle*{1}}

\put(78,27.4){\vector(4,1){18}} \put(78,26.6){\vector(4,-1){18}}


\put(100,32){\circle{2.1}} \put(100,32){\circle{1.0}}
\put(100,32){\circle*{0.5}}

\put(100,22){\circle{2.5}} \put(100,22){\circle{1.4}}

\put(103,32){\vector(4,-1){18}} \put(103,22){\vector(4,1){18}}



\put(125,27){\circle*{1.4}}


\put(80,4.5){\makebox(0,8)[bl]{(b) series-parallel scheme}}


\end{picture}
\end{center}


  In addition,
  it is reasonable to point out constraint-based approaches
 (e.g., \cite{fle98}, \cite{mail98},
 \cite{stum98})
  including
 composite constraint satisfaction problems
 and AI-based solving approaches
 (e.g.,
 \cite{sabin98}, \cite{stefik95}).


\section{Conclusions}

  In the article,  several
 MA-based system design approaches
 were described.
%
  Generally, it can be very useful and prospective to extend studies
  in the examination and usage of the MA based approaches in engineering,
  computer science, and management.
 For example,
 the following significant applied domains may be pointed out:
 (i) usage of morphological methods in allocation
 (layout, positioning)
 problems
 (e.g., \cite{lev98}, \cite{lev09});
 (ii) usage of morphological methods in combinatorial evolution and forecasting of modular
 systems (e.g., \cite{levdan05}, \cite{levkr09}).
  The future research directions can include the following:

 {\it 1.} continuation of the analysis, evaluation, comparison
 of MA-based system design methods;

 {\it 2.} consideration of uncertainty in all modifications of MA;

 {\it 3.} extension of "method of closeness to ideal point"
 while taking into account the following:
 (i) a set of ideal points, (ii) various kinds of proximity
 (e.g., functions, vector functions);

 {\it 4.} analysis, investigation, and modification
 of morphological methods based on
 multiple choice problem and its generalizations
 including special constraints for system elements compatibility;

 {\it 5.} design and investigation of special computer-aided systems based on
 morphological approaches;

 {\it 6.} investigation of special versions of morphological approaches
 which involve experts into a solving process (i.e., interactive approaches);

 {\it 7.} investigation of dynamical versions of morphological approaches
 while taking into account changes of system requirements;

%
%
%

 {\it 8.} usage of morphological system design methods
 for integration of heterogeneous networks;

  {\it 9.} usage of morphological system design methods
 in embedded systems for configuration and reconfiguration
 (including online mode)
 of hardware and/or software;

 {\it 10.} generation of engineering benchmarks
 for evaluation and analysis
 of MA-based system design methods;
 and

 {\it 11.}
  usage of MA and its modifications in
 engineering, IT/CS, and mathematical education
 (e.g., \cite{lev06a},
 \cite{lev09edu},
  \cite{lev10},
 \cite{levkhod07},
 \cite{levleus09},
  \cite{levshar09},
  \cite{levfim10},
 \cite{levsaf10}).

%
%
%


\end{document}